\documentclass[preprint,12pt]{elsarticle}
\usepackage{amssymb}
\usepackage{amsmath}
\usepackage{url} % MM: URL wrapping
\usepackage{makecell} % MM: Needed for wrapping "This survey" in the novelty table.

\journal{Information Systems}

\begin{document}

\begin{frontmatter}



\title{Data-Driven Prescriptive Analytics Applications: A Comprehensive Survey}


\author[1]{Martin Moesmann\corref{cor1}}
    \ead{mmoes@cs.aau.dk}

\author[1]{Torben Bach Pedersen}
    \ead{tbp@cs.aau.dk}

\affiliation[1]{organization={Department of Computer Science, Aalborg University},
                addressline={Selma Lagerløfs Vej 300}, 
                city={Aalborg Ø},
                postcode={DK-9220}, 
                country={Denmark}}
\cortext[cor1]{Corresponding author}

%% Abstract
\begin{abstract}
\textit{Prescriptive Analytics (PSA)}, an emerging business analytics field suggesting concrete options for solving business problems, has seen an increasing amount of interest after more than a decade of multidisciplinary research. This paper is a comprehensive survey of existing applications within PSA in terms of their use cases, methodologies, and possible future research directions. To ensure a manageable scope, we focus on PSA applications that develop data-driven, automatic workflows, i.e., \textit{Data-Driven PSA (DPSA)}. Following a systematic methodology, we identify and include 104 papers in our survey. As our key contributions, we derive a number of novel taxonomies of the field and use them to analyse the field’s temporal development. In terms of use cases, we derive \textit{10 application domains} for DPSA, from Healthcare to Manufacturing, and subsumed problem types within each. In terms of \textit{individual} method usage, we derive \textit{5 method types} and map them to a comprehensive taxonomy of method usage within DPSA applications, covering \textit{mathematical optimization}, \textit{data mining and machine learning}, \textit{probabilistic modelling}, \textit{domain expertise}, as well as \textit{simulations}. As for \textit{combined} method usage, we provide a statistical overview of how different method usage combinations are distributed and derive \textit{2 generic workflow patterns} along with subsumed workflow patterns, combining methods by either sequential or simultaneous relationships. Finally, we derive \textit{5 possible research directions} based on frequently recurring issues among surveyed papers, suggesting new frontiers in terms of methods, tools, and use cases.
\end{abstract}


% \begin{highlights}
% \end{highlights}

%% Keywords

\end{frontmatter}


\section{Introduction}\label{sc:introduction}

''The best way to predict the future is to create it'', as a quote commonly attributed to one of the founders of modern management theory says \cite{createthefuture}. Yet more than a willingness to act might be required for successful organizational decision making in a world marked by risk, uncertainty, and complexity \cite{knight1921risk}. Enter \textit{business analytics}, a field seeking to provide decision support in a variety of business domains through statistical analysis of historical data, predictive modelling, mathematical optimization, and more \cite{holsapple2014unified,davenport2017competing}. 

One of the current frontiers within business analytics is \textit{Prescriptive Analytics (PSA)}, which arrived as a term in the 2010s \cite{lustig2010analytics,maoz2013should}. Going beyond understanding the \textit{past} or predicting the \textit{future}, the purpose of PSA is to suggest concrete courses of actions in the \textit{present} \cite{siksnys2018prescriptive}. With PSA, proactive decision making is supported by combining, e.g., most archetypically, mathematical optimization with data-driven machine learning (ML) \cite{vsikvsnys2016solvedb}. 

Such workflows are especially useful within enterprises where success hinges on a harmonious marriage between \textit{prediction} and \textit{planning}. Such is the case in many seemingly unrelated business domains, from advertising to healthcare wherein, e.g., customer and patient treatment plans alike must be weighted by prognostic data \cite{lepenioti2020prescriptive}. From conversations with our own industrial contacts and collaborators within retail, energy trading, and beyond, we indeed find that many real-world companies are \textit{already} doing PSA, even though they might be unaware of the ''PSA'' term within their organization. To them, it goes without saying that one must of course, e.g., estimate future user demand in order to manage inventory logistics, or likewise, to make bids on the European electricity market a day before delivery. 

PSA also has seen an increasing level of interest among researchers in recent years \cite{lepenioti2020prescriptive}, especially due to the inherent technical challenges of combining, e.g., data-driven ML and mathematical optimization technologies \cite{frazzetto2019prescriptive}. Indeed, the 2025 ACM SIGMOD conference, the most important scientific conference within data management systems, now has PSA as a dedicated topic in their call for papers for the first time ever \cite{sigmod2024callforpapers}. With more than a decade of multidisciplinary research activity to comprehend, along with an increasing level of interest among researchers, there is currently a need for \textit{synthesis} and \textit{guidance} on how to progress PSA further as a field. Yet as our analysis in Section \ref{sec:related-work} shows, no \textit{comprehensive} survey on PSA applications, providing complete coverage of problem and solution types encountered across all application domains, exists as of yet. This crudely shaped research gap is arguably a curious predicament. As others have previously observed, applications indeed make up the vast majority of the PSA research field in terms of the number of papers published, compared to the number of papers discussing general theoretical or technical issues \cite{frazzetto2019prescriptive}.

The primary objective of this survey is to provide a comprehensive analysis of where the many applications of PSA have been, and where they are going. To ensure a manageable survey scope aligned with a precise, foundational definition of business analytics \cite{davenport2007competing}, we focus on a prominent subset of PSA that we call \textit{Data-Driven PSA} (DPSA, pronounced ''deepsa''). As our analysis in Section \ref{sc:scope-and-definition} shows, DPSA uniquely encompasses the fully realized \textit{data-driven}, \textit{automated} decision workflows envisioned by founding figures within business analytics. The pursuit of our primary objective of providing a comprehensive survey is decomposed into the following three research questions:

\begin{enumerate}[RQ1:]
    \item \textit{Which problems has DPSA previously been applied to?}
    \item \textit{Which methodological trends can be found in DPSA applications?}
    \item \textit{Which possible research directions can be derived from existing DPSA application research?}
\end{enumerate}

Both existing problems and solutions as well as promising unexplored directions will thus be addressed in this paper, utilizing a systematic survey methodology. We highlight the following key contributions, all derived from our set of 104 papers included in this survey:

\begin{itemize}
    \item We derive \textbf{10 application domains} and subsumed \textit{problem types} and analyse their temporal development, forming a comprehensive, updated overview of current DPSA use cases. 
    \item We derive \textbf{5 method types} along with a \textit{taxonomy} of subsumed methods and analyse their temporal development, forming a comprehensive, updated overview of methods found within existing DPSA applications.
    \item We derive \textbf{2 generic workflow patterns} and subsumed \textit{concrete workflow patterns}, and provide a statistical overview of different \textit{method type usage combinations}, including their temporal development, forming a comprehensive, updated overview of how DPSA application methodologies are structured. 
    \item Finally, we derive \textbf{5 possible research directions} reflecting the current state of the art, based on an analysis of frequently recurring issues and temporal trends within DPSA applications.
\end{itemize}

The remainder of this paper is structured as follows: First, we provide a systematic analysis of how the present survey marks a relevant scientific contribution compared to existing PSA surveys in Section \ref{sec:related-work}. Section \ref{sc:background} covers relevant background for readers unfamiliar with PSA. Then Section \ref{sc:scope-and-definition} provides a definition of DPSA, to delimit the scope of this survey. Section \ref{sc:method} moves on to present the chosen survey methodology, choices made as part of the survey process, along with a statistical overview of included papers. Section \ref{sc:rq1}, Sections \ref{sc:method-types}-\ref{sc:rq2-synthesis}, and Section \ref{sc:rq3} address each research question in turn, providing a comprehensive framework of problem and solution types found within existing application research, along with a number of possible future research directions. We provide concluding remarks in Section \ref{sc:conclusion}.

\section{Comparison to existing PSA surveys}\label{sec:related-work}
Several surveys about PSA have been published within recent years. An analysis of feature overlaps between existing surveys and this one can be found in Table \ref{tab:survey-overlap-table}, with papers sorted by publication year from left to right. 

\begin{table*}[ht]
  \resizebox{\linewidth}{!}{%
    \begin{tabular}{|l|c|c|c|c|c|c|c|c|c|c|c|c|c|c|}
    \hline
    \textbf{Feature / Paper}         & \cite{stefani2018constituent} & \cite{frazzetto2019prescriptive} & \cite{lepenioti2020prescriptive} & \cite{poornima2020survey} & \cite{raeesi2021prescriptive} & \cite{fox2022review} & \cite{kubrak2022prescriptive} & \cite{soeffker2022stochastic} & \cite{mishra2023prescriptive} & \cite{hall2024systematic} & \cite{mendoza2024prescriptive} & \cite{niederhaus2024technical}   & \cite{wissuchek2024prescriptive} & \makecell{\textbf{This} \\ \textbf{survey}} \\ \hline
    Papers from 2020s included       &                               &                                  &                           &                                  & \checkmark                    & \checkmark           & \checkmark                    & \checkmark                    & \checkmark                    & \checkmark                & \checkmark                     & \checkmark                       & \checkmark                       & \checkmark                                  \\ % \hline
    Applications the primary focus   &                               &                                  &                           & \checkmark                       & \checkmark                    & \checkmark           &                               & \checkmark                    & \checkmark                    & \checkmark                & \checkmark                     &                                  &                                  & \checkmark                                  \\ % \hline
    Covers more than one sector  & \checkmark                    & \checkmark                       & \checkmark                & \checkmark                       &                               &                      & \checkmark                    &                               & \checkmark                    &                           &                                & \checkmark                       & \checkmark                       & \checkmark                                  \\ % \hline
    Analysis of application domains  &                               &                                  & \checkmark                & (\checkmark)                     &                               &                      &                               &                               & (\checkmark)                  &                           &                                &                                  & (\checkmark)                     & \checkmark                                  \\ % \hline
    Analysis of problem types        &                               &                                  &                           & (\checkmark)                     & \checkmark                    & \checkmark           &                               & \checkmark                    & (\checkmark)                  &                           &                                &                                  & (\checkmark)                     & \checkmark                                  \\ % \hline
    Analysis of utilized methods     &                               & \checkmark                       & \checkmark                &                                  & \checkmark                    & \checkmark           & \checkmark                    & \checkmark                    &                               & \checkmark                & \checkmark                     & \checkmark                       & (\checkmark)                     & \checkmark                                  \\ % \hline
    Analysis of workflow patterns    &                               &                                  &                           &                                  &                               &                      &                               &                               &                               &                           &                                &                                  &                                  & \checkmark                                  \\ % \hline
    Analysis of temporal trends      &                               &                                  &                           &                                  &                               & \multicolumn{1}{l|}{}&                               &                               &                               &                           &                                &                                  &                                  & \checkmark                                  \\ \hline
    \end{tabular}%
    }
  \caption{This survey's overlaps with existing PSA surveys, with papers sorted by publication year from left to right. (\checkmark) denotes significantly limited fulfilment. \cite{wissuchek2023survey} and \cite{lepenioti2019prescriptive} are respectively older versions of \cite{wissuchek2024prescriptive} and \cite{lepenioti2020prescriptive}, and are therefore left out for the sake of brevity.}
  \label{tab:survey-overlap-table}
  \end{table*}

First, we note that a number of existing surveys don't cover any papers from this decade \cite{stefani2018constituent,frazzetto2019prescriptive,lepenioti2020prescriptive,poornima2020survey}. As our analysis in Section \ref{sc:method:statistics} shows, 76.9\% of our surveyed papers were published within the 2020-2024 year range, with the number of papers having accelerated significantly over the past three years. Looking solely at the recency of content, this survey therefore provides \textit{substantially updated knowledge}. 

Furthermore, only a few past surveys have had a primary focus on PSA applications, with others focusing on systems and tools \cite{frazzetto2019prescriptive,niederhaus2024technical}, methodologies \cite{lepenioti2020prescriptive,kubrak2022prescriptive}, or sociotechnical system analysis \cite{wissuchek2024prescriptive,stefani2018constituent}. Multiple surveys indeed focus primarily on applications, but they either limit their scope to studying specific application domains or analysis methods utilized therein \cite{raeesi2021prescriptive,fox2022review,soeffker2022stochastic,kubrak2022prescriptive,mendoza2024prescriptive,hall2024systematic}. 

Remaining application-oriented papers limit their scope to a significantly limited subset of applications by more or less explicit survey-methodological choices \cite{poornima2020survey,mishra2023prescriptive}. One of these papers survey ''various'' PSA applications, as its title aptly conveys, yet only 21 in total, and without specifying how these papers were found or selected \cite{poornima2020survey}. The other paper explicitly limits its search query to only include PSA papers intersecting with keywords under the topic of ''sustainable operations research'' \cite{mishra2023prescriptive}. These methodological choices, excluding the larger portion of existing PSA papers, severely limit how comprehensive analyses of PSA application domains and problem types can be, which is reflected by parenthesized check marks in Table \ref{tab:survey-overlap-table}. 

The most recent PSA survey paper \cite{wissuchek2024prescriptive} only covers the ''more prevalent'' (in terms of the number of papers) application domains and problem types, or ''affordances'' in the chosen nomenclature, which is also reflected in Table \ref{tab:survey-overlap-table}. The aforementioned survey also mentions a number of overall PSA method categories and example methods. These are however not derived from its own set of survey papers, but from Lepenioti et al.'s survey from 2020 \cite{lepenioti2020prescriptive}.

In addition to being comprehensive unlike previous endeavours, this survey marks a relevant research contribution by introducing novel analysis dimensions, namely an analysis and taxonomy of how solution methods are combined into \textit{PSA workflows}. Although others have considered the relevance of studying this topic \cite{wissuchek2024prescriptive}, only individual workflow patterns have been discussed fragmentarily within PSA research so far (e.g., \cite{elmachtoub2022smart}). Furthermore, while existing surveys sometimes include statistics on the number of papers published per year (e.g., \cite{lepenioti2020prescriptive}), none provide deeper \textit{temporal} insights into the PSA application research field's development over time in terms of encountered problem types or solution methods - another gap closed by the present survey.

\section{Background on PSA}\label{sc:background}

PSA arose as a term within the field of business analytics in the previous decade \cite{lustig2010analytics}. A broad, yet comprehensive definition of business analytics, derived from a survey of existing definitions within the field \cite{holsapple2014unified}, goes as follows: 

\newdefinition{definition}{Definition}
\begin{definition}[Business Analytics]\label{def:BA}
Evidence-based problem solving within the context of business situations.
\end{definition}

As described in in Davenport and Harris' seminal book on the subject, \textit{Competing on Analytics: The New Science of Winning} \cite{davenport2007competing}, problem solving being \textit{evidence-based} more concretely entails that business decisions should be driven by the ''extensive use of data, statistical and quantitative analysis, explanatory and predictive models, and fact-based management'', enabled by a set of technologies and processes \cite[~p. 7]{davenport2007competing}. 

The introduction of the ''PSA'' term can be attributed to IBM \cite{frazzetto2019prescriptive}. In a 2010 paper \cite{lustig2010analytics}, they divide business analytics into three distinct areas. \textit{Descriptive}, \textit{Predictive} and \textit{Prescriptive} Analytics. Each area has a specific problem-solving orientation (cf. Definition \ref{def:BA}), which is popularly conveyed by a set of \textit{core questions}, such as:

\begin{itemize} 
    \item \textbf{Descriptive}: What happened, and why?
    \item \textbf{Predictive}: What will happen, and why?
    \item \textbf{Prescriptive}: How to make it happen, and why?
\end{itemize}

The specific wording of these questions vary among authors, and the explanatory ''whys'' are not always included (e.g.,  \cite{soltanpoor2016prescriptive,lepenioti2020prescriptive,hagerty2017planning}). Each area is now briefly characterized in turn.

\textit{Descriptive Analytics} is about understanding the \textit{past}. It synthesizes information about historical performance to stakeholders, enables users to monitor their business processes, and more \cite{lustig2010analytics}. In practice, it involves producing dashboards, reports, and various ad-hoc information queries from historical data, e.g., in a data warehouse setting \cite{hagerty2017planning,delen2013data,deshpande2019predictive}. 

\textit{Predictive Analytics} is about using an understanding of the past to predict the \textit{future}. It identifies causal relationships in temporal data, answers ''what-if'' questions about future trends, expected outcomes, and more \cite{lustig2010analytics}. In practice, it involves the usage of ML, simulation models, and more in order to provide forecasting functionality \cite{soltanpoor2016prescriptive,deshpande2019predictive,frazzetto2019prescriptive}. 

Finally, \textit{Prescriptive Analytics} is about using an understanding of the connection between the \textit{past} and the \textit{future} in order to make decisions in the \textit{present}. Complex problem domains and the sheer number of situational factors to consider can make some decision problems infeasible to solve for human beings \cite{lustig2010analytics}. Consider for instance a large-scale delivery route planning problem for a logistics company, which must take into account local traffic levels, the risk of failed deliveries, etc. while minimizing fuel consumption \cite{soeffker2022stochastic}. Such problems might at best be tedious to solve manually.

PSA \textit{suggests} (i.e., \textit{prescribes}) the best course of action in a given scenario with respect to certain goals, constrained by factors of the problem domain, economic resources, etc. \cite{frazzetto2019prescriptive,stefani2018constituent}. Often but not always, PSA involves solving optimization problems, using various approaches and tools \cite{lustig2010analytics,delen2013data,soltanpoor2016prescriptive}. 

The following dictionary definition of PSA narrows down the scope of Definition \ref{def:BA} according to characterizations given so far \cite{siksnys2018prescriptive}:

\begin{definition}[Prescriptive Analytics]\label{def:PA} 
An area of business analytics dedicated to finding and suggesting the best decision options for a given situation.
\end{definition}

\begin{figure}[!ht]
       \centering
       \includegraphics[width=0.8\linewidth]{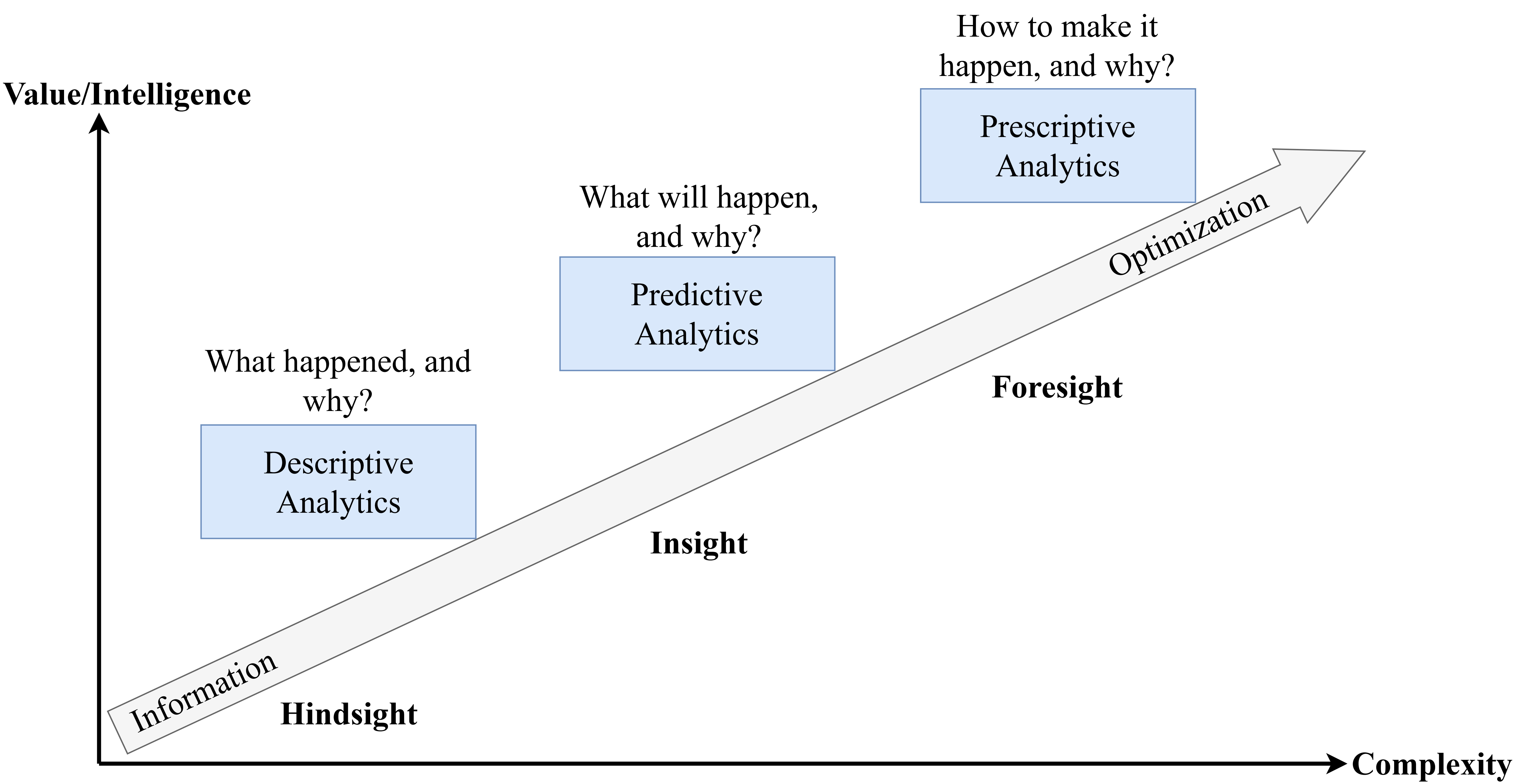}
       \caption{An Analytics Ascendancy Model, adapted and modified for this paper from \cite{siksnys2018prescriptive}. Modifications include the omission of Diagnostic Analytics as a separate area from Descriptive Analytics (cf. \cite{lepenioti2019prescriptive}), preferring \textit{Complexity} over the more subjective \textit{Difficulty} axis, as well as core question wordings.}
       \label{fig:gaam}
\end{figure}

As popularly envisioned, PSA is an advancement on top of Predictive Analytics, which is an advancement on top of Descriptive Analytics - each progressively incorporating techniques from their predecessors and possibly offering higher levels of competitive advantage or value for a business \cite{lepenioti2020prescriptive, oesterreich2020understanding,siksnys2018prescriptive}. This idea is illustrated by Gartner's \textit{Analytics Ascendancy Model}, commonly attributed to a closed-access 2013 paper \cite{maoz2013should} by researchers (e.g., \cite{banihashem2022systematic,petropoulos2023operational}), and adapted for this paper in modified form in Figure \ref{fig:gaam}.

\section{PSA definitions and survey scope}\label{sc:scope-and-definition}

To ensure a well-defined scope, it was ultimately deemed necessary to use a more specific version of Definition \ref{def:PA} in this survey. Throughout history, many prescriptive approaches have been used within human organizational endeavours: from ancient astronomy using the constellations to time when to harvest the crops \cite{evans1998history}, to modern operations research arriving to support military efforts during the Second World War \cite{petropoulos2023operational}. Even disregarding definitional issues, a single survey paper cannot realistically aspire to cover everything written within the scope of Definition \ref{def:PA} since the dawn of history, let alone entire fields like operations research, which came to be more than 80 years ago \cite{hillier2021introduction}, with a corresponding number of papers to boot. 

Instead, we have chosen to refine our scope to focus on prescriptive approaches following a \textit{data-driven}, \textit{automated} paradigm in terms of solution methods. With this focus, we answer recent calls for more data-driven PSA research \cite{sigmod2024callforpapers,lepenioti2020prescriptive,giesecke2018call}. As we will argue next, the aforementioned PSA paradigm is furthermore characteristic of the business analytics and technological context in which the ''PSA'' term originally arose during the past decade (cf. Section \ref{sc:background}). 

To start, consider the defining characteristics of analytics given in Davenport and Harris' seminal book \cite{davenport2007competing}. First, analytics is literally understood as a subset of \textit{business intelligence}: ''a set of technologies and processes that use \textit{data} to understand and analyse business performance'' (our emphasis) \cite[~p. 8]{davenport2007competing}. Analytics is thus defined as a fundamentally \textit{data-driven} endeavour. Second, while it is recognized in the same book that analytics may technically either be input to human decisions, and thus subjected to manual intervention, or be fully automated, it is also stated that any ''rational individual'' (our paraphrase) using analytics in the current age would prefer to rely on information technology over pen and paper \cite[~p. 8]{davenport2007competing}, signifying a preference for \textit{automation} when possible. Apart from this singular perspective on the matter, many existing definitions of and rationales behind business analytics explicitly mention some combination of being data-driven and leveraging automation as part of its key traits, as summarized in \cite{holsapple2014unified}. 

In addition, note how the ''PSA'' term arose within the technological context of the 2010s (cf. Section \ref{sc:background}): a time period during which increasingly scalable computing systems for (Big) data-driven, automated analytics had a momentous impact on industry and society \cite{kleppmann2017designing,zaharia2016apache}, and a time period during which the data-driven analytics ecosystem underwent rapid expansion in terms of user base and available tools \cite{chollet2021deep}, offering increasing levels of workflow automation \cite{he2021automl}.

Based on what has been discussed in this section, we thus present a definition of \textit{Data-Driven PSA} (i.e., DPSA), which helps delimit our survey scope, as follows:

\begin{definition}[Data-Driven Prescriptive Analytics]\label{def:DPSA}
    Prescriptive Analytics done by combining \textit{data-driven prediction} with \textit{automatic prescription}.
\end{definition}

While we will avoid pontificating on whether DPSA is in fact the ''proper'' definition of PSA, DPSA definitely covers a proper subset of Definition \ref{def:PA}.

''Prescriptive Analytics'' refers to Definition \ref{def:PA}. The remainder of Definition \ref{def:DPSA} reuses as its basis an existing characterization of PSA from another survey with a particular focus on methodologies \cite{lepenioti2020prescriptive}, in which PSA is characterized by combining \textit{predictive} and \textit{prescriptive} methods. 

\textit{Data-driven prediction}, exemplified by data mining and ML methods, accords with original characterizations of business analytics discussed in this section, while \textit{automatic prescription}, exemplified by mathematical optimization, accords with how the value proposition of PSA was originally formulated by IBM (cf. Section \ref{sc:background}): finding the best decision options within problem spaces otherwise too large or complex to search manually. 
 
Noting the open-ended wording used in Definition \ref{def:DPSA}, omitting universal quantifiers like ''only'' or ''all'', the required usage of said prediction-prescription combination is \textit{existentially quantified} by default. That is, the described combination must simply be \textit{present} in an application, if not ubiquitously so. 

\section{Survey methodology}\label{sc:method}
Sections \ref{sc:method:dbs}-\ref{sc:method:criteria} outline the utilized survey methodology, taking their onset in recommendations followed by previous surveys within PSA \cite{wissuchek2024prescriptive,lepenioti2020prescriptive}. Sections \ref{sc:method:procedure} and \ref{sc:method:statistics} provide a statistical summary of the search process and identified papers, respectively. Throughout, we follow the PRISMA 2020 guidelines for systematic literature review reporting \cite{page2021prisma}.

\subsection{Databases}\label{sc:method:dbs}
A total of five scientific literature databases are utilized, of which three (\textit{DBLP}, \textit{ACM Digital Library}, and \textit{IEEE Xplore}) specialize in computer science and adjacent fields, while two (\textit{Web of Science and Scopus}) have a larger, interdisciplinary scope. This way, applications with a sufficient level of technical rigour from a broad variety of application domains are discoverable in the search results. To obtain as many relevant search results as possible, the complete \textit{ACM Guide to Computing Literature} is searched within the ACM Digital Library, as opposed to the more limited \textit{ACM Full-Text Collection}. The default collection is utilized for all other databases. Each database was last consulted on September 3rd, 2024.

\subsection{Search queries}\label{sc:method:queries}
To obtain the initial set of records, \textit{prescriptive analytics} is utilized as the search string. This generic term is preferred to a more opinionated block-structured search string calling for, e.g., particular methodologies or application domains, in order to avoid biased results. 

To help filter away records in which PSA-related topics only had a passing relevance, search queries are only executed on \textit{metadata fields} (title, abstract, and keywords) as opposed to full-text in all databases. 

Note that as an exception to this rule, DBLP queries are locked into searching within additional metadata fields \cite{dblp2024howto}. However, no search results ultimately obtained from DBLP matched purely on the basis of these additional fields. Thus the scope of the DBLP search ended up being effectively the same as the other databases.

\subsection{Selection criteria}\label{sc:method:criteria}
To narrow down initial search results to a subset adequately addressing the research questions (cf. Section \ref{sc:introduction}), a set of selection criteria was developed, used for screening the initial set of records in several phases. Within the context of this survey, we will refer to the remaining set of records after screening with these criteria as the set of \textit{surveyed papers}, or just \textit{papers} for short when unambiguous.
The set of selection criteria is as follows, to be explained in the remainder of this section:

\begin{enumerate}[C1:]
    \item The full text must be assessable.
    \item The record must be a peer-reviewed \textit{journal paper}, \textit{conference paper}, or \textit{book chapter}.
    \item The outlet cannot employ a \textit{permissive review process} as part of its business model.
    \item The outlet must demonstrate a minimum level of \textit{research impact}.
    \item The paper must develop an \textit{end-to-end DPSA application} according to Definition \ref{def:DPSA}.
    \item The paper cannot have the development of generic methods or tools as its main focus.
\end{enumerate}

Criteria C1-C4 relate to other factors than paper content, while C5 and C6 address the latter.

C1 requires that the paper is written in English and either available online (i) free of charge, (ii) by institutional access, or (iii) by personal request from the authors on ResearchGate, within a response time of 3 weeks. Without the full text at hand, it would simply be impossible to assess the entire set of criteria and address the research questions. 

Regarding (ii), our affiliated university library at Aalborg University granted full-text access to 106 non-open-access e-book and e-journal databases at the time of doing the survey \cite{aub2024databases}, including ACM Digital Library, IEEE Xplore, ScienceDirect, SpringerLink, Taylor \& Francis Online, Wiley Online Library, SAGE Journals, among others, which we deem to have satisfactory  coverage of the most significant outlets.

The reasoning behind C2 is that these channels tend to publish \textit{original results} already verified by \textit{empirical experiment}. Note that, unlike in other PSA surveys (e.g., \cite{lepenioti2020prescriptive}), workshops are \textit{not} regarded as regular conference proceedings and are therefore excluded here, due to the generally preliminary nature of ideas presented in this setting. 

C3 and C4 were both motivated by experiences from a pilot version of this survey, in which many papers were obtained, exhibiting widely inconsistent levels of scientific rigour and quality, while technically adhering to all present criteria. 

C3 concretely excludes \textit{predatory} and \textit{mega outlets}, as identified by \cite{predatory2024thelist,bjork2018evolution}, respectively. While it should be emphasized that these outlet types have fundamental differences, making total conflation unwarranted, both by definition employ a more large-scale, permissive review process than others, generally subsidized by publication fees. For instance, mega journals might emphasize ''technical soundness'', regardless of scientific novelty or impact \cite{bjork2018evolution}.

C4 introduces a specific requirement per outlet type mentioned in C2, based on the same rationale as C3. For journal papers, a minimum level of research impact is enforced by requiring a Q1 \textit{SCImago Journal Rank (SJR)} within at least one subject area \cite{scimago2024ranking}, as of the most recent ranking during September 2024. SJR accounts for both the number of citations received by the ranked journal and the number of citations received by papers citing it. The Q1 requirement ensures that only journals with an SJR within the \textit{top 25\% quartile} of \textit{at least one subject area} are included in this survey.

For conference papers, where existing ranking systems (e.g., \cite{core2024ranking,conference2024ranking}) were deemed too limited to uniformly assess the highly multidisciplinary set of surveyed papers, the requirement of only including \textit{international conferences} was introduced instead, meaning that participants must represent an international set of institutions. The rationale behind this is that, all things being equal, international participation entails more wide-reaching \textit{knowledge dissemination}, and thus impact, of the presented research. For analogous reasons, book chapters were subjected to a similar requirement: that the total set of authors in the book must represent an international set of institutions.

As for C5, the requirement of developing an \textit{end-to-end} application means that the paper must analyse a \textit{concrete problem} solved with \textit{concrete methods}. %, verifiable by \textit{empirical testing}. 
For instance, it is insufficient for a paper to simply discuss hypothetical applications of PSA within some application domain on a purely conceptual level, or to propose ''machine learning'' as a generic solution to a specific real-world problem without further details. This criterion is introduced in order to be able to assess adherence to Definition \ref{def:DPSA} and adequately address the research questions (cf. Section \ref{sc:introduction}).

C5 effectively helps exclude many PSA papers where describing elaborate applications is not a significant focus, such as theoretical papers, call-for-papers, survey papers, extended abstracts, and more. Still, C6 is utilized in order to make sure to exclude papers mainly developing generic methods or tools for PSA. These papers will sometimes technically develop end-to-end applications according to C5 \cite{siksnys2021solvedb+}, but these generally play an auxiliary role as example problems for the sake of demonstration. We therefore deemed it problematic to conflate them with application-oriented papers. 

\subsection{Survey process}\label{sc:method:procedure}
\begin{figure*}[!ht]
       \centering
       \includegraphics[width=\linewidth]{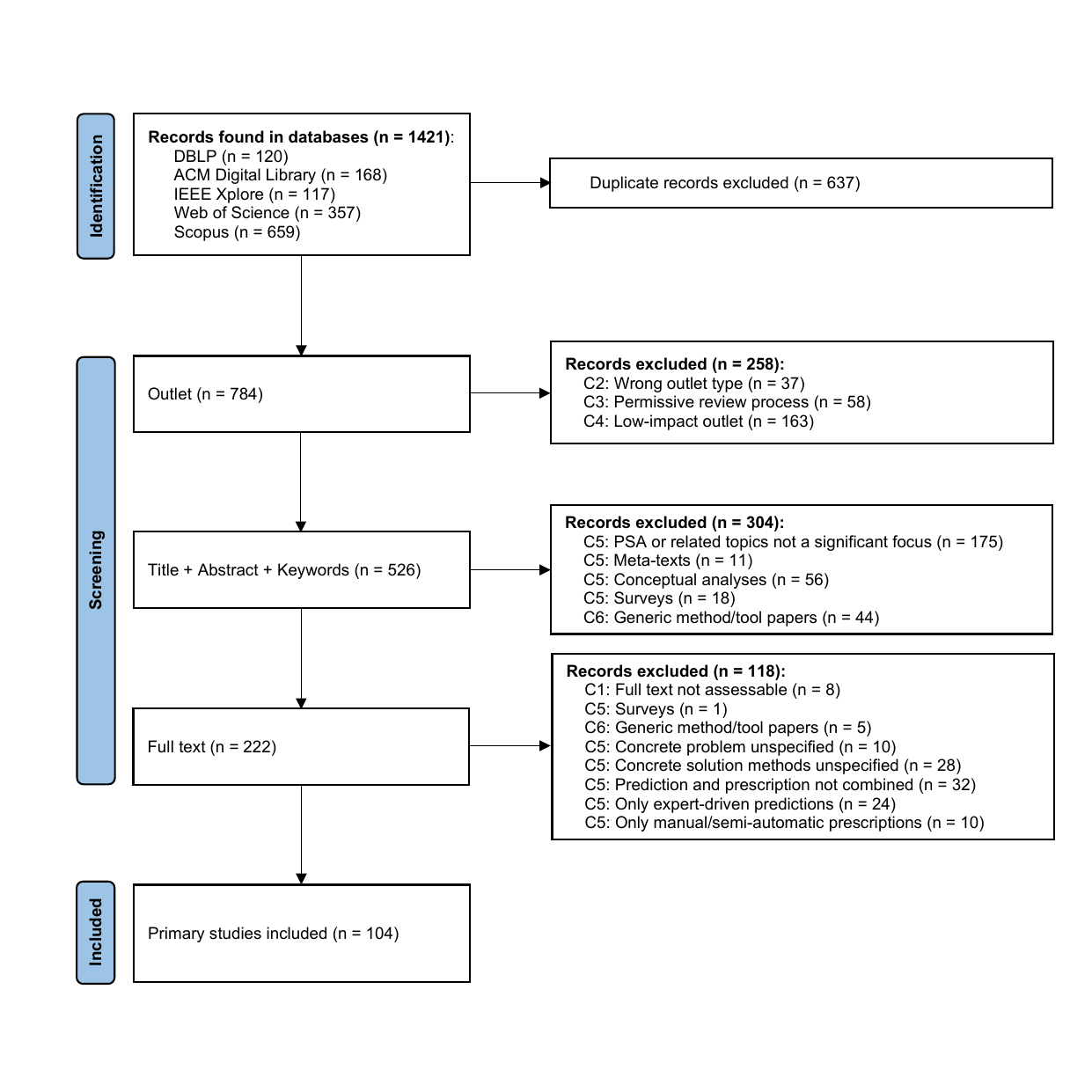}
       \caption{Flow diagram of the executed survey procedure, using the PRISMA 2020 format \cite{page2021prisma}. Each box in the left chain represents a phase in a sequential screening workflow, along with its number of input records. The boxes to the right summarize how many records were excluded in each screening phase, as well as the reasons why, with respect to the selection criteria of Section \ref{sc:method:criteria}.}
       \label{fig:prisma}
\end{figure*}

This section describes the process of how the survey procedure was executed, including choices made during screening. Using the databases, queries and selection criteria of Sections \ref{sc:method:dbs}, \ref{sc:method:queries}, and \ref{sc:method:criteria}, the screening process for obtaining the final set of surveyed papers proceeded as depicted in Figure \ref{fig:prisma}. This figure depicts a flow diagram according to the PRISMA 2020 guidelines for systematic reviews  \cite{page2021prisma}, summarizing the number of records and choices made during each screening phase. 

After executing the search query in each database and deduplicating records, screening proceeded in three phases, during which the \textit{outlet}, \textit{metadata} (title, abstract, and keywords), as well as the \textit{full text} of each record were assessed, respectively. One author executed each screening phase end-to-end, while another participated as an independently working co-reviewer during the outlet and metadata phases, flagging records for possible exclusion, based on a complete list of references without abstracts. Disagreements on screening choices were resolvable by subsequent discussion. 

A few notes on choices made during each phase are in order, and the reader is advised to confer Figure \ref{fig:prisma} as needed. For the \textit{outlet screening phase}, records excluded due to C2 were PhD theses, workshop papers, and papers from the Computing Research Repository (CoRR) which were not peer-reviewed. Exclusions due to C3 and C4 followed the objective metrics described in Section \ref{sc:method:criteria} by looking up the relevant information in the chosen online registers, with the larger part of records being excluded due to the outlet being low-impact (C4). 

During the \textit{metadata screening phase}, a large number of papers only cursorily mentioned ''PSA'' in the abstract, e.g., within the context of broad discussions about 21st century technological trends, which were then excluded due to C5, because they did not even contribute directly to the PSA field. Other papers were excluded in this phase due to the same criterion, on the basis of not putting any emphasis on the topic of DPSA applications in their metadata. Some of these were meta-texts, i.e., call-for-papers and extended abstracts, which only discuss original research indirectly, while others were conceptual analyses or surveys on PSA-related topics. 

While the metadata phase mainly concerned itself with screening papers in terms of their main focus, the \textit{full text screening phase} mainly concerned itself with evaluating more fine-grained methodological constraints (cf. Definition \ref{def:DPSA}) in the actual paper content. Eight conference papers could not be obtained by any means allowed by C1, and were therefore excluded from the survey. In addition, a few papers turned out to be surveys or generic method/tool papers upon closer inspection, and were therefore excluded due to C5 and C6, respectively. 

In Figure \ref{fig:prisma}, the remaining five reasons for exclusion with respect to C5 concern whether an end-to-end DPSA application has been described. To aid understanding, all reasons for exclusion are presented in a \textit{prioritized order} in the figure. That is, papers excluded for reasons further down were not eligible for exclusion due to reasons further up, but might have been eligible for exclusion for reasons presented even further down.

\subsection{Survey statistics}\label{sc:method:statistics}

\begin{figure}[!ht]
       \centering
       \includegraphics[width=0.8\linewidth]{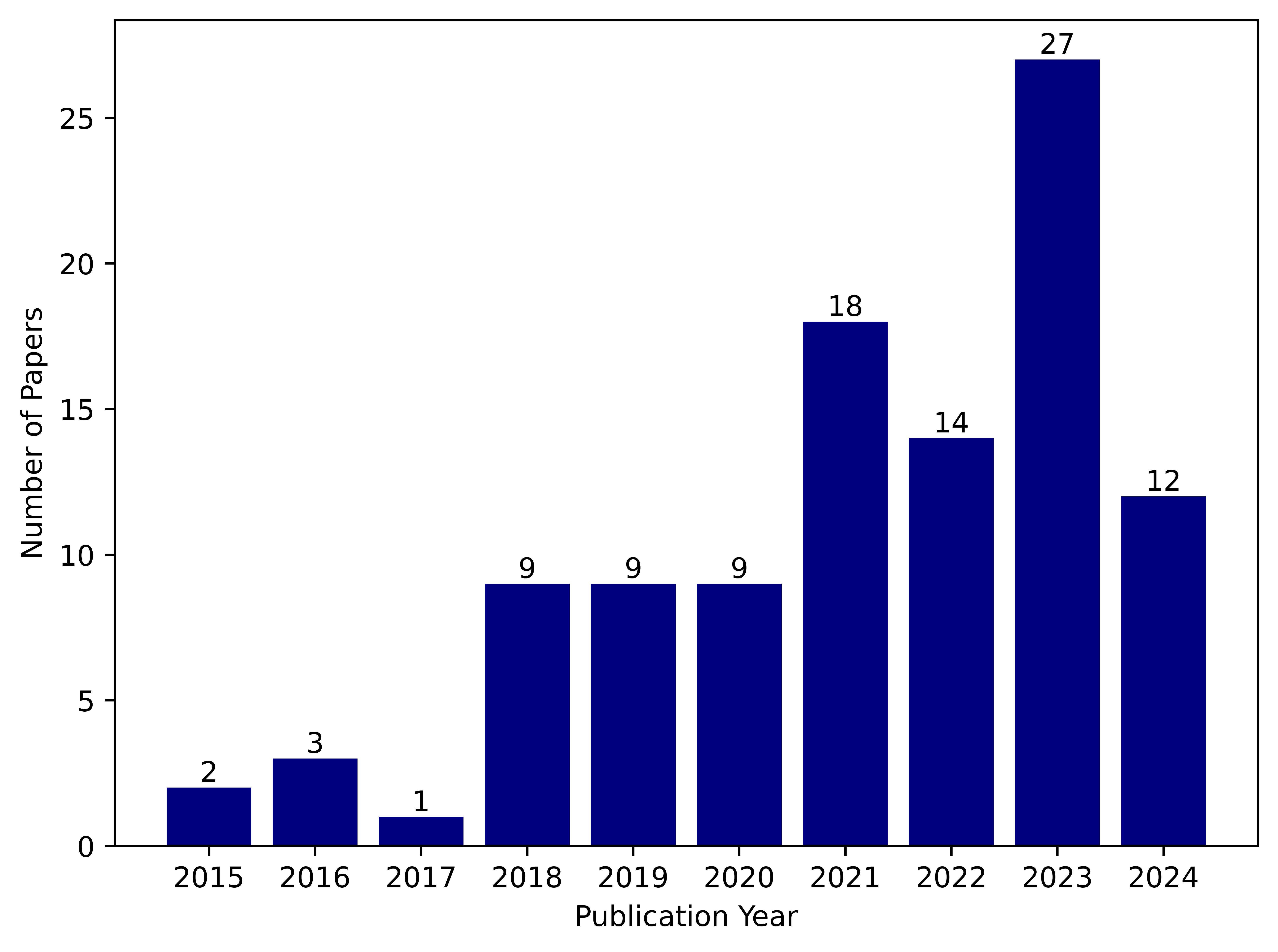}
       \caption{Number of papers per year in the set of surveyed papers.}
       \label{fig:paper-year-bar}
\end{figure}

A total of 104 papers were obtained from executing the survey procedure. Figure \ref{fig:paper-year-bar} summarizes the distribution of papers per year, which spans the entire 2015-2024 range. 
There is evidently a pronounced skew towards more recent years, which might be attributed to the increasing popularity and impact of DPSA applications within PSA research, with the 2020-2024 range making up about 76.9\% of all papers. Supplementary survey statistics can be found in \ref{app:a}.

\section{Problem types}\label{sc:rq1}
In order to address RQ1 (cf. Section \ref{sc:introduction}), problem types handled by the set of surveyed papers are summarized in Table \ref{tab:domains}. For ease of comprehension, problem types addressed in the set of surveyed papers were grouped under ten categories based on their targeted application domain. An extended discussion on each individual use case found within each application domain can be found in \ref{app:b}. 

\begin{table*}[ht]
    \centering
    \resizebox{\linewidth}{!}{%
    \begin{tabular}{ll}
    \hline
    \textbf{Application domain}  & \textbf{Problem types} \\ \hline
    Academia                     & Student guidance \cite{osakwe2024towards,ara2024collaborative,Susnjak2023,Yanta2021,Du2016,Kiaghadi2023}\\ %\hline
                                 & Researcher career guidance \cite{Cho2015} \\
    Advertising and Marketing    & Targeted advertising \cite{Jabr2023,Ferreira2022,Borenstein2023,Chaudhry2023}\\ %\hline
                                 & Uplift analysis \cite{Gubela2021,Caigny2021,Devriendt2021,Sanisoglu2023,singh2023machine} \\
    Healthcare                   & Clinical decision making \cite{Fang2019,Meng2020,Khan2021,Bertsimas2022,Bertsimas2021a,Zhou2023}\\ %\hline
                                 & Inventory and personnel logistics \cite{Galli2021,Williams2022,Tan2021,Shi2021,Bertsimas2021b} \\
                                 & Appointment scheduling \cite{Srinivas2018,Salah2022} \\
    Human Resources              & Recruitment decision making \cite{Pessach2020,Ramannavar2018,zhu2024optimal}             \\ %\hline
                                 & Staffing capacity management \cite{Notz2022,Notz2023,Bischhoffshausen2015}\\
                                 & Employee attrition prevention \cite{Brockett2019}\\
    Infrastructure               & IT network orchestration \cite{John2019,Ceselli2018,Ceselli2019}              \\ %\hline
                                 & Urban planning \cite{Brandt2021,Sinha2023}\\
                                 & Building construction \cite{Li2019}\\
                                 & Power grid management \cite{Goyal2016,stratigakos2024interpretable,chen2024towards}\\
    Logistics and Transportation & Airline operations efficiency \cite{Achenbach2018,Ayhan2018,Jacquillat2022,Birolini2023a,Birolini2023b}             \\ %\hline
                                 & Port state control \cite{Tian2023a,Yan2023,yang2024prescriptive,oudani2023prescriptive} \\
                                 & Ship maintenance planning \cite{Tian2023b,mohd2022prescriptive} \\
                                 & Railway maintenance planning \cite{Amiel2023,Consilvio2019,consilvio2024data} \\
                                 & Ground vehicle route planning \cite{Grzegorowski2022,Kandula2021} \\
     Manufacturing               & Proactive maintenance and error correction \cite{Kuzyakov2020,Oberdorf2021,Stein2018,Vater2019,Vater2020,Tham2023,Mohan2023,karakaya2024sensor} \\
                                 & Procurement and inventory decision making \cite{Lee2022,Thammaboosadee2018,Rakhmasari2018,Yu2021,Mandl2021,Mandl2023} \\
                                 & Production flow efficiency \cite{Cakir2023,Kumar2023,Sangwan2023,reisch2023prescriptive} \\
                                 & Production parameter design \cite{Ribeiro2023,Suvarna2022} \\
    Recreational Activities      & Social media follower recommendation \cite{Dash2022} \\ %\hline
                                 & Music learning \cite{razgallah2024using} \\
                                 & Sports league planning \cite{hassanzadeh2024conclude} \\
    Retail and Services          & Inventory management \cite{Shi2020,Heide2020,Keskin2022,Punia2020,Adulyasak2023,Caro2019} \\
                                 & Portfolio management \cite{Chen2022,Qu2020,Lash2016,Huang2019,Mehrotra2020,han2024identifying} \\
                                 & Business health monitoring \cite{Hauser2021,Ravi2021,kurniawan2023prescriptive} \\
    Social Policy                & Criminal justice system guidance \cite{Brandt2022,Bahulkar2018,Delen2021} \\ 
                                 & ESG pillars implementation \cite{sariyer2024predictive} \\
                                 & Child services intervention planning \cite{Schwartz2017} \\ \hline
    \end{tabular}}
    \caption{Papers and problem types associated with each application domain. ''ESG'' abbreviates Environmental, Social, and Governance.}
    \label{tab:domains}
\end{table*}

Developing the aforementioned categories was done \textit{iteratively} during the full text screening phase (cf. Section \ref{sc:method:procedure}), using previous efforts as the initial expectation (e.g., \cite{poornima2020survey,lepenioti2020prescriptive}), yet ultimately ending up with an original taxonomy that would fit the set of surveyed papers - merging, splitting, deleting and inventing categories as necessary. Specifically, the heuristic followed was that a category should contain at least three papers, in order to avoid an overly baroque set of categories, yet at most 20\%, or 20 papers, in order for the categories to work as a more useful index within the context of this survey.

As Table \ref{tab:domains} demonstrates, DPSA has been applied to a wide variety of application domains and problem types. To explain some of the more specialized terms, \textit{Uplift analysis} is about designing a treatment, e.g., a promotional campaign, including which customers to target, in order to maximize increases (i.e., ''lifts'') in positive metrics, e.g., profit, due to the treatment \cite{Devriendt2021}. \textit{Environmental, Social, and Governance (ESG)} pillars refer to a set of responsible organizational practices, which may be subject to legal regulations or be highly prioritized by some investors \cite{sariyer2024predictive}.

\begin{figure*}[!htbp]
       \centering
       \includegraphics[width=0.8\linewidth]{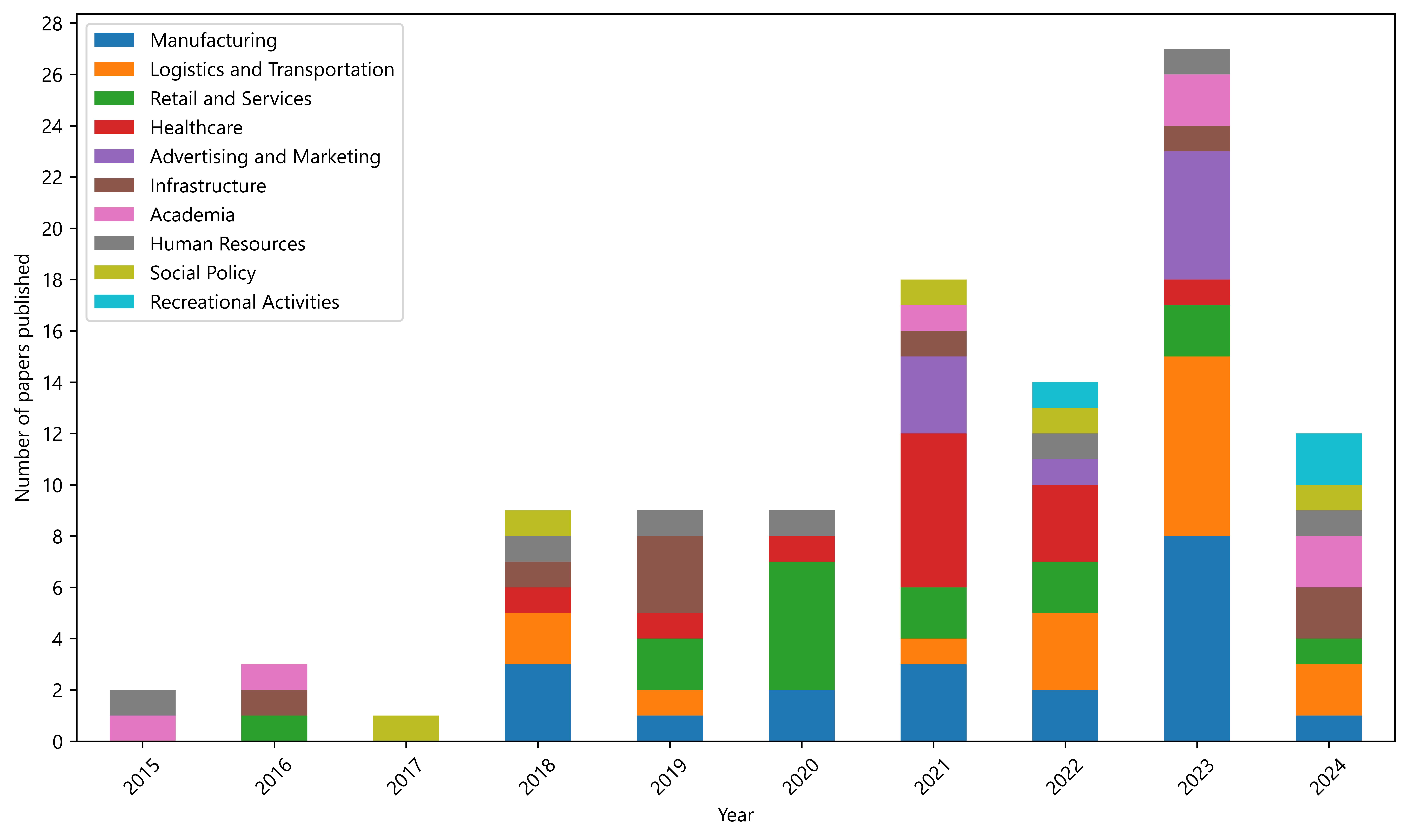}
       \caption{Stacked bar chart illustrating the number of papers published within each problem domain over time, as well as the distribution of published papers per year.}
       \label{fig:temporal-domains}
\end{figure*}

As Figure \ref{fig:temporal-domains} illustrates, the relative activity levels among problem domains have varied over time. Notable trends include the larger number of Healthcare papers published in connection to the COVID-19 pandemic in 2021, along with notable increases in activity for a select few domains in 2023: Manufacturing, Logistics and Transportation, along with Advertising and Marketing. This phenomenon could possibly be related to post-COVID shifts in global supply chains and customer behaviours. In any case, DPSA use cases seem to have been motivated by real-world developments over the years.

The sheer variety of problem types covered by Table \ref{tab:domains} suggests that a broad notion of ''business problem solving'' (cf. Definition \ref{def:BA}) is reflected by the literature as a whole, concerning \textit{organizational value creation} \cite{osterwalder2010business} for many different kinds of stakeholders, be it in hospitals, factories, universities, social services, grocery stores, or elsewhere. Still, due to their common core methodology (cf. Definition \ref{def:DPSA}), patterns in terms of choice of methods and workflows can be identified across use cases and application domains, as Sections \ref{sc:method-types}-\ref{sc:rq2-synthesis} explore further.

\section{Method types}\label{sc:method-types} 

This section addresses part of RQ2 by summarizing the array of \textit{individual} methods used for DPSA in the set of surveyed papers. Later on, Section \ref{sc:method-patterns} addresses RQ2 in terms of how these methods are \textit{combined} in different workflow patterns. In both Sections \ref{sc:method-types} and \ref{sc:method-patterns}, there is a focus on offering new \textit{taxonomies} and \textit{conceptual models} to form a structured understanding of existing DPSA methodologies. 

The array of different methods used among the set of surveyed papers can be divided into five identified \textit{method types}: \textit{Data Mining and Machine Learning (DM/ML)}, \textit{Mathematical Optimization (OPT)}, \textit{Probabilistic Modelling (PROB)}, \textit{Domain Expertise (EXP)}, and \textit{Simulation (SIM)}. Going forward, their abbreviations will be used for conciseness.

\begin{figure}[!ht]
       \centering
       \includegraphics[width=0.8\linewidth]{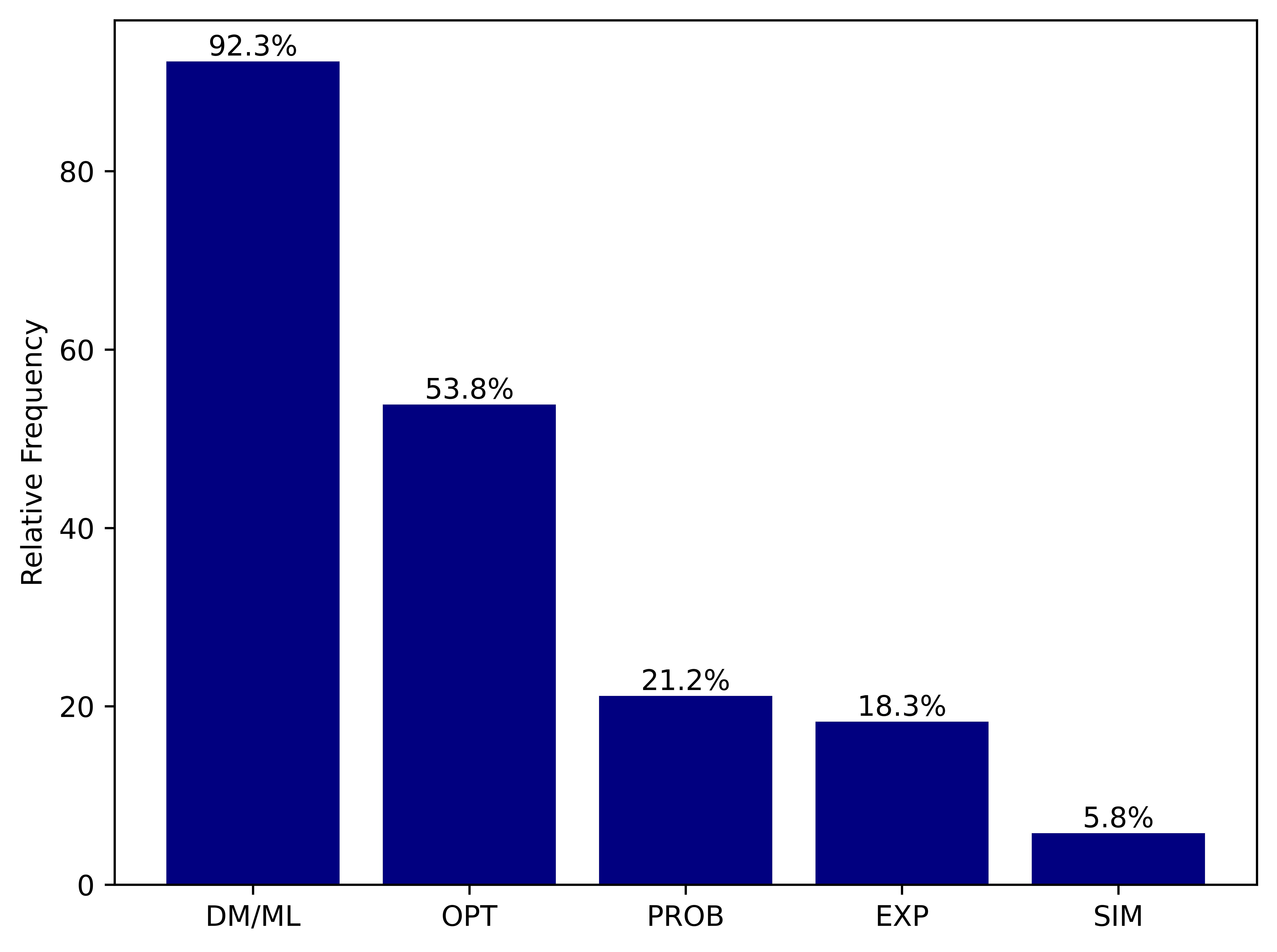}
       \caption{Bar chart summarizing the relative frequency of the different method types in the set of surveyed papers. Note that this aggregation does not form distinctions based on the role played by each method on the DPSA workflow in terms of providing predictions or prescriptions.}
       \label{fig:method-bar}
\end{figure}

\begin{figure}[!ht]
       \centering
       \includegraphics[width=\linewidth]{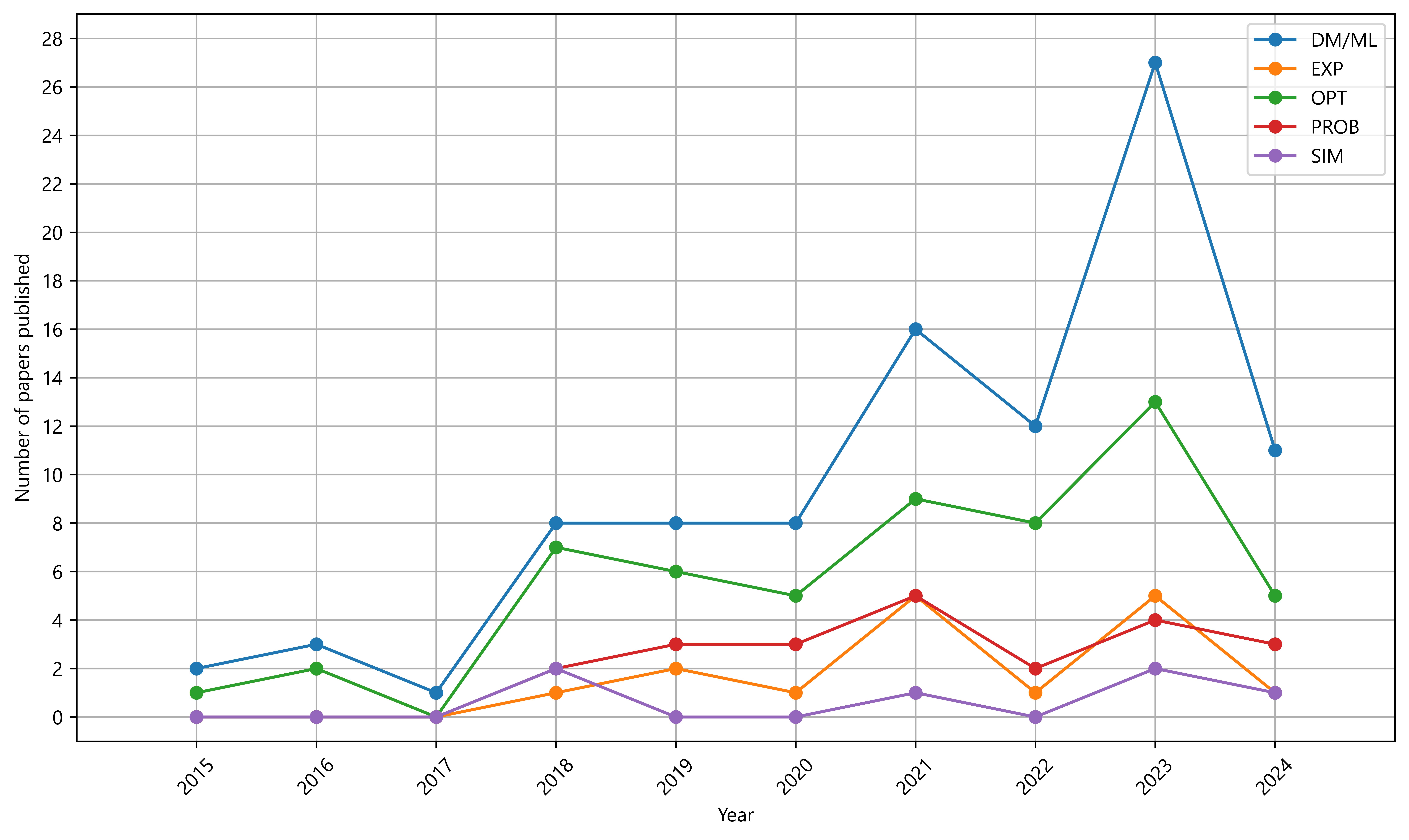}
       \caption{Line chart illustrating the number of papers using each method type published over time, as well as the distribution of published papers per year.}
       \label{fig:method-temporal}
\end{figure}

Figure \ref{fig:method-bar} illustrates the relative frequency of each method type in the set of surveyed papers. As is apparent from the bar chart, DM/ML and OPT dominate. As Figure \ref{fig:method-temporal} furthermore shows, this trend has only become increasingly pronounced in recent years. As we shall see in subsequent analyses, more intricate patterns hide behind this heavily aggregated visual representation however.  

\begin{figure*}[!htbp]
       \centering
       %\resizebox{!}{\linewidth}
       \includegraphics[height=0.93\textheight,width=\linewidth]{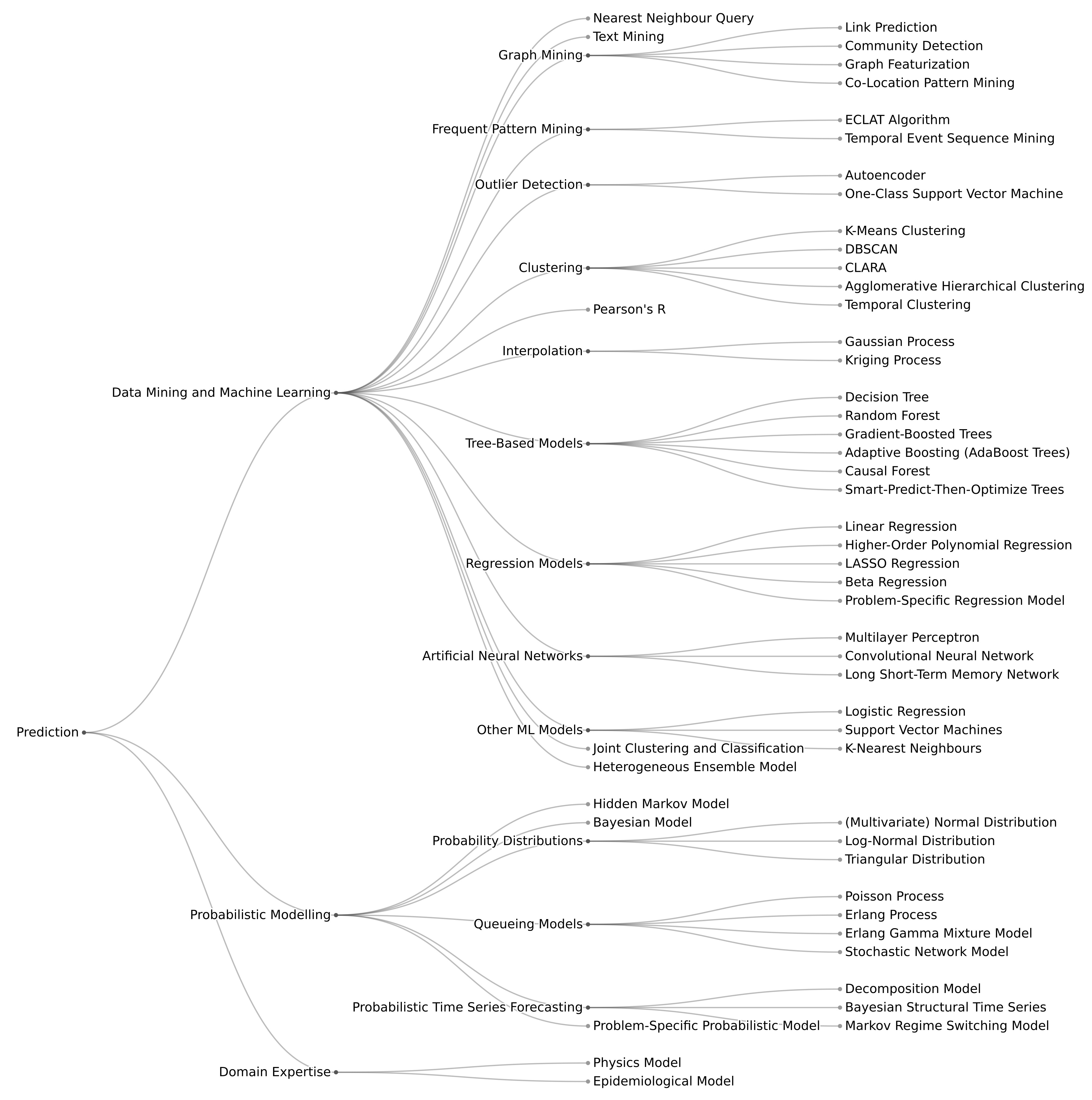}
       \caption{Overview of methods used alone or in combination with others to provide predictions in DPSA workflows in this survey.}
       \label{fig:prediction-tree}
\end{figure*}

\begin{figure*}[!htbp]
       \centering
       \resizebox{!}{0.93\textheight}{\includegraphics[angle=90]{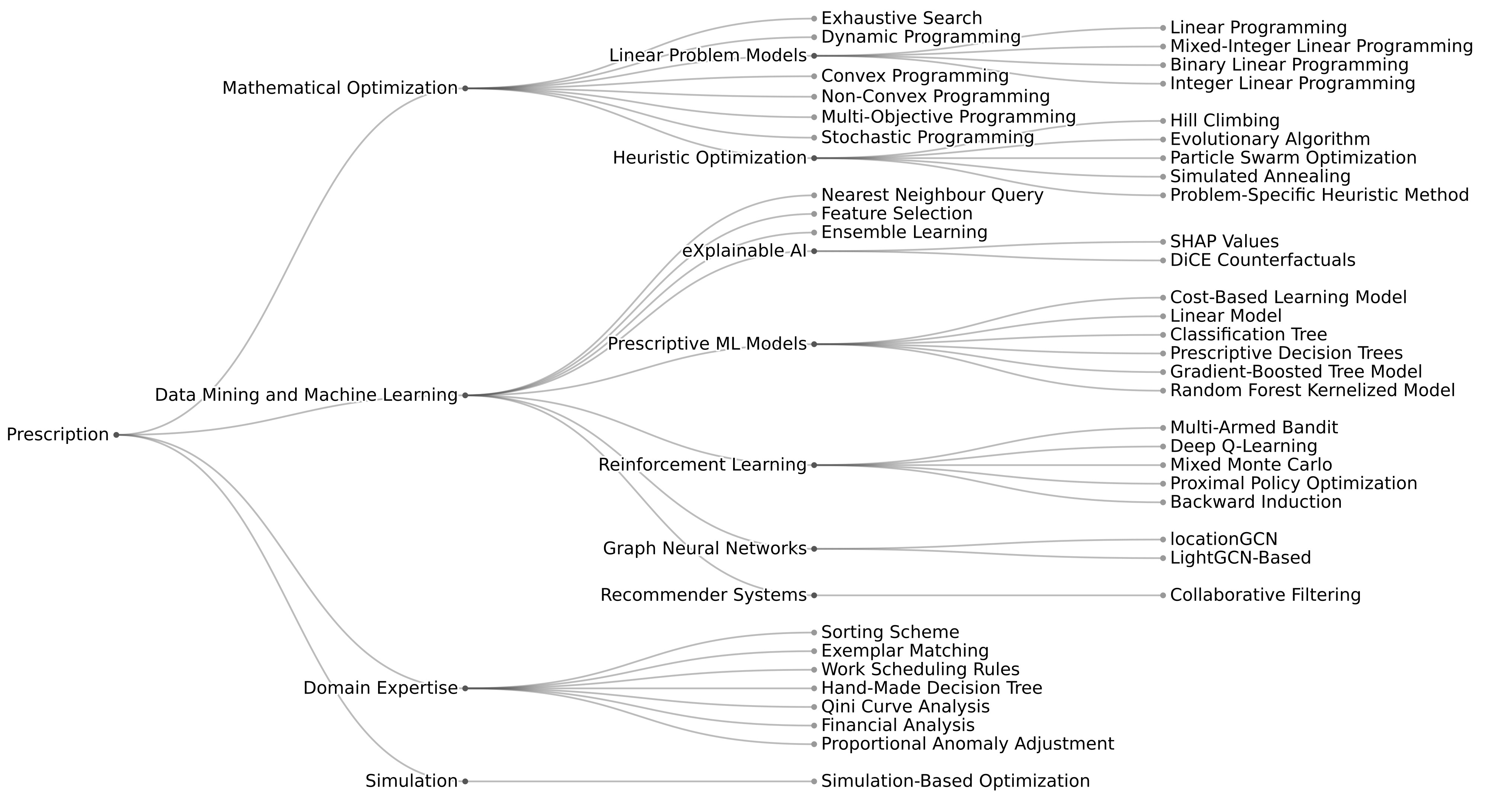}}
       \caption{Overview of methods used alone or in combination with others to provide prescriptions in DPSA workflows in this survey.}
       \label{fig:prescription-tree}
\end{figure*}

With respect to the different method types, Figures \ref{fig:prediction-tree} and \ref{fig:prescription-tree} summarize all methods used in the set of surveyed papers for prediction and prescription, respectively (cf. Definition \ref{def:DPSA}), whether they are used alone or in combination with others. Throughout, we use the convention of capitalizing method names when they refer to the content of Figures \ref{fig:prediction-tree} or \ref{fig:prescription-tree}. 

The following sections describe each method type, the array of options they provide for prediction and prescription, as well as main trends found in the set of surveyed papers. An extended discussion on each individual utilized method and how it maps to specific application scenarios can be found in \ref{app:c}. Readers needing detailed information on individual methods beyond what the main text supports are advised to refer to this appendix. 

We observe in the extended discussion that methodological choices and considerations vary widely due to the practical constraints of individual applications, even within the same use case and problem domain. For this very reason, we avoid making generic, under-fitted, methodological recommendations. Still, we can derive a few general trends per method type, as presented in the main text.

\subsection{Data Mining and Machine Learning (DM/ML)}\label{sc:methodtype:dmml}
This method type covers statistical learning and knowledge discovery from historical datasets \cite{han2022data}. As shown in Figure \ref{fig:prediction-tree}, predictive DM/ML encompasses a large array of methods. Indeed, one might find that this subtree reads like a data mining or ML textbook, with comprehensive coverage of most popular algorithms and approaches. 

While many authors use staple methods off-the-shelf, others opt for solution methods customized to the needs of the individual application. For instance, some authors design non-linear \textit{Problem-Specific Regression Model}s to, e.g., estimate customer demand \cite{Caro2019,Shi2021,Heide2020}. Others use \textit{Smart-Predict-Then-Optimize Trees}, with which the downstream decision problem is incorporated into the training objective, in order to minimize \textit{regret} from suboptimal predictions, as opposed to mere predictive error \cite{Yan2023,Tian2023b}. Still, the most popular methods used for predictive DM/ML are variations of tree-based ensembles, including \textit{Random Forest} (e.g., \cite{Birolini2023a, delen2013data}), \textit{Gradient-Boosted Trees} (e.g., \cite{Adulyasak2023, Galli2021}), \textit{AdaBoost Trees} \cite{oudani2023prescriptive}, among others, being used in 26\% of all papers for either regression and classification. The runner-up was \textit{Logistic Regression}, used in 7.7\% of all papers (e.g., \cite{Chaudhry2023,Chen2022}). 

As shown in Figure \ref{fig:prescription-tree}, prescriptive DM/ML also covers a significant number of approaches. \textit{Reinforcement Learning}, was used in several papers to solve sequential decision problems (e.g., \cite{Ferreira2022,Zhou2023}), for instance, to plan fishing routes based on predicted hauls at each location \cite{Cakir2023}. Techniques from \textit{eXplainable AI} (XAI) were also used by some authors to map various measures of feature importance to prescriptions, e.g., to suggest remedial actions to students at risk of dropping out \cite{Susnjak2023}. Still other authors find ways to generate prescriptions by providing upstream predictions as input to standard DM/ML methods, e.g., a set of \textit{Ensemble Learning} models \cite{Bertsimas2022,Bertsimas2021a}, \textit{Recommender Systems} \cite{ara2024collaborative}, \textit{Graph Neural Networks (GNNs)} \cite{han2024identifying,razgallah2024using} or \textit{Nearest Neighbour Queries} \cite{Kuzyakov2020,Ramannavar2018}. For instance, one paper used a \textit{Co-Location Pattern Mining} algorithm to first predict viable restaurant locations, which were fed as input to a convolutional GNN making the final suggestion \cite{han2024identifying}. \textit{Prescriptive ML Models}, a newer approach on the rise, unify prediction and prescription by training custom ML models, typically with problem-specific loss functions, which can then output prescriptions in a single step, based on input problem parameters (e.g., \cite{Mandl2023,Notz2022,Notz2023,Punia2020}). For example, in one paper staffing capacity management decisions were provided by a model using a Random Forest Kernel function at its core to weigh the similarity between input points and a dataset of known points \cite{Notz2022}.

\subsection{Mathematical Optimization (OPT)}\label{sc:methodtype:opt}
This method type covers the area of mathematical optimization. While optimization is a practically ubiquitous phenomenon in nature, the OPT method type rather refers to a set of techniques developed within fields related to mathematics. These methods all address the problem of how to pick assignments to \textit{decision variables} from a \textit{feasible set}, sometimes delimited by \textit{constraints}, such that one or more \textit{objective functions} are maximized or minimized \cite{kochenderfer2019algorithms}. As found in this survey, OPT is solely used as a prescriptive method in DPSA workflows.

Depending on the complexity of the optimization model, some optimization problem types are NP-hard in the general case, e.g., when some decision variables are constrained to be \textit{integers} \cite{genova2011linear}, or when objective functions or feasible sets are \textit{non-convex} \cite{danilova2022recent}. Ensuring reasonably fast solutions for non-trivially-sized real-world problems is therefore often a matter of operating with careful assumptions and simplifications in the optimization model \cite{hillier2021introduction}. This leads to what might be called an \textit{approximation trade-off} between the model and solution of the optimization problem at hand: Only an approximate solution, found by heuristics, might be expected with an accurate optimization model, while an accurate solution, found by an exact optimization algorithm, might have required an approximate optimization model. OPT thus encompasses a multitude of imperfect solution methods with different approximation trade-offs, and one's mileage might vary depending on application requirements \cite{kochenderfer2019algorithms}. 

The challenge of dealing with the aforementioned approximation trade-off is reflected in the breadth of methods used in the set of surveyed papers (cf. Figure \ref{fig:prescription-tree}), which covers a broad array of problem structures and solution approaches, including advanced extensions such as \textit{Stochastic} (e.g., \cite{Shi2021,Birolini2023a}) and \textit{Multi-Objective Programming} (e.g., \cite{Bertsimas2021b,Jacquillat2022}). Still, variants of \textit{Linear Problem Models} were the most pervasive by far, accounting for 26\% of surveyed papers. With only a few exceptions \cite{Bahulkar2018,Borenstein2023,Rakhmasari2018}, these papers all used linear methods involving integer decision variables, such as \textit{Integer} (e.g., \cite{Sinha2023,Brandt2022}) and \textit{Mixed-Integer Programming} (e.g., \cite{Ceselli2019,Consilvio2019}), with \textit{Binary Programming} being a special case of the former. A staple within Operations Research \cite{hillier2021introduction}, linear methods can efficiently solve many types of real-world resource allocation problems with many variables. However, coping with limited scalability due to NP-hardness and the restrictions of a linear model is a recurrent theme among surveyed papers (e.g., \cite{Ceselli2019,Ceselli2018,Consilvio2019,Birolini2023b}). \textit{Heuristic Optimization} methods, the runner-up used in 11.5\% of papers (e.g., \cite{Grzegorowski2022,Lee2022,Ribeiro2023}), partially circumvents these issues by easing problem assumptions, at the cost of hard solution guarantees (cf. the approximation trade-off). An objective function requiring black-box ML prediction for evaluation is the typical use case for heuristic methods, e.g., to optimize delivery plans \cite{Grzegorowski2022} or biodiesel properties \cite{Suvarna2022}.

\subsection{Probabilistic Modelling (PROB)}
This method type covers techniques modelling uncertain causal relationships. While some prescriptive methods already discussed, such as Stochastic Programming and Reinforcement Learning (cf. Sections \ref{sc:methodtype:opt} and \ref{sc:methodtype:dmml}) overlap with PROB conceptually, this category is ultimately more closely associated with a set of predictive methods (cf. Figure \ref{fig:prediction-tree}).

Most methods in use under this type are relatively well-known within the broader computer science community, e.g., a variety of \textit{Probability Distributions} (e.g., \cite{Lee2022,Qu2020,Shi2020,Srinivas2018}), \textit{Hidden Markov Models} (e.g., \cite{Ayhan2018}), and \textit{Decomposition Models} (e.g., \cite{Adulyasak2023,consilvio2024data}) under \textit{Timeseries Forecasting}. 

On the other hand, \textit{Queueing Models}, which are quite common among surveyed papers \cite{Birolini2023a,Caro2019,Hauser2021,Notz2023,Srinivas2018,Tan2021,Jabr2023,Shi2021}, are perhaps more closely associated with Operations Research \cite{hillier2021introduction}. They form probabilistic models of the efficiency of service systems given the expected demand and service rates along with other factors \cite{denning1978operational}. Consider for instance how to model the number of customers serviced by an ice cream shop with three active employees on a hot summer day, where a line of potential customers might form. Different stochastic models may be used for modelling service and arrival times. \textit{Poisson and Erlang Processes}, or variants thereof, are common choices (cf. Figure \ref{fig:prediction-tree}), respectively predicting the number of events occurring within a particular time span, and the time span required for a particular number of events occurring, thus in a sense being complementary \cite{hanley2019more}. To give examples, one paper used a queueing model to predict the distribution of incoming calls for an outpatient appointment system in a clinical setting \cite{Srinivas2018}, while another used an elaborate \textit{Stochastic Network Model}, involving several different stages and probability distributions to predict expected workloads at hospitals within a COVID-19 setting, from diagnosis to surgery and intensive care \cite{Shi2021}.

\subsection{Domain Expertise (EXP)}\label{sc:methodtype:exp}

This method type covers the application of reasoning, knowledge, and heuristics with respect to the problem domain. While it is not unusual to apply a bit of EXP when, e.g., doing feature selection in ML \cite{theng2024feature}, this method type rather refers to the application of EXP as a comprehensive approach to prediction or prescription. As seen in Figures \ref{fig:prediction-tree} and \ref{fig:prescription-tree}, EXP was used for both aforementioned purposes, and quite commonly so, as is evident from Figure \ref{fig:method-bar}. 

Two studies used predictive EXP, formulating \textit{Epidemiological} models for COVID-19 \cite{Bertsimas2021b} and \textit{Physics} models of electrical grid wear based on heat exchange \cite{Goyal2016}, respectively.

Readers are not expected to recognize most EXP method names in Figure \ref{fig:prescription-tree}. We had to make up most of these terms ourselves, since prescriptive EXP in the literature tends to be applied in a tacit common-sense fashion, with a dash of ''goes-without-saying''. Some of these methods can indeed be regarded as generic common-sense heuristics. For example, if an ML model predicts the probability of defects in a set of ships, one can simply use a \textit{Sorting Scheme}, picking the top k most likely defective ships for inspection \cite{Yan2023}. In \textit{Exemplar Matching}, remedial actions are prescribed based on role model cases, e.g., a higher salary is prescribed for an employee, if an otherwise similar loyal colleague happens to be paid more \cite{Brockett2019}. Other generic methods include encoding domain rules into \textit{Hand-Made Decision Tree}s (e.g., \cite{Vater2019,Vater2020}) or procedural \textit{Work Scheduling Rule}s (e.g., \cite{Oberdorf2021,Salah2022,Srinivas2018}) to output prescriptions automatically. 

Remaining methods are more application-specific in nature, with one paper using \textit{Financial Analysis} formulas along with some thresholds for desired yields in order to provide bank loaning strategies \cite{kurniawan2023prescriptive}, and another calculating \textit{Proportional Anomaly Adjustment}s based on detected irregularities and material properties, in order to increase reliability of 3D printing in aerospace manufacturing \cite{reisch2023prescriptive}. Surveyed papers doing uplift analysis for Advertising and Marketing (cf. Section \ref{sc:rq1}) typically used \textit{Qini Curve Analysis}, a problem-specific technique analyzing a cumulative curve of predicted uplift per customer to determine which ones to target (e.g., \cite{Caigny2021,Devriendt2021,Sanisoglu2023}).

\subsection{Simulation (SIM)}\label{sc:methodtype:sim}
This method type covers the application of simulation models to capture how systems develop over time. The rarest method type among DPSA applications (cf. Figure \ref{fig:method-bar}), papers utilizing SIM all use it for providing prescriptions (cf. Figure \ref{fig:prescription-tree}), each using their own problem-bespoke implementations \cite{Stein2018,Tan2021,Srinivas2018,Adulyasak2023,Jabr2023,sariyer2024predictive}. Providing prescriptions by evaluating different solution candidates in simulations, i.e., by using \textit{Simulation-Based Optimization} \cite{gosavi2015simulation}, is the strategy followed by these papers. The aforementioned technique has similar advantages to Heuristic Optimization techniques discussed previously (cf. Section \ref{sc:methodtype:opt}), since it is able to handle black-box predictive models. In addition, it has a natural way of handling \textit{stochastic} problem models, in that expected outcomes under complex conditions can be reasoned about by using Monte Carlo approximations \cite{harrison2010introduction}. Indeed, all surveyed papers utilizing SIM also use stochastic and/or black-box models as part of their methodology. In one example, the authors simulated different scenarios with respect to a stochastic model, in order to determine the minimum replenishment quantity needed for critical groceries to prevent retailer shortages during COVID-19 panic buying \cite{Adulyasak2023}.

\section{Method patterns}\label{sc:method-patterns}

While Section \ref{sc:method-types} only considered RQ2 in terms of individual method usage, this section considers methodological trends in terms of the \textit{combination} of methods in current DPSA solutions.

\subsection{Method type combinations}\label{sc:method-combinations}
\begin{figure*}[!ht]
       \centering
       \includegraphics[width=\linewidth]{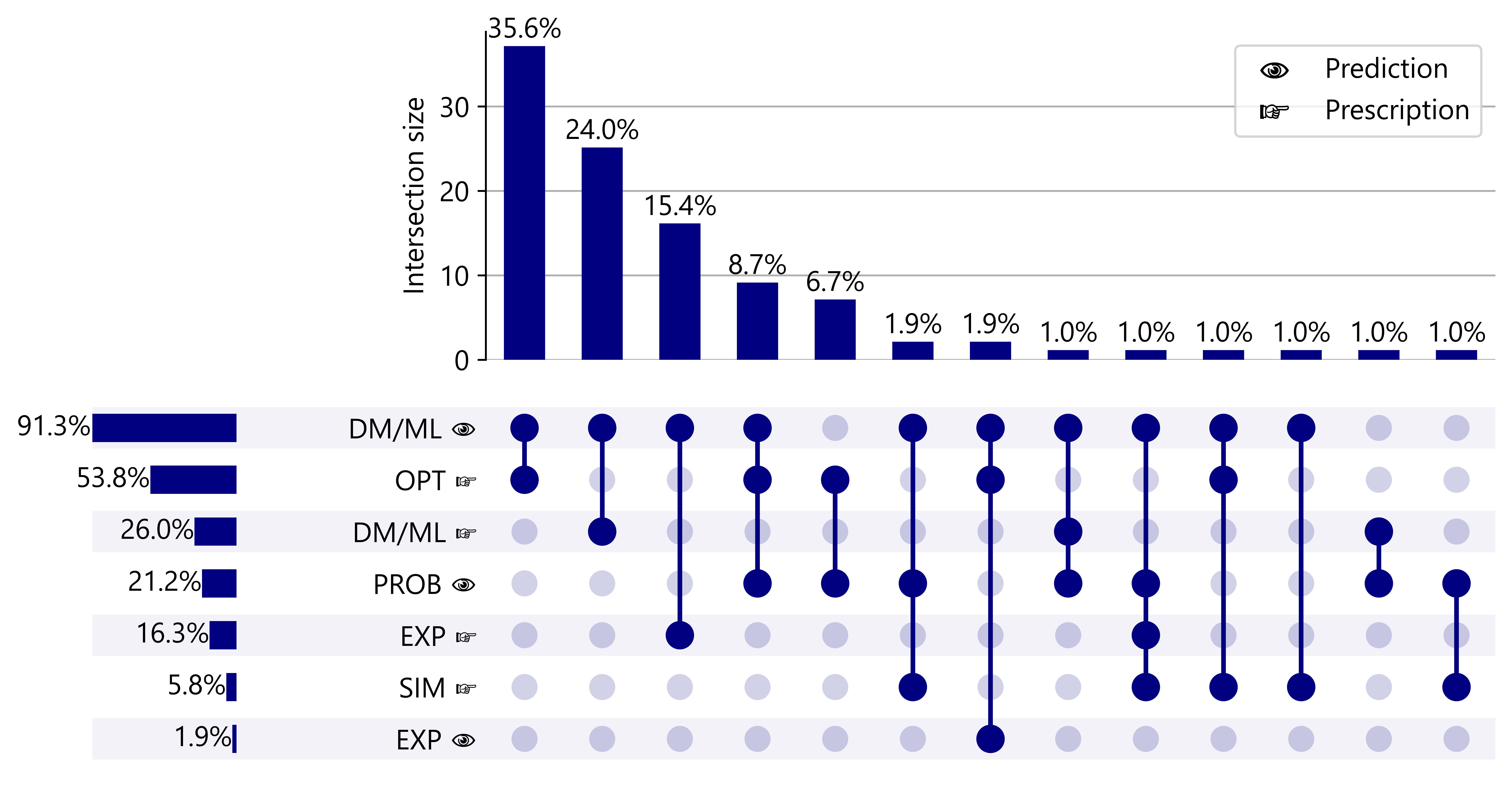}
       \caption{UpSet plot using a matrix layout to summarize the usage of different method types to perform DPSA in the set of surveyed papers. The foreseeing eye (\raisebox{-.55ex}[0pt][0pt]{\includegraphics[height=1em]{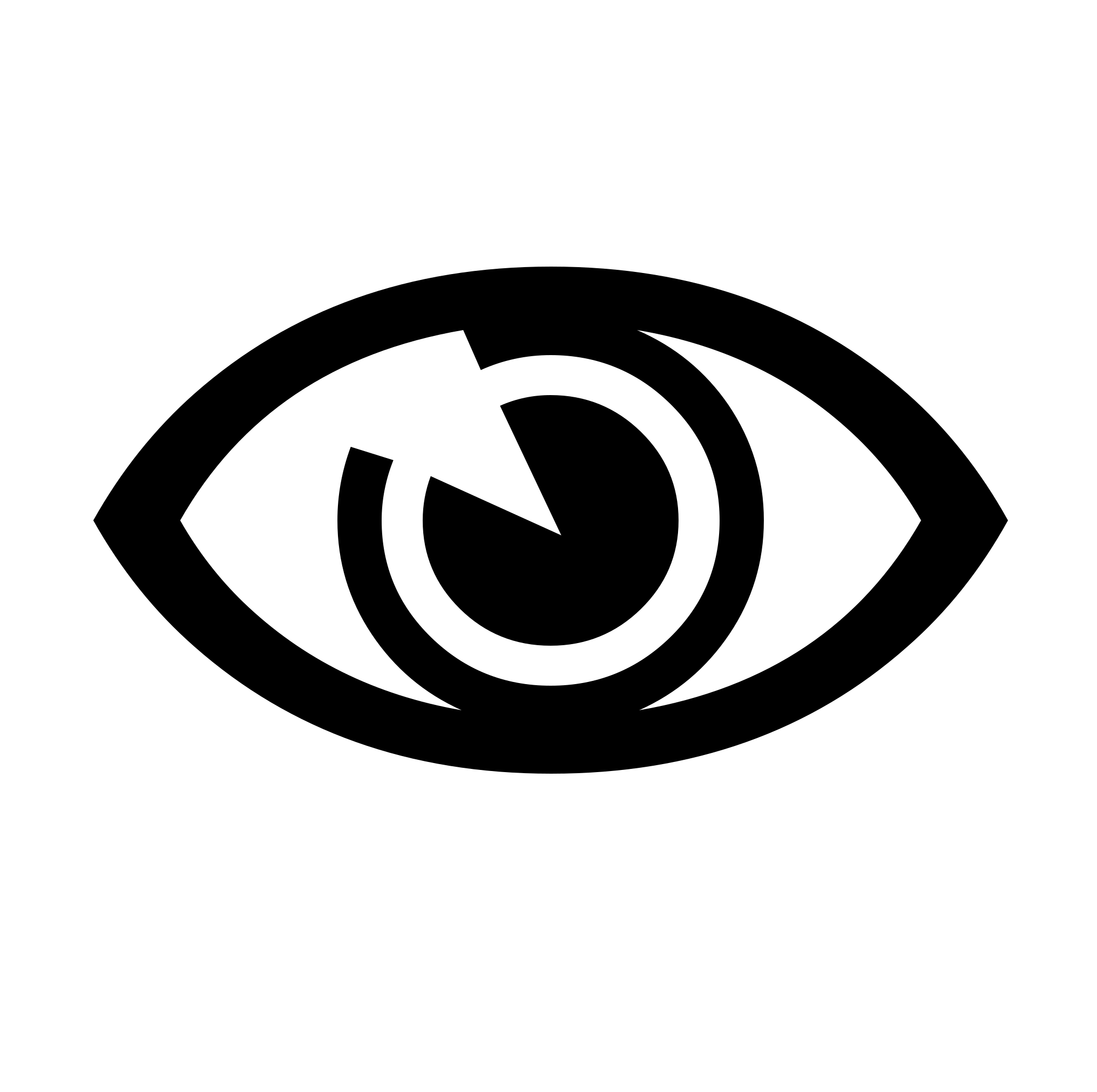}}) and directing finger (\raisebox{-.75ex}[0pt][0pt]{\includegraphics[height=1.2em]{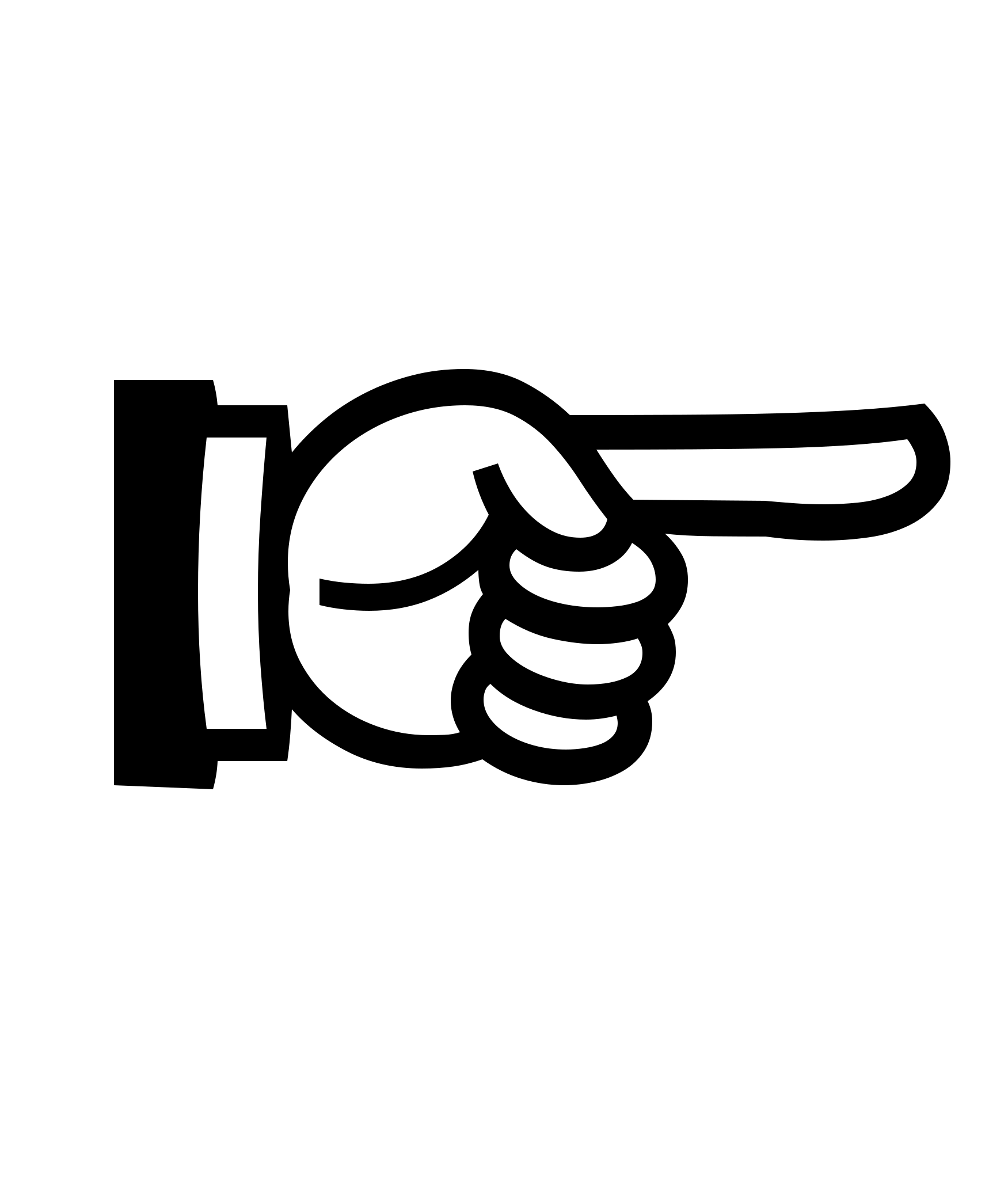}}) denote predictive and prescriptive usage of the different method types, respectively.}
       \label{fig:upset-plot}
\end{figure*}

Figure \ref{fig:upset-plot} provides an UpSet plot \cite{lex2014upset}, which uses a matrix layout to summarize how the different method types for prediction and prescription (cf. Section \ref{sc:method-types}) are combined in the set of surveyed papers. Unlike Figure \ref{fig:method-bar}, this visual representation makes a distinction in terms of whether the different method types were used for providing predictions or prescriptions in the DPSA workflow (cf. Definition \ref{def:DPSA}). How individual papers were associated with each method combination is described in \ref{app:d}.

The bottom row-wise bar chart in Figure \ref{fig:upset-plot} summarizes the relative frequency of different method types for prediction and prescription \textit{in isolation}, with the foreseeing eye (\raisebox{-.55ex}[0pt][0pt]{\includegraphics[height=1em]{images/eye}}) and directing finger (\raisebox{-.75ex}[0pt][0pt]{\includegraphics[height=1.2em]{images/finger}}) denoting predictive and prescriptive usage for each type, respectively. The top column-wise bar chart summarizes the relative frequency of different \textit{combinations} of method types for prediction and prescription. Zero-frequency rows and columns are left out in the plot.

Several conclusions can be drawn from the results encoded in Figure \ref{fig:upset-plot}. For one, the popularity of individual method type usages and combinations, as encoded in the horizontal and vertical bar charts, respectively, both seem to follow a \textit{power law}, with five method combinations accounting 90.2\% of all surveyed papers. 

Now looking to the horizontal bar chart, note that DM/ML as well as EXP are used for both prediction and prescription, although predictive usage is more than three times as common for both method types. The three remaining method types are used solely for either prediction or prescription. 

For predictions, DM/ML dominates, with (necessarily data-driven, cf. Definition \ref{def:DPSA}) PROB being its replacement of choice in the remaining percentage of papers not using it for predictions. Predictive EXP only occasionally played a supplementary role together with predictive DM/ML. 

As for prescriptions, OPT dominates, being significantly more popular than DM/ML and EXP, its closest runner-ups in terms of popularity. Combinations of several prescriptive methods were confined to SIM applications, with one application solving a travelling salesman problem (OPT) during simulation to suggest optimal search paths for shop floor fault detection \cite{Stein2018}, and another using SIM to evaluate different clinical scheduling policies common to the domain (EXP) \cite{Srinivas2018}.

Now looking to the vertical bar chart, first note that the combination of DM/ML and OPT is the most popular methodological mix by far - arguably not a surprising result, given that this solution might indeed be deemed archetypal for PSA as such \cite{frazzetto2019prescriptive}. Note however, that 52.8\% of all surveyed papers, i.e., more than half of them, does \textit{not} use this method combination or any superset thereof. This result challenges the notion that DPSA is only about combining ML and optimization.

\begin{figure*}[!ht]
       \centering
       \includegraphics[width=\linewidth]{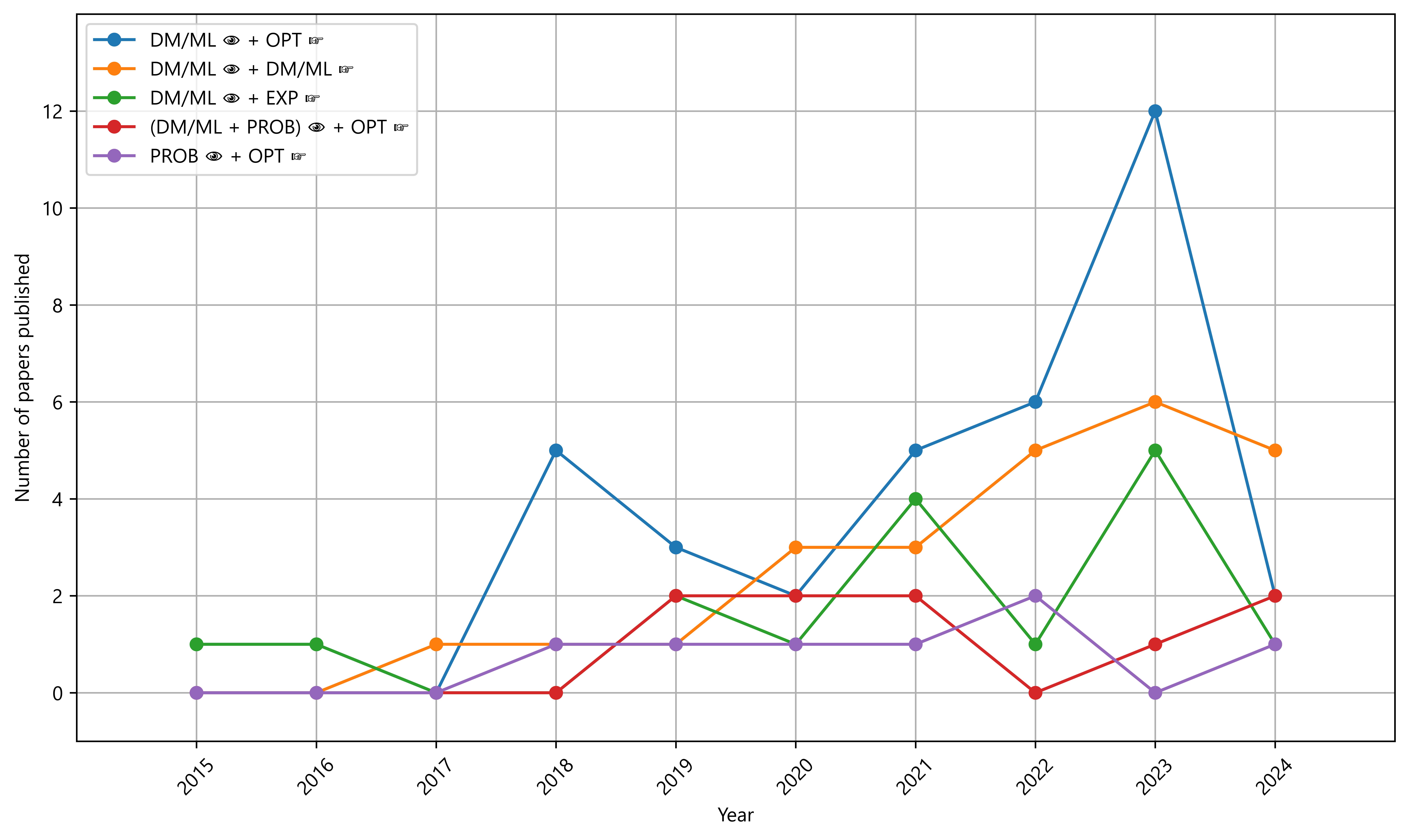}
       \caption{Line chart illustrating the relative popularity of the top five most popular method combination over time in terms of the number of published papers per year.}
       \label{fig:combination-temporal}
\end{figure*}

Similarly, Figure \ref{fig:combination-temporal} shows that while combinations of DM/ML and OPT have become increasingly dominant over time among the top five most popular method combinations within DPSA, solutions instead using DM/ML or EXP for prescriptions have also become increasingly popular. This implies that methodological exploration has \textit{not} yet stagnated within the field, despite what the increasing dominance of the DM/ML and OPT method types in Figure \ref{fig:method-temporal} might suggest in a vacuum.

\subsection{Two generic DPSA workflow patterns}\label{sc:generic-dpsa-workflows}
Understanding analytics workflows through method combinations only provides a relatively abstract view of DPSA methodologies. To form a conceptual model on \textit{how} these methods are combined, we consider more or less method-agnostic \textit{workflow patterns} in this and the subsequent two sections.

\begin{figure}[!ht]
       \centering
       \includegraphics[width=0.8\linewidth]{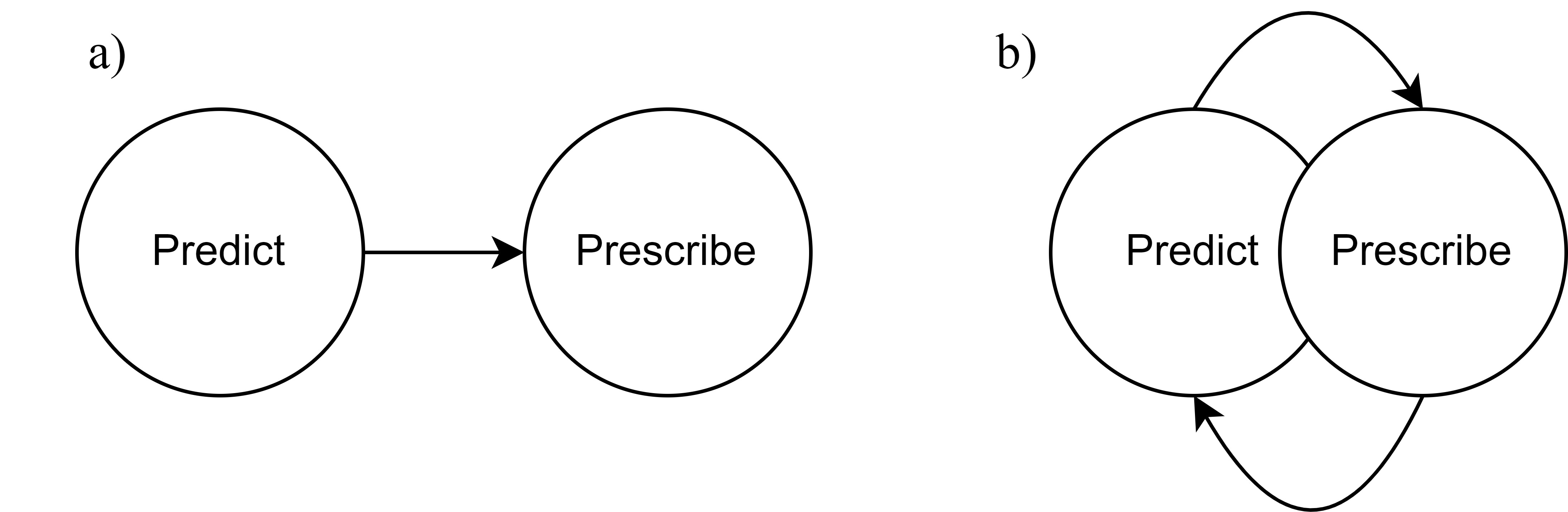}
       \caption{A visual illustration of how generic a) PTP and b) PWP workflows are structured, being purely sequential in the case of PTP, and including overlaps or mutual transitions in the case of PWP. An attempt to illustrate an idea, and neither graphs nor sets.}
       \label{fig:ptp-vs-pwp}
\end{figure}

First, there is a need for an overarching framework to help structure this analysis, provided by two method-agnostic workflow patterns presented in this section. As it turns out, a useful piece of information about different types of DPSA workflows is whether they follow what we call a \textit{Predict-Then-Prescribe} (PTP) or a \textit{Predicting-While-Prescribing} (PWP) workflow pattern. In the former generic pattern, prediction and prescription take place in a \textit{sequential} flow, from prediction to prescription, with prediction effectively ending once prescription begins. In the PWP pattern, prediction and prescription form \textit{simultaneous} relationships. More specifically, this either means that analysis iterates between solving predictive and prescriptive subproblems, or that predictive and prescriptive problem solving are handled by unified methods. Figure \ref{fig:ptp-vs-pwp} provides an attempt to illustrate this distinction visually.

\begin{figure*}[!htbp]
       \centering
       \includegraphics[width=0.8\linewidth]{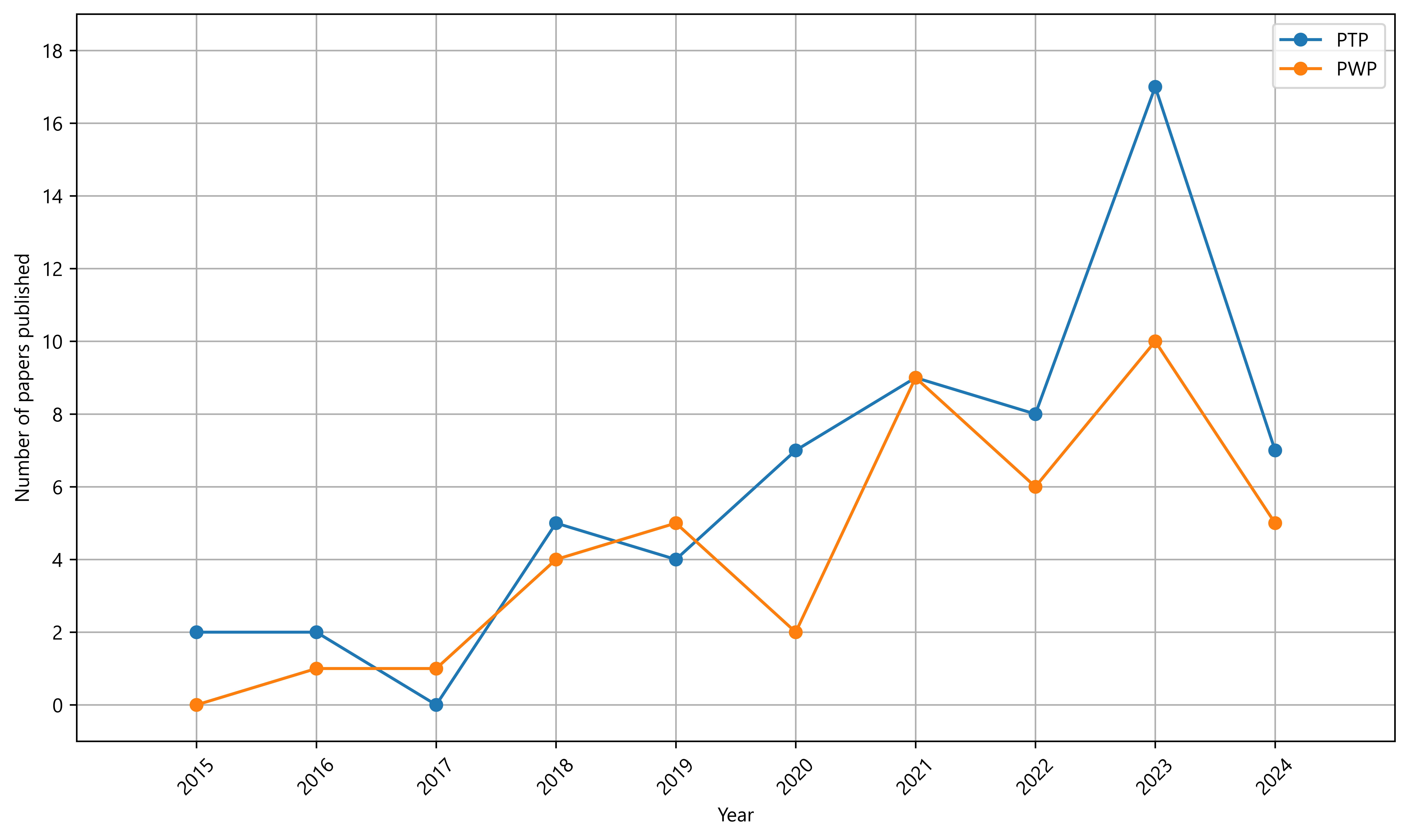}
       \caption{Line chart illustrating the relative popularity of the PTP vs. the PWP pattern over time in terms of the number of published papers using each pattern.}
       \label{fig:temporal-workflows}
\end{figure*}

The PTP/PWP distinction is deemed useful in an information-theoretical sense, since it forms an approximate 50/50 percentage-wise split of the set of surveyed papers. As illustrated in Figure \ref{fig:temporal-workflows}, the relative popularity of the PTP vs. the PWP pattern has been close over the years, with only 2023 being an outlier in terms of PTP pattern dominance. Furthermore, as we shall see, several concrete workflow patterns and methodological choices are furthermore \textit{exclusive} to either generic workflow pattern. This bisection is therefore in a sense a ''clean cut'' and useful for wrapping our heads around an otherwise untamed territory of methodologies. 

As the reader may be aware, the notion of a \textit{Predict-Then-Optimize} workflow pattern already exists within the methodological PSA literature \cite{elmachtoub2022smart}, with the same meaning as PTP for mathematical optimization specifically. PTP is simply a generalization to prescriptive method types beyond OPT, according to the diversity of prescriptive methods found in this survey (cf. Section \ref{sc:method-types}). The PWP pattern has to our knowledge never been explicitly named and subjected to formal analysis - possibly since such workflows form newer and rarer contributions to PSA research. 

\begin{figure*}[!htbp]
       \centering
       \includegraphics[width=\linewidth]{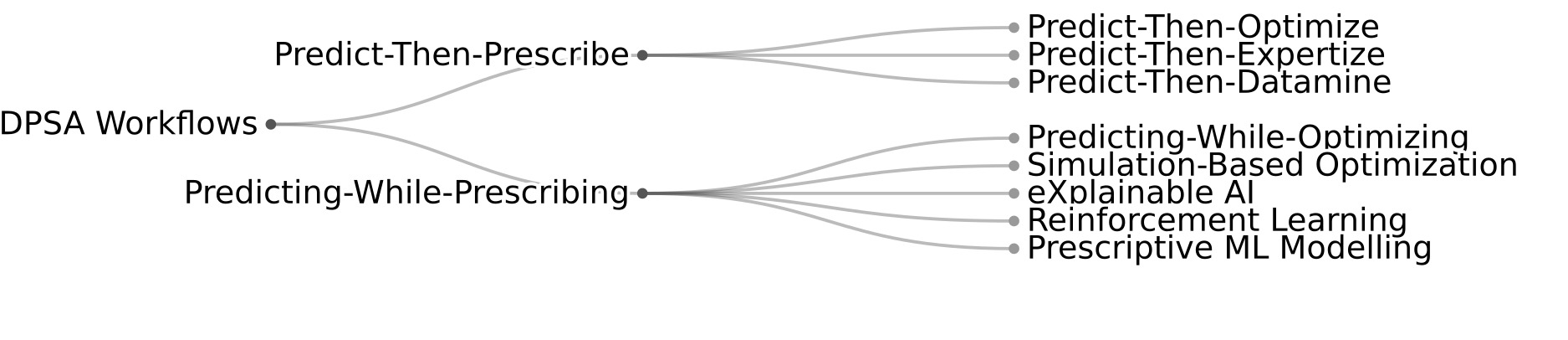}
       \caption{Overview of (generic) workflow patterns encountered within DPSA application research.}
       \label{fig:workflow-tree}
\end{figure*}

Figure \ref{fig:workflow-tree} provides a taxonomy of workflow patterns encountered within DPSA research. The two subsequent sections will discuss each generic workflow pattern further in turn.

\subsection{Predict-Then-Prescribe (PTP) workflows}
61 surveyed papers, or 58.7\% of all surveyed papers, use a PTP workflow in their solutions. Accounting for 39 of these papers, the most widespread instance of PTP by far is the well-known \textit{Predict-Then-Optimize} variant \cite{elmachtoub2022smart}, in which the role of predictive models, usually based on DM/ML (cf. Figure \ref{fig:upset-plot}), is to provide inputs to a subsequent prescriptive OPT phase, using Exhaustive Search \cite{Chen2022,Goyal2016}, Dynamic Programming \cite{Ayhan2018,Yu2021}, Linear Programming \cite{Bahulkar2018,Borenstein2023,Rakhmasari2018}, Mixed-Integer Programming \cite{Ceselli2019,Heide2020,Bertsimas2021b,Williams2022,Consilvio2019,karakaya2024sensor,yang2024prescriptive}, Integer Programming \cite{Sinha2023,Brandt2022,Ceselli2018,Jacquillat2022,Bischhoffshausen2015,Kiaghadi2023,Birolini2023b,Mehrotra2020,Pessach2020,Galli2021,Birolini2023a}, Convex Programming \cite{Kumar2023,Qu2020,Sangwan2023,Shi2020,hassanzadeh2024conclude}, Non-Convex Programming \cite{Keskin2022,Amiel2023}, or Heuristic Optimization \cite{Kandula2021,consilvio2024data,oudani2023prescriptive} (cf. Section \ref{sc:methodtype:opt}). In concrete terms, this means that the role of the predictive model is to fill in initial blank coefficients in the optimization model, representing demand forecasts or otherwise, before an optimization algorithm is utilized. 

''Smart-Predict-Then-Optimize'' is a subvariant of the Predict-Then-Optimize pattern, which seeks to solve the initial prediction task to minimize the loss (or decision regret) from making the wrong decision in the downstream prescription task \cite{Tian2023a,Tan2021,Yan2023,chen2024towards}. Such workflows provide more integration between prediction and prescription, without introducing the temporal overlaps characteristic of the PWP pattern. 

Aside from certain advanced extensions falling under the PWP pattern, Linear Programming generally implies a PTP workflow in the set of surveyed papers. The technical reason behind this phenomenon might be that popular, useful forecasting models, like tree-based ensembles or Logistic Regression for instance (cf. Section \ref{sc:methodtype:dmml}) tend to have non-linear, black-box, and/or non-parametric properties that are \textit{incompatible} with the restrictions of linear modelling. The only sure-fire way to combine such forecasting models with Linear Programming is therefore to pre-compute forecasts as linear optimization model coefficients. 

Remaining surveyed papers using PTP rely on either EXP or DM/ML for providing prescriptions, with the corresponding workflow patterns respectively being \textit{Predict-Then-Expertize} and \textit{Predict-Then-Datamine} in Figure \ref{fig:workflow-tree}. 

After having obtained exemplars \cite{Brockett2019,Cho2015,Du2016}, scheduling requirements \cite{Oberdorf2021,Salah2022}, found defects to be remediated \cite{Vater2019,Vater2020,reisch2023prescriptive}, possible yields from new employees or loans \cite{kurniawan2023prescriptive,zhu2024optimal}, or potential uplift per customer \cite{Caigny2021,Devriendt2021,Gubela2021,Sanisoglu2023,singh2023machine} from predictive technologies like DM/ML, heuristics and problem domain reasoning can be applied to provide prescriptions in a subsequent phase (cf. Section \ref{sc:methodtype:exp}), analogously to the Predict-Then-Optimize pattern. As it turns out, \textit{all} surveyed papers relying solely on EXP for prescriptions utilize PTP workflows. This could perhaps be explained by the notion that implementing sound domain reasoning and heuristics with respect to complex forecast models generally is easier, if the input for this analysis is simply the forecast value as is.

As for prescriptive DM/ML, prescriptions through Nearest Neighbour Queries \cite{Kuzyakov2020,Ramannavar2018}, feature selection algorithms \cite{Mohan2023}, or trained ML models \cite{Bertsimas2021a,Bertsimas2022,Dash2022,han2022data} can be obtained similarly to the EXP case, after upstream predictive models have provided their input (cf. Section \ref{sc:methodtype:dmml}). As we shall see shortly, prescriptive DM/ML however also comes in different shapes within the PWP pattern.

\subsection{Predicting-While-Prescribing (PWP) workflows}
43 surveyed papers, or 41.3\% of all surveyed papers, use a PWP workflow in their solutions. 
Accounting for 17 of surveyed papers using PWP workflows, the most common scenario is that prediction is utilized as an optimization subroutine, meaning that outputs from predictive models are required \textit{online}, i.e., during the optimization process (cf. \textit{Predicting-While-Optimizing} in Figure \ref{fig:workflow-tree}). One reason for why this might be necessary is that forecasts in the optimization model might depend on the assignment of decision variables, and that it is often impractical or inefficient to generate forecasts for all possible assignments before running optimization for non-trivially-sized search spaces. 

Consider for instance one paper in which the goal was to optimize three fatty acid properties of a biodiesel product in concert \cite{Suvarna2022}. The authors used three ML models to predict each of these fatty acid properties, each taking a number of continuous decision variables as input concerning the chosen acid composition of the product. Two tree-based ensembles and a Support Vector Machine, i.e., all non-parametric methods, were used. In such cases, the best option might be to use an optimization routine leveraging online predictions from these models.

The need for online predictions from what is usually predictive DM/ML models (cf. Figure \ref{fig:upset-plot}), somewhat limits the array of applicable optimization methods able to cope with these cases. As found in this survey, Heuristic Optimization methods \cite{Brandt2021,Grzegorowski2022,Lash2016,Lee2022,Ravi2021,Ribeiro2023,Suvarna2022,Thammaboosadee2018,Yanta2021,Huang2019} were strongly associated with the PWP pattern, yet Exhaustive Search was also utilized \cite{Achenbach2018,Chaudhry2023,Li2019,Hauser2021}, and one paper used online gradient estimations of a Support Vector Regression model to solve a convex optimization problem \cite{John2019} (cf. Section \ref{sc:methodtype:opt}).

Certain extensions of Stochastic Programming also require forming online predictions. \textit{Stochastic programming with recourse} \cite{hansotia1980stochastic} is useful for modelling multi-stage decisions, in which stochastic elements of the optimization model have to be instantiated in initial optimization stages, before \textit{recourse decisions} can be made in subsequent optimization stages. One paper used this technique to optimize nurse transshipment decisions, by generating initial demand estimates followed by recourse decisions during optimization \cite{Shi2021}. Another paper used a recourse mechanism to estimate product inventory levels in a sequence of mixed-integer programs \cite{Caro2019}.

Analogously to PWP solutions using prescriptive OPT, SIM solutions using \textit{Simulation-Based Optimization} (cf. Section \ref{sc:methodtype:sim}) also exclusively follow a PWP pattern \cite{Adulyasak2023,Jabr2023,Srinivas2018,Stein2018,Tan2021,sariyer2024predictive}.

As seen in Figure \ref{fig:workflow-tree}, remaining PWP workflows all rely on some form of DM/ML. Similarly to the aforementioned cases, papers using techniques from \textit{eXplainable AI} to form prescriptions using SHAP values or DiCE \cite{Delen2021,Susnjak2023} (cf. Section \ref{sc:methodtype:dmml}) need to run online predictions to obtain the required output from the aforementioned methods, in that these methods are designed to work for a variety of black-box models in need of explainability \cite{lundberg2017unified,mothilal2020explaining}.

In a sense, \textit{Reinforcement Learning} embodies the PWP pattern. Some reinforcement learning algorithms solve predictive and prescriptive problems \textit{concurrently} - the value and policy function, respectively estimating the prospect of being in a state and the optimal way to navigate from it, being ''moving targets'' for each other until convergence \cite{sutton2020introduction}. Whether they use data-driven reinforcement learning as a stand-alone approach \cite{Ferreira2022,Khan2021,Meng2020,Tham2023,osakwe2024towards} or supplement it with additional predictive methods used in an online fashion \cite{Cakir2023,Mandl2021,Zhou2023}, papers using reinforcement learning all at least embody the PWP pattern in such a sense.

Other applications solely using DM/ML follow the PWP pattern by training a data-driven predictive model to provide prescriptions \cite{Fang2019,Mandl2021,Notz2022,Notz2023,Schwartz2017,Punia2020,mohd2022prescriptive,stratigakos2024interpretable,ara2024collaborative,razgallah2024using}, thus using \textit{Prescriptive ML Modelling} (cf. Figure \ref{fig:workflow-tree}). In other words, such papers have unified data-driven prediction and automatic prescription in one model, used in a similar way to policy functions from Reinforcement Learning. 

To give a concrete example, one paper trained the coefficients for their prescriptive model, which calculates Random Forest kernelized weights for the input vector with respect to a set of historical data points (cf. Section \ref{sc:methodtype:dmml}), by solving a convex optimization problem involving said Random Forest kernel \cite{Notz2022}. In another paper, it was proposed to solve several instances of the core optimization problem of the application at hand, and then use this dataset of problem-solution pairs to train a tree-based ensemble to predict optimal solutions for new input vectors, which experiments indicated to be an viable approach compared to alternatives \cite{Notz2023}.

\section{Synthesis on methods}\label{sc:rq2-synthesis}
As Sections \ref{sc:method-types} and \ref{sc:method-patterns} have covered while addressing RQ2, several definite methodological trends characterize the state of current DPSA application research, both in terms of individual methods and their combinations.

DPSA applications can be said to pivot around \textit{five method types}, or ''clusters'' if such imagery is preferred, among which DM/ML and OPT dominate (cf. Figure \ref{fig:method-bar}), and has become increasingly dominant in recent years (cf. Figure \ref{fig:method-temporal}), yet with several viable alternatives in the shape of PROB, EXP, and SIM. While a pronounced emphasis on \textit{tree-based ensembles} for predictions and \textit{Linear Programming} methods for prescriptions was found (cf. Sections \ref{sc:methodtype:dmml} and \ref{sc:methodtype:opt}), individual method choices varied widely between applications, exhibiting a textbook level of coverage for the most popular method types. To our knowledge, the various prescriptive EXP methods described in Section \ref{sc:methodtype:exp} have not been covered explicitly elsewhere, despite the popularity of prescriptive EXP (cf. Figure \ref{fig:upset-plot}).

In terms of how methods were combined, we found that DPSA applications pivot around a relatively small number of method combinations among the aforementioned types (cf. Section \ref{sc:method-combinations}), utilizing primarily DM/ML and secondarily PROB for predictions, and primary OPT, supplemented by DM/ML and EXP, for prescriptions. We observe from Figure \ref{fig:upset-plot} that there is seemingly a \textit{power law} at play in terms of popularity of individual method types for prediction/prescription, as well as their combinations, with (combinations of) DM/ML and OPT clearly being the most popular, increasingly over time as well (cf. Figure \ref{fig:combination-temporal}), yet contested by EXP for prescriptions and PROB for predictions. 

As for workflow patterns, we identified \textit{two generic patterns} for structuring DPSA workflows, dividing the set of surveyed papers into two evenly-sized parts (cf. Section \ref{sc:generic-dpsa-workflows}). The \textit{Predict-Then-Prescribe} pattern is already well-known within the methodological PSA literature in some form \cite{elmachtoub2022smart}, as a \textit{sequential} pattern with a clear-cut separation between prediction and prescription. The alternative \textit{Predicting-While-Prescribing} pattern, as introduced in this survey, breaks the mold of PTP, bringing prediction and prescription closer together by \textit{simultaneous} relationships (iterative and/or overlapping), with recent methodological developments even unifying data-driven prediction and automatic prescription in one model. 

All in all, while we find that DM/ML and OPT, as well as their combination in the PTP pattern, is a staple solution among DPSA applications, a substantial number of methodological alternatives with viable niches is also being utilized in real-world DPSA, forming flourishing meadows in its methodological wilderness.

\section{Promising research directions}\label{sc:rq3}
Based on pervasive patterns found in the set of surveyed papers, we derive five promising research directions (RD1-5) to address RQ3, presented in the following subsections.

\subsection{Under-explored application domains}
As found in connection to RQ1 (cf. Section \ref{sc:rq1}), DPSA has already found use in a variety of identified application domains, and our temporal analysis demonstrates an ebb and flow in terms of relative prominence among these domains over time (cf. Figure \ref{fig:temporal-domains}). Still, several prominent real-world business domains have been left (almost) unexplored by the set of surveyed papers as a whole. Using UN's \textit{International Standard Industrial Classification (ISIC)}, revision 4, \cite{united2008international}, we have identified eleven business domains in which at most three DPSA applications were found in this survey. Leaving out the miscellaneous sections S, T, and U \cite{united2008international}, the eight remaining business domains and section letters, along with associated surveyed papers, if any, are the following: 

\begin{itemize}
    \item Agriculture, forestry and fishing (A) \cite{Cakir2023}
    \item Mining and quarrying (B)
    \item Water supply; sewerage, waste management and remediation activities (E)
    \item Construction (F) \cite{Li2019}
    \item Accommodation and food service activities (I) \cite{han2022data}
    \item Financial and insurance activities (K) \cite{Ravi2021,Heide2020,kurniawan2023prescriptive}
    \item Real estate activities (L)
    \item Arts, entertainment and recreation (R) \cite{razgallah2024using,hassanzadeh2024conclude}
\end{itemize}

These sections touch upon a variety of areas, including primary industries (A, B, E), public infrastructure (E, F), the service industry (I, K, L), as well as creative enterprises (R). 

Several reasons could underlie the sparsity of DPSA applications within these specific areas. For traditionally asset-heavy, field-based industries like agriculture (A), mining (B), water/waste (E), and construction (F), \textit{data collection}, e.g., through sensors, might still generally be too fragmented, manual, or sparse to properly support DPSA workflows. Within construction (F), arts and entertainment (R), along with real estate (L), decision processes traditionally being highly \textit{non-routine, subjective, or context-dependent}, and thus harder to generalize, might have hindered adoption. Within real estate (L) along with finance and insurance (K), adoption of DPSA might be hindered by highly regulated, high-stakes decision processes with a need for higher \textit{auditability and explainability} than current DPSA solutions support. For accommodation and food services (I), arts and entertainment (R), along with agriculture (A), the lack of adoption might be a matter of \textit{resource-efficiency}: These industries might contain many small and medium-sized enterprises who lack the necessary data basis, IT infrastructure, employee skills, etc. to adopt current DPSA solutions. Note that all of the above highlighted issues are merely \textit{hypothetical possibilities}. The real-world challenges of implementing DPSA within these domains still remain to be uncovered and solved. This is an interesting direction for future research in itself.

To further the proliferation of DPSA applications towards new frontiers, cross-fertilizing an already interdisciplinary field with new problems and solutions, we thus propose as a possible research direction to: 

\noindent \textit{RD1: Develop impactful DPSA applications within under-explored business domains.}

\subsection{Methodological alternatives to (mixed-)integer linear programming}\label{sc:rd3}
While covering OPT methods in Section \ref{sc:methodtype:opt}, we found that Linear Programming, and in particular (Mixed-)Integer Linear Programming ((M)ILP) algorithms, were the most dominant class of methods by far. What we didn't dwell on, for the sake of brevity, is the frequently recurring observation that \textit{one does not simply use (M)ILP for DPSA}. Among DPSA applications there is a strong tendency of various difficulties arising whenever authors try to apply (M)ILP to real-world problems, calling for improvised modelling tricks, heuristics, and workflows. 

While several surveyed papers get by with relatively small problem sizes (e.g., \cite{Kiaghadi2023,Sinha2023,Tian2023a,Bischhoffshausen2015}), the NP-hardness of (M)ILP means that its scalability with respect to an increasing number of decision variables is a recurring concern (e.g., \cite{Heide2020,Birolini2023b,Shi2020,Brandt2022,Bischhoffshausen2015,Kandula2021}). Some researchers thus resort to simplifying their original optimization model by various heuristics and modelling tricks \cite{Ceselli2018,Jacquillat2022,Tian2023a,Caro2019,Shi2020,Pessach2020,Heide2020}, or they leverage knowledge about ''what goes'' within the problem domain and split the optimization model into smaller subproblems to be solved iteratively in custom workflows \cite{Ceselli2018,Jacquillat2022,Tian2023a,Caro2019,Birolini2023a,Ceselli2019,Consilvio2019,Mehrotra2020,Heide2020}. 

Another challenge with (M)ILP when considering DPSA specifically, is the popularity of predictive DM/ML (cf. Figure \ref{fig:upset-plot}), in particular non-parametric or non-linear models (cf. Section \ref{sc:methodtype:dmml}). Limited options for embedding such models into linear optimization models makes (M)ILP less \textit{expressive}, to the possible detriment of solution quality, e.g., as found in \cite{Huang2019}. 

A different concern is about \textit{usability}: Does the lack of alternatives to (M)ILP, which as discussed tends to be the very opposite of a plug-and-play method, limit the reach of DPSA to a broader range of users than, e.g., researchers with highly specialized skill sets? 

We recognize that it might be impossible to design a \textit{perfect} alternative to current prescriptive methods, providing hard guarantees without hard costs. Still, in our temporal analysis of Figure \ref{fig:combination-temporal} we observe increasing efforts to find good alternative prescriptive methods to rigid OPT strategies piloted by domain experts. Specifically, leveraging DM/ML directly for prescriptions, as manifested in various PTP/PWP workflows (cf. Section \ref{sc:generic-dpsa-workflows}), seems like a promising direction - especially due to how data-driven methods like Deep Learning have already made great strides in solving hard prediction problems while being relatively easy to pilot with high-level tools \cite{chollet2021deep}. There is indeed already a considerable body of work within ML investigating how to approximately solve (especially) combinatorial optimization problems, e.g., Travelling Salesman problems, with neural networks \cite{bengio2021machine}. Yet to our knowledge, none of these works have considered how to incorporate these methods into complete DPSA workflows. Therefore, we propose as a possible research direction to: 

\noindent \textit{RD2: Develop scalable, expressive, usable, and reliable prescriptive DM/ML methods for DPSA applications.}

\subsection{Hybrid processing solutions for Big DPSA}
Extracting value from large datasets, i.e., Big Data, can be said to form a significant part of a \textit{narrative} surrounding PSA, in which PSA, extending the reach of its descriptive and predictive counterparts (cf. Figure \ref{fig:gaam}), is the third and final piece to unlock maximum Value from Big Data in a fully automated, large-scale analytics infrastructure \cite{basu2013five,deshpande2019predictive,soltanpoor2016prescriptive}. As shown by our temporal analyses in Figures \ref{fig:method-temporal} and \ref{fig:combination-temporal}, DM/ML has indeed played an increasingly prominent role within the DPSA field in recent years, both for predictions and prescriptions, and a substantial number of surveyed papers correspondingly discuss the potentials of Big Data in their opening or closing sections \cite{Achenbach2018,Adulyasak2023,Brandt2021,Caigny2021,Notz2023,Oberdorf2021,Punia2020,Tham2023,Vater2019,Notz2022,Stein2018}. Nonetheless, hardly any DPSA applications in this survey work with what we would even \textit{consider} to be Big Data. 

To clarify our own definition of the phenomenon, we prefer to avoid arbitrary distinctions counted in bytes or cardinalities and instead define \textit{data-intensive processing} by the need for \textit{distributed processing} - as exemplified by the primary use cases of distributed data ingestion, analytics and storage systems like Apache Kafka, Spark, and Cassandra \cite{thein2014apache,zaharia2016apache,chebotko2015big}. In other words, if a dataset can feasibly be analysed on one workstation/server with respect to application requirements, as done in most surveyed papers (e.g., \cite{Ceselli2018,Grzegorowski2022,Kumar2023,Ribeiro2023,Salah2022,Bischhoffshausen2015}), then any standard analytics solution like Excel and MySQL might suffice, and said dataset is not considered Big.

A few surveyed papers describe their own work as involving Big Data or evidently use distributed processing to handle large datasets \cite{Goyal2016,Mandl2023,Mehrotra2020,Ramannavar2018}. Yet these papers tend to omit implementation details, in some cases possibly for confidential reasons \cite{Mehrotra2020}, while others only analyse single-node-sized datasets \cite{Ramannavar2018}. 

The promise of Big Data being a loud refrain inside PSA \textit{narratives} while only being a faint whimper inside \textit{actual} DPSA applications is surprising on the surface. We shall not speculate on human motives and ascribe this finding to, e.g., hypocritical marketing or a lack of industrial buy-in. We rather hypothesize that there might be a simple, technical reason behind this situation: \textit{With current methods, DPSA is actually predominantly compute-intensive}. As Section \ref{sc:rd3} discusses, applications currently struggle with the scalability of popular prescriptive methods like (M)ILP with respect to the number of decision variables and constraints - becoming a computational bottleneck way before any Big Data issues attain relevance.  

Like previous surveys on PSA \cite{lepenioti2020prescriptive,frazzetto2019prescriptive}, we still find that the promise of Big Data has not come to fruition within DPSA application research. Unlike previous works, we however propose that unlocking the Value of Big DPSA requires \textit{hybrid processing}, handling \textit{both} data- and compute-intensiveness. Creating such solutions introduces compounded scalability challenges. Namely, it requires integrating high-throughput data pipelines with complex optimization and search tasks that are notoriously difficult to parallelize efficiently - each with their own compute profiles and fault tolerance mechanisms \cite{Ralphs2016,kleppmann2017designing}. In sum, we propose as a research direction to: 

\noindent \textit{RD3: Develop novel architectures, methods, and tools that support hybrid processing of Big DPSA workflows.}

\subsection{Dedicated tools for emerging DPSA workflow patterns}
From our analyses concerning RQ2 (cf. Sections \ref{sc:method-types} and \ref{sc:method-patterns}), we note that DPSA applications display a high degree of methodological diversity, both in terms of individual methods and how they are combined, having diversified further over time (cf. Figure \ref{fig:combination-temporal}). Accommodating such a wide variety of workflows poses a challenge to developers of dedicated DPSA tools: As is evident from the aforementioned sections, a PTP pattern with ML for predictions and Linear Programming for prescriptions is but one way to structure a DPSA workflow among many. As for (M)ILP specifically, we furthermore saw in Section \ref{sc:rd3} that significant workflow variations even exist within this one category.

Dedicated DPSA tools, integrating predictive and prescriptive technologies with a data management layer, might nonetheless help their users avoid improvised ''duct-tape-style'' workflows that are both error-prone to develop and exhibit poor performance \cite{frazzetto2019prescriptive}. Solutions like SolveDB+, a relational database management system for PSA, have already been developed in this direction \cite{vsikvsnys2016solvedb,siksnys2021solvedb+}, yet they only target a particular subset of workflows discussed in this survey, namely the ones combining predictive DM/ML with prescriptive OPT.

As seen in Figure \ref{fig:combination-temporal}, applications relying on prescriptive DM/ML or EXP in their workflows have been on the rise in recent years. We therefore expect a correspondingly growing tool gap within this niche. As described in Section \ref{sc:methodtype:dmml}, prescriptive DM/ML solutions are still quite application-bespoke and in need of generalization, e.g., as part of a tool, in order to be applicable to a broader set of different, but similar problems. As for prescriptive EXP workflows, domain-specific tools offering strong modelling and reasoning capabilities at a high level of abstraction might be a viable path forward. 

How to best offer tool support for emerging DPSA workflow patterns, be it in one unified tool, in several tools with good mutual integration, or otherwise, we leave as a possible research direction: 

\noindent \textit{RD4: Develop flexible and efficient tools for DPSA workflow patterns relying on prescriptive DM/ML or EXP.}

\subsection{Solutions for managing DPSA in production}

In our selection criteria of this survey, we require under C5 that applications must be ''end-to-end'' - i.e., that they must solve concrete problems with concrete methods and not just present hypotheticals (cf. Section \ref{sc:method:criteria}). A stricter notion of ''end-to-end'' concerns the entire \textit{life cycle} of an application, including how the solution is deployed, monitored, and maintained in a production setting under changing requirements. One would expect that managing DPSA solutions in the wild is non-trivial. To mention a few example possibilities: They might need to interface with real-time control systems and/or people under strict time constraints; they might need safeguards and fallback mechanisms to prevent infeasible, dangerous, or untimely prescriptions; they might need specialized logging, instrumentation, and metrics to detect any performance deviations over time.  

We are regrettably only able to give hypothetical examples, since our surveyed papers barely provided \textit{any} information on how their solutions were, or might have been, deployed in a production setting, beyond simply stating that the solution was, will or could be ''deployed'' (e.g., \cite{Brockett2019,Du2016,Borenstein2023}). Other papers only consider individual issues that relate to the broader application life cycle - for example, the design of a dashboard to interface with their solution (e.g., \cite{Susnjak2023,Kumar2023}), or the automation of their ML pipeline (e.g., \cite{Ribeiro2023}). Still, deployment details are surprisingly \textit{absent} in our surveyed papers.

\textit{MLOps} concerns how to properly deploy, monitor, and maintain ML solutions in production \cite{kreuzberger2023machine}, which has been a major roadblock for the adoption of ML in industry \cite{sculley2015hidden}. Within the DPSA field, a narrow data scientist's focus on how to combine various methods into novel \textit{workflows} would leave it blind to important engineering issues related to the \textit{life cycle} of its solutions, preventing them from broader adoption in industry. We therefore find that DPSA could benefit from exploring engineering-based approaches to ensure practical, maintainable, and robust deployments of its solutions - i.e., ''\textit{DecisionOps}'', in a similar vein to MLOps. More concretely, we propose as a research direction to:

\noindent \textit{RD5: Develop novel architectures, methods, and tools to deploy, monitor, and maintain DPSA solutions in production.}

\section{Conclusion}\label{sc:conclusion}
The primary objective of this survey was to investigate where the many applications of DPSA have been and where they are going (cf. Section \ref{sc:introduction}). In this endeavour, we decided to delimit our scope to \textit{Data-Driven PSA}, a field deemed to have a sound basis in the historical origins of PSA as well as recent calls for papers (cf. Section \ref{sc:scope-and-definition}). 

In answering our three research questions in Sections \ref{sc:rq1}-\ref{sc:rq3}, we have provided new conceptual models and temporal analyses of the field as a whole, both in terms of \textit{problem types} (cf. Section \ref{sc:rq1}), \textit{method types} (cf. Sections \ref{sc:method-types}), and \textit{workflow patterns} (cf. Section \ref{sc:method-patterns}). Based on temporal trends and pervasive Gordian Knots found within the set of surveyed papers, we have furthermore provided a foundation for future work by proposing a number of possible \textit{research directions} (cf. RD1-5 from Section \ref{sc:rq3}), both in terms of new methods, tools, and use cases. We can mnemonically summarize the key contributions in our conceptual framework as \textit{10 applications domains}, \textit{5 method types}, \textit{2 generic workflow patterns}, and \textit{5 research directions}. These contributions form a comprehensive model of the current state of the DPSA application research field and where it might be going in the future.

Looking ahead, we believe that this survey can be a useful guide for new recruits and veterans alike. As found in Section \ref{sc:rq3}, there are still ample opportunities for developing new methods, tools, and use cases for DPSA applications with a positive impact on the real world. Revisiting our opening quote in Section \ref{sc:introduction}, we hope that upcoming DPSA applications will continue to find new ways to create the future by predicting it.

\section*{Author contributions: CRediT}
\textbf{Martin Moesmann:} Conceptualization, Data Curation, Formal Analysis, Investigation, Methodology, Validation, Visualization, Writing – original draft, Writing – review \& editing \textbf{Torben Bach Pedersen:} Conceptualization, Funding acquisition, Methodology, Project administration, Supervision, Validation, Writing – review \& editing.

\section*{Acknowledgments}
This work was supported in part by the MORE project funded by the European Commission under grant agreement no. 957345 and the Digital Energy Hub project funded by the Association of Danish Industry.

\appendix

\section{Additional survey statistics}\label{app:a}
\begin{figure}[!ht]
       \centering
       \includegraphics[width=0.6\linewidth]{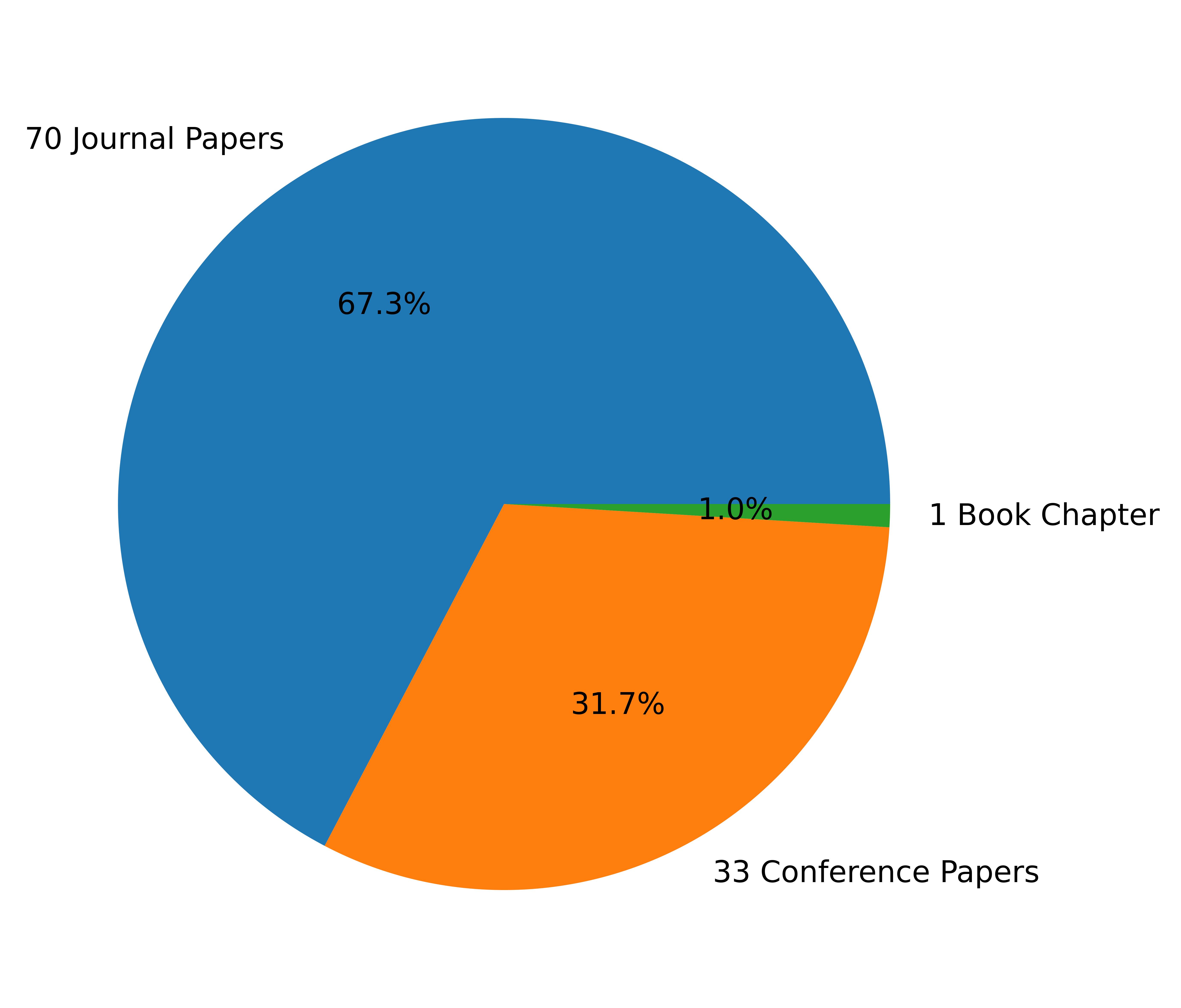}
       \caption{The representation of different outlet types in the set of surveyed papers.}
       \label{fig:paper-type-pie}
\end{figure}

% Appendix one text goes here. You can choose not to have a title for an appendix if you want by leaving the argument blank
Figure \ref{fig:paper-type-pie} summarizes the representation of different outlet types in the set of surveyed papers. Of note, the number of book chapters found is comparatively meagre, there simply being few found books written on PSA-related topics in general. The abundance of journal papers compared to conference papers might in part be attributed to the fact that this survey selects for fully developed applications (cf. Section \ref{sc:method:criteria}). All things being equal, journal papers tend to be further down this pipeline than conference papers. As discussed, the associated selection criterion C5 was introduced in order to be able to adequately address the research questions and assess adherence to Definition \ref{def:DPSA}. 

Table \ref{tab:popularoutlets} summarizes all outlets with more than one paper in the set of surveyed papers. As implied by the table data, there was a high level of diversity among outlets. The most popular ones were however all scientific journals, pivoting around the overlapping topics of decision support, operations research, management science, information systems, and more, with a few journals honing in on specific application domains. That DPSA application papers would be found in these types of outlets is perhaps an unsurprising revelation.

\begin{table}[ht]
    \centering
    \resizebox{\columnwidth}{!}{%
    \begin{tabular}{ll}
    \hline
    \textbf{Outlet}                                & \textbf{Papers (\#)} \\ \hline
    Decision Support Systems                       & 4                         \\ %\hline
    European Journal of Operational Research       & 4                         \\ %\hline
    Management Science                             & 4                         \\ %\hline
    Manufacturing \& Service Operations Management & 4                         \\ %\hline
    Production and Operations Management           & 4                         \\ %\hline
    
    Transportation Research                        & 3                         \\ %\hline
    Computers \& Operations Research               & 3                         \\ %\hline
    
    Computers \& Industrial Engineering            & 2                         \\ %\hline
    Decision Sciences                              & 2                         \\ %\hline
    Health Care Management Science                 & 2                         \\ %\hline
    Journal of Business Research                   & 2                         \\ %\hline
    Journal of Management Information Systems      & 2                         \\ %\hline
    Journal of Marketing Analytics                 & 2                         \\ %\hline
    MT-ITS                                         & 2                         \\ %\hline
    Transportation Science                         & 2                         \\ %\hline
    \textbf{Sum:}                                  & \underline{42}           \\ \hline
    \end{tabular}}
    \caption{Number of papers for all outlets with more than one element in the set of surveyed papers. ''MT-ITS'' abbreviates \textit{International Conference on Models and Technologies for Intelligent Transportation Systems}.}
    \label{tab:popularoutlets}
\end{table}

\section{Extended discussion on individual use cases per application domain}\label{app:b}
The following subsections summarize the work done within each application domain, as identified in Section \ref{sc:rq1}.

\subsection{Academia} 
The focus of these applications is to improve academic outcomes for students or researchers at educational institutions. 

Papers about students generally seek to improve individual \textit{student guidance}. This might involve recommending particular learning activities or modules in order to maximize learning gains \cite{osakwe2024towards,ara2024collaborative}. Some surveyed papers investigate how to reduce risk of drop-out \cite{Susnjak2023,Yanta2021}, and how to help students align their studies and extra-curricular activities with their academic goals \cite{Du2016}. Prescriptions in these works were given as suggested \textit{remedial actions}, i.e., as minimum changes required in the student's current behaviour for a satisfactory predicted outcome. Another paper implemented a solution to help future students decide on the best university to apply to depending on budget and preferences (e.g., host country) \cite{Kiaghadi2023}, maximizing their chance of admission.

One paper targeting \textit{researchers} sought to provide individual recommendations on which journals to submit papers to over time, when to submit, and which researchers to collaborate with \cite{Cho2015}. This was done by using domain knowledge, e.g., about journal impact, and by using exemplar researchers similar to the user as a guideline.

\subsection{Advertising and Marketing}\label{sc:advertising-and-marketing}
The focus of these applications is to improve the value gained by a company from promoting its products. As an overall trend, these applications all have the prescriptive orientation of targeting \textit{individual} clients in the most effective way. 

Some surveyed papers investigate how to provide effective \textit{targeted advertising} through various channels: Using clickstream data to optimize webshop designs (choice of product categories, ranking and average price of highlighted products, etc.) for revisiting and purchasing \cite{Jabr2023,Ferreira2022,Borenstein2023}, or designing coupon bundles for individual households \cite{Chaudhry2023}.

The remaining applications concern themselves with the issue of \textit{uplift analysis} \cite{Gubela2021,Caigny2021,Devriendt2021,Sanisoglu2023,singh2023machine}. Uplift analysis is about designing a treatment, e.g., a promotional campaign, including which customers to target, in order to maximize increases (i.e., ''lifts'') in positive metrics, e.g., profit, due to the treatment \cite{Devriendt2021}. Targeting all customers with gratuitous discounts might however lead to suboptimal uplift. As illustrated in Table \ref{tab:uplift}, some customers (Do-Not-Disturbs and Lost Causes) will not change their behaviour for the better, with Do-Not-Disturbs indeed changing for the \textit{worse}. Sure Things would have engaged with the company regardless, and are thus in a sense wasteful to target. The goal of uplift analysis is to design a campaign, e.g., with the right discount levels, to get the maximum return of investment from Persuadables. The applications in question design campaigns to maximize profit \cite{Gubela2021,Sanisoglu2023} or reduce customer churn \cite{Caigny2021,Devriendt2021}, within E-commerce, telecommunication, B2B software development, and financial institution settings, respectively.

\begin{table}[ht]
    \centering
    \resizebox{0.8\linewidth}{!}{%
    \begin{tabular}{llll}
    \multicolumn{1}{c}{\textit{Engages when treated?}} & \textbf{No}  & \multicolumn{1}{l|}{Do-Not-Disturbs}   & Lost Causes   \\ \cline{2-4} 
    \newline (treatment group)                         & \textbf{Yes} & \multicolumn{1}{l|}{Sure Things}       & Persuadables  \\
                                                       &              & \multicolumn{1}{l|}{\textbf{Yes}}      & \textbf{No}   \\
                                                       &              & \multicolumn{2}{c}{\textit{Engages when not treated?}} \\
                                                       &              & \multicolumn{2}{c}{(control group)} \\
    \end{tabular}}
    \caption{Customer categories considered in uplift analysis \cite{Caigny2021,Devriendt2021}.}
    \label{tab:uplift}
\end{table}

\subsection{Healthcare}
The focus of these applications is to improve the quality of healthcare services. As an overall trend, large-scale, complex, healthcare challenges arising during the recent COVID-19 pandemic seems to have motivated several new applications of DPSA \cite{Bertsimas2021a,Bertsimas2021b,Shi2021}.

In terms of problem types, the larger part of healthcare applications address the use case of \textit{clinical decision making}: How to provide the best possible treatment options when clinical factors and outcomes are uncertain. Concrete decisions include when to use mechanical respiration support systems or diagnose patients with breast cancer \cite{Fang2019}, which long-term treatment strategy to use for diabetes 2 patients \cite{Meng2020}, how to manage sepsis in Intensive Care Units (ICU) \cite{Khan2021}, and which medical drugs to prescribe for hypertensive hospitalized patients with and without COVID-19 \cite{Bertsimas2022,Bertsimas2021a}. As a satellite to this use case focusing more on healthcare decisions of laypersons, another paper investigated how to provide personalized recommendations for healthy activities on online healthcare platforms to keep users engaged \cite{Zhou2023}. 

Other authors investigate how to provide manage \textit{clinical inventory and personnel logistics} over time, predicting the amount of different resources needed for an expected patient workload - e.g., drug types, personnel, and number of beds or ventilators, possibly under pandemic conditions (e.g., \cite{Bertsimas2021b}), and then providing resource management strategies based on this information \cite{Galli2021,Williams2022,Tan2021,Shi2021,Bertsimas2021b}.  

Remaining papers on healthcare investigate the issue of \textit{scheduling clinical appointments} in large numbers, using predictive methods to account for uncertain factors such as the risk of no-show patients during outpatient appointments \cite{Srinivas2018,Salah2022}.

\subsection{Human Resources}
The focus of these applications is to improve the management of professional workers according to their individual or collective abilities.

Some papers found within this domain concern how to prescribe the best \textit{recruitment decisions}: how to hire the best set of individuals according to the needs of the company, based on their resume, personality traits, skill overlap with other applicants, among other factors \cite{Pessach2020,Ramannavar2018}. Some of these approaches also take into the account an identified need to provide equity for under-represented groups during the selection process \cite{zhu2024optimal}. 

Other applications concerned \textit{staffing capacity management}, i.e., determining the right staffing levels (capacity) according to demand factors, within a German postal service company for sorting letters and parcels according to the expected workload \cite{Notz2022}, and within a large airline company for staffing multiple shifts each workday according to predicted customer arrival rates \cite{Notz2023}. Another application addressed the optimal allocation of a set of employees to a set of high-priority customers, based on employee skill sets, predicted customer satisfaction and profit, among other factors within a sales setting \cite{Bischhoffshausen2015}. 

A final application concerned remediating risks of \textit{employee attrition}, identifying employees at risk of leaving the company and suggesting remedial HR interventions, based on similar employees not at risk \cite{Brockett2019}.

\subsection{Infrastructure}
The focus of these applications is to improve the design and maintenance of structures and facilities needed for the basic operation of companies and societies. In this abstract sense, Infrastructure concerns both IT and urban architectures.

Some of these applications concern how to do \textit{IT network orchestration}, including how to improve the performance of communication channels in large-scale enterprise microservice architectures \cite{John2019}, and how to orchestrate access points and perform network optimization within the context of Mobile Edge Computing and Clouds (MEC), ensuring optimal service levels without violating capacity constraints of the network \cite{Ceselli2018,Ceselli2019}. 

Other applications address \textit{urban planning}, including where to place parking spots and electric vehicle charging stations for optimal availability \cite{Brandt2021}, and how to plan the construction of proper infrastructure for future air taxi operations in New York City, USA \cite{Sinha2023}. 

The final applications within this domain concerned how to properly monitor for soil slope failure when excavating during \textit{building construction} \cite{Li2019}, and how to provide proactive asset health management \cite{Goyal2016}, optimal power flow plans \cite{stratigakos2024interpretable}, or day-ahead unit commitment strategies \cite{chen2024towards} for the \textit{electric power grid}.

\subsection{Logistics and Transportation}
The focus of these applications is to improve the management of how transportation vehicles deliver their payloads.

Several issues within \textit{airline operations} have previously been the subject of DPSA applications: how to plan for cost-efficient and safe flights \cite{Achenbach2018,Ayhan2018}, and how to provide flight schedules and passenger slot allocations to minimize passenger wait time, while giving passengers enough time to catch their plane \cite{Jacquillat2022,Birolini2023a,Birolini2023b}. 

Within maritime transportation, some applications concern \textit{port state control}, i.e., ship inspection under time constraints, selecting which ships to inspect for possibly risky faults and in which order, in order to maximize the number of problems found \cite{Tian2023a,Yan2023,yang2024prescriptive,oudani2023prescriptive}. Other papers addressed how to proactively design \textit{ship maintenance plans} to prevent such problems from arising in the first place while minimizing operational costs \cite{Tian2023b,mohd2022prescriptive}. 

Within urban transportation, other applications concern how to proactively design \textit{railway maintenance plans}, prioritizing limited resources while ensuring passenger safety and comfort \cite{Amiel2023,Consilvio2019,consilvio2024data}.

A final set of applications address \textit{ground vehicle route planning}, especially within the context of order delivery, where the application must take uncertainties, such as the risk of failed deliveries, into account \cite{Grzegorowski2022,Kandula2021}.

\subsection{Manufacturing}
The focus of these applications is to improve efficiency in the industrial production of goods. 

\textit{Proactive maintenance and error correction} is one archetypal use case within this domain. The idea behind such prescriptive maintenance is to leverage predictions about remaining tool life etc., to proactively prevent faults from arising and halting production. Specific applications take place within natural gas \cite{Kuzyakov2020}, mechatronic \cite{Oberdorf2021}, steel component \cite{Stein2018}, and automotive \cite{Vater2019,Vater2020} manufacturing settings. Other more generic applications address a variety of factory machinery \cite{Tham2023,Mohan2023,karakaya2024sensor}.

Another set of applications address how to make sound \textit{procurement and inventory decisions} for raw materials with respect to downstream customer or quality demands in the production of solar cells \cite{Lee2022}, LEDs \cite{Thammaboosadee2018}, herbal medicine \cite{Rakhmasari2018}, food snacks from a large Chinese manufacturer \cite{Yu2021}, and within gas-fired energy production \cite{Mandl2021,Mandl2023} specifically.

Other applications addressed issues within the improvement of \textit{production efficiency} in the end-to-end production flow. Specifically, how to choose harvesting locations for fish trawlers \cite{Cakir2023} and how to pick the right parameters for 3D printing machines to optimize, among other metrics, tool life, defect rate, carbon footprint, and printing time \cite{Kumar2023,Sangwan2023,reisch2023prescriptive}.

Finally, a few papers address how to find the right \textit{production design parameters} with respect to product quality: textile compositions for optimal fabric designs \cite{Ribeiro2023} and fatty acid compositions for optimal biodiesel properties \cite{Suvarna2022}, specifically.

\subsection{Recreational Activities}

The focus of these applications is to improve the quality of services supporting entertainment, hobbies, and other pastimes.

One paper seeks to improve the system for recommending relevant people to follow on Instagram \cite{Dash2022}. Another investigates the problem of how to overcome choice overload on Tomplay, a digital music learning platform, and provide personal recommendations of relevant music tracks to learn \cite{razgallah2024using}. Yet another paper investigates how to best handle cancelled professional sports leagues, e.g., due to pandemic lockdowns - specifically how to select the most relevant subset of the remaining matches to play, e.g., in terms of outcome uncertainty, in order to satisfyingly conclude the season in a shortened time frame \cite{hassanzadeh2024conclude}.

\subsection{Retail and Services} 
The focus of these applications is to improve how retailers and service providers manage their inventories, product portfolios, and business health. 

A number of papers concern \textit{inventory management}: Under uncertain customer demand over time, determining the right inventory level for each product in a predetermined portfolio in order to minimize waste and maximize profit. Such applications cover a variety of specific settings and motivations, from stocking overseas women's clothing warehouses, facing uncertain demand and cross-border taxes \cite{Shi2020}, to replenishing denomination mixes of Automated Teller Machines (ATMs), according to the needs of bank clients and with minimum operational costs \cite{Heide2020}. A recurrent variant is the \textit{newsvendor problem}, which introduces the complication of product value being expirable and thus a deadline for earning revenue, as found with daily, paper-form newspapers \cite{qin2011newsvendor}. Newsvendor problem papers concern the inventory levels of food and grocery items from retailers who sell such expirable products \cite{Keskin2022,Punia2020}. Other applications within inventory management concern how to distribute inventory levels across different stores according to local demand patterns, specifically within the context of grocery panic buying during the COVID-19 pandemic \cite{Adulyasak2023} and clearance sales at Zara, a large fashion retailer \cite{Caro2019}.

Although some of the aforementioned applications also cover pricing decisions (e.g., \cite{Keskin2022,Caro2019}), other applications address the more general issue of \textit{portfolio management} front and centre, i.e., how to design a product line-up and make pricing decisions in order to maximize profit according to predicted demand \cite{Mehrotra2020,Chen2022,Qu2020,Huang2019,Lash2016}, including, e.g., decisions on which stores to sell a car maintenance add-on product \cite{Huang2019}, and Walmart as a general retailer of many different product categories \cite{Mehrotra2020}. Yet another paper investigated the problem of \textit{business location recommendation} within the restaurant industry \cite{han2024identifying}.

Additional papers within this application domain consider more general issues linked to monitoring \textit{business health}, i.e., configuring Radio Frequency Identification (RFID) movement tracking in brick and mortar stores in order to improve service and security levels \cite{Hauser2021} and optimizing the financial reliability of banks \cite{Ravi2021,kurniawan2023prescriptive}.

\subsection{Social Policy}
The focus of these applications is to shape human behaviour on an institutional level. 

Some of these papers concern improving the functioning of the \textit{criminal justice system}, from guiding urban police patrols based on predicted hot spots for crime \cite{Brandt2022}, to efficiently cutting the information flow of transnational criminal organizations \cite{Bahulkar2018}, as well as guiding drug courts on when to recommend rehabilitation versus imprisonment \cite{Delen2021}. Another paper investigated how to provide recommendations for companies on which Environmental, Social, and Governance (ESG) pillars (considered in responsible investing) should be improved \cite{sariyer2024predictive}. The final paper seeks to improve the choice of interventions during child services risk assessment \cite{Schwartz2017}.

\section{Extended discussion on individual methods}\label{app:c}
The following section summarize individual methods used for prediction and/or prescription under their respective method type. Throughout, we use the convention of capitalizing method names when they refer to the content of Figures \ref{fig:prediction-tree} or \ref{fig:prescription-tree}.

\subsection{Data Mining and Machine Learning (DM/ML)}\label{sc:app:methodtype:dmml}

Possibly the most straightforward predictive DM/ML methods used by DPSA authors rely on \textit{case-based reasoning}, making \textit{Nearest Neighbour Queries} in historical databases to predict whether a known problem has re-emerged in a manufacturing setting \cite{Kuzyakov2020}, or to find exemplary employees or researchers with respect to a query object in order to inform human resources or academic practices \cite{Brockett2019,Cho2015}. Another paper predicted the amount of Baltic Sea cod at possible harvesting locations based on the average historical catch amounts of the k nearest triangular sections on the map \cite{Cakir2023}.

Other authors utilized a \textit{Text Mining} approach, extracting information from job applicant CV's and mapping it to an organizational ontology in order to reason about the fit of the applicants for a particular open employment position \cite{Ramannavar2018}.

\textit{Graph Mining} techniques was used by others, \textit{Link Prediction} and \textit{Community Detection} to discover criminal information networks \cite{Bahulkar2018}, and a variety of \textit{Graph Featurization} techniques, including the Jaccard coefficient, cosine distance, Katz centrality, and weight features to predict the probability that two people know each other in a social network \cite{Dash2022}. \textit{Co-Location Pattern Mining} was utilized by in one paper to predict a candidate range of possible locations for restaurant before doing recommendation \cite{han2022data}.

A couple of surveyed papers furthermore used \textit{Frequent Pattern Mining} algorithms - in combination with \textit{Pearson's R} correlation coefficient and clustering to identify employees at risk of attrition \cite{Brockett2019}, and using \textit{Event Temporal Sequence Mining} on historical student data to provide academic guidance for university students \cite{Du2016}.

\textit{Outlier Detection} was also used in a couple of papers \cite{consilvio2024data,reisch2023prescriptive}, a \textit{One-Class Support Vector Machine} to detect anomalous statuses of railway assets, and an \textit{Autoencoder}-based model to help calculate an anomaly score for possible defects during 3D printing.

\textit{Interpolation Models}, which as opposed to regression models require a perfect fit to the set of training points \cite{conn2009introduction}, are perhaps best known within the ML community through Bayesian hyperparameter optimization, which typically uses a Gaussian Process as its underlying interpolant \cite{snoek2012practical}. \textit{Gaussian Processes}, and the closely related \textit{Kriging Processes}, which in a sense generalizes the former, are however also quite commonly used for simulation-based design optimization within fields like aerospace engineering, in which simulations require complex computations possibly taking several days complete \cite{forrester2009recent}. 
Within the set of surveyed papers, researchers used these models to predict the location of holes in vacuum-sealed components during manufacturing \cite{Stein2018} and as an alternative approach to predict the possible catch size of Baltic Sea cod \cite{Cakir2023}.

The remaining predictive DM/ML methods encompass more or less well-known supervised and unsupervised ML methods. 

One paper used a \textit{Joint Clustering and Classification} approach, partitioning a customer dataset into clusters with each their own logistic regression model for demand prediction by minimizing the aggregate prediction loss \cite{Borenstein2023}.

Stand-alone \textit{Clustering} is used for a number of different purposes in DPSA applications, from identifying areas on subway train tracks with an unusually large number of registered bumps during transit with DBSCAN \cite{Amiel2023}, to identifying different demand patterns on the Mobile Edge Cloud with temporal clustering analysis \cite{Ceselli2019}, for instance. 

Among all supervised ML methods, ensemble \textit{Tree-Based Models} like \textit{Random Forest} and \textit{Gradient-Boosted Trees}, were the most common by far, for both classification and regression tasks. In total, they were used in 26\% of all papers for prediction, routinely being the identified top performers during model selection \cite{Adulyasak2023,Brandt2022,Delen2021,Galli2021,Grzegorowski2022,Kandula2021,Pessach2020,Schwartz2017,Stein2018,Susnjak2023,Suvarna2022,Thammaboosadee2018,Yanta2021,Birolini2023a,Birolini2023b,Devriendt2021,Jacquillat2022,Kiaghadi2023,Lash2016,Notz2022,Notz2023,Oberdorf2021,Punia2020,Shi2020,oudani2023prescriptive,singh2023machine,oudani2023prescriptive}. The closest contender was in fact \textit{not} Heterogeneous Ensemble Models, combining different model types, but bi- and multinomial \textit{Logistic Regression}, used in 7.7\% of all papers for prediction \cite{Chaudhry2023,Chen2022,Caigny2021,Devriendt2021,Jabr2023,Qu2020,Ravi2021,Sanisoglu2023}. 

While there is indeed No Free Lunch in terms of ML model selection \cite{wolpert1997no}, tree-based ensembles often have a strong showing during ML competitions, through implementations like XGBoost \cite{chollet2021deep}, so the found dominance of tree-based ensembles is perhaps not a surprising finding. Authors however still tend to rely on Deep \textit{Artificial Neural Networks} for dealing with non-tabular data, such as images for fault detection during electronic vehicle manufacturing \cite{Vater2019,Vater2020}.

Less well-known tree-based models used in DPSA applications include \textit{Causal Forests}, an extension of Random Forest for estimating conditional treatment effects \cite{Borenstein2023}, and \textit{Smart-Predict-Then-Optimize Trees}, specifically designed for PSA, trained to minimize downstream decision regret from the prescriptive method as opposed to solely maximizing predictive accuracy \cite{Yan2023,Tian2023b}.

As for \textit{Regression Models}, some authors stick to generic methods with attractive analytical properties, such as \textit{Linear}, \textit{LASSO} and \textit{Higher-Order Polynomial Regression} (e.g., \cite{Brandt2022,Lash2016,Kumar2023,chen2024towards,kurniawan2023prescriptive}), while others opt for special \textit{Problem-Specific Regression Models}, first and foremost designed to accurately capture dynamics within the problem domain \cite{Caro2019,Shi2021,Heide2020}, with one paper for instance opting for a custom, non-smooth, regression model, including several exponential and logarithmic terms, to estimate demand during clearance sales for their industrial collaborator, Zara \cite{Caro2019}.
\newline
\noindent As is evident from Figure \ref{fig:prescription-tree}, DM/ML techniques are also utilized to provide prescriptions. One way to do this is to use a prescriptive \textit{Nearest Neighbour Query} \cite{Ramannavar2018,Kuzyakov2020}, as used in one paper to suggest an employee to hire based on the degree of match with a set of desired attributes \cite{Ramannavar2018}. One paper aggregated several \textit{Feature Selection} and elimination techniques to identify which input features of an ML model accounted for a predicted factory machine failure due to tool wear, thus suggesting possible targets for repair \cite{Mohan2023}.

A couple of surveyed papers utilized \textit{Ensemble Learning} to obtain prescriptions in a clinical setting \cite{Bertsimas2022,Bertsimas2021a}. Having trained several regression models for a number of treatment-dosage combinations, predicting the effect of antihypertensive medication on the blood pressure of individual patients, prescriptions are found in one paper \cite{Bertsimas2022} with the following ensemble learning approach: first, the treatment-dosage combination that the majority of regression models predict the smallest blood pressure levels for is found. If such a majority exists, then the mean blood pressure prediction of the majority is calculated. If a majority of models are in agreement, and if the expected improvement in blood pressure according to the predicted mean is more than 20\%, then a change in medicine regimen is recommended for the patient. The other paper follows a very similar voting scheme to aggregate risk scores and suggest treatments for hypertensive COVID-19 patients \cite{Bertsimas2021a}.

Other papers used techniques from \textit{eXplainable AI (XAI)} to provide prescriptions. \textit{SHapley Additive exPlanations (SHAP)} can provide a linear model of how much each input feature value contributed to a particular prediction of a black-box model, and this measure of feature importance was used to give recommendations for drug courts on whether it would be beneficial to prefer rehabilitation over imprisonment in a given case, based on a prediction model of social outcomes \cite{Delen2021}. \textit{Diverse Counterfactual Explanations (DiCE)} is another XAI technique, which can effectively suggest small changes to input feature values of an ML model to change its output, used for suggesting remedial actions for students at risk of dropping out in one paper \cite{Susnjak2023}.

An alternative way to do prescriptive DM/ML is to frame the prescriptive question as a predictive one, using a traditional ML model to output prescriptions. Such \textit{Prescriptive ML Models} can take a variety of shapes. \textit{Cost-Based Learning} trains an ML model to predict the right decision to make, e.g., through a classification problem, minimizing the cost of wrong predictions, which was used for training a decision tree to recommend binary clinical decisions in one paper \cite{Fang2019}. Another paper simply trained a Gradient-Boosted Tree on featurized graph data in order to do link prediction, suggesting other users to follow on Instagram \cite{Dash2022}. Other surveyed papers solved specially designed optimization problems to train prescriptive Linear or tree-based models \cite{Mandl2023,Notz2022,Notz2023,Punia2020}. For instance, one paper prescribed work capacity management decisions with a model using a Random Forest Kernel function at its core to weigh the similarity between input points and a dataset of known points \cite{Notz2022}.

\textit{Reinforcement Learning} also formed a recurrent prescriptive DM/ML approach \cite{Ferreira2022,Zhou2023,Tham2023,Khan2021,Meng2020,osakwe2024towards}, e.g., using a \textit{Multi-Armed Bandit} approach to continuously learn which assortment of products to highlight in an online commerce setting \cite{Ferreira2022}, or \textit{Mixed Monte Carlo} to recommend subsequent steps to take in order to manage sepsis, given a partially completed case in an intensive hospital care setting \cite{Khan2021}.

Other papers used \textit{Graph Neural Networks} to provide prescriptions \cite{han2024identifying,razgallah2024using}, e.g., the convolutional \textit{locationGCN} architecture to recommend restaurant locations \cite{han2024identifying}. A final paper used a traditional \textit{Collaborative Filtering} approach, based on an Autoencoder model, to recommend learning modules to students \cite{ara2024collaborative}.

\subsection{Mathematical Optimization (OPT)}\label{sc:app:methodtype:opt}
As found in this survey, OPT is solely used as a prescriptive method in DPSA workflows (cf. Figure \ref{fig:prescription-tree}).

The most straightforward approach used for OPT is \textit{Exhaustive Search}, in which all valid candidate solutions are enumerated and evaluated. Given enough time, this brute-force approach always provides an optimal solution. It might be useful when the set of solution candidates is relatively small, easy to enumerate, and cheap to evaluate. A considerable number of DPSA applications used such an approach \cite{Achenbach2018,Chaudhry2023,Chen2022,Lash2016,Li2019,Hauser2021,Goyal2016}, with one paper for instance enumerating all possible solutions to obtain optimal cost indices for airline operations, minimizing delay and fuel costs \cite{Achenbach2018}.

The remaining array of utilized OPT methods provide more involved solution approaches, leveraging different levels of assumptions regarding the optimization model. \textit{Dynamic Programming}, leveraging optimal subproblem structure, was for instance used for resolving possible aircraft path conflicts in one paper \cite{Ayhan2018}. 

\textit{Linear Problem Models}, only including linear objective functions and constraints, are analytically attractive due to efficient solution algorithms and solvers (using, e.g. the simplex method), even for problems with possibly millions of decision variables, at the cost of linear modelling restrictions \cite{hillier2021introduction}. They turned out to be a staple among DPSA applications, with 26\% of surveyed papers utilizing such approaches. This makes \textit{Linear Programming} \cite{Bahulkar2018,Borenstein2023,Rakhmasari2018} and extensions thereof the most popular OPT approach by far. 

One significant extension used among surveyed papers is to use some or only integer variables in the problem model, problem classes named \textit{Mixed-Integer} \cite{Caro2019,Bertsimas2021b,Birolini2023a,Ceselli2019,Consilvio2019,Heide2020,Shi2020,Shi2021,chen2024towards,karakaya2024sensor,yang2024prescriptive} and \textit{Integer Linear Programming} \cite{Sinha2023,Brandt2022,Ceselli2018,Jacquillat2022,Kiaghadi2023,Bischhoffshausen2015,Tian2023a,Galli2021,Williams2022,Tian2023b,Birolini2023b,Mehrotra2020,Pessach2020}, respectively. \textit{Binary Programming}, constraining all variables to be either 0 or 1, is a special case of the latter (e.g., \cite{Tian2023b}). As implied previously, inclusion of integer variables makes Linear Programming NP-hard \cite{genova2011linear}. Having to circumvent this challenge with a variety of solution heuristics and modelling tricks to ensure timely solutions is thus a recurrent theme in the set of surveyed papers (e.g., \cite{Ceselli2019,Ceselli2018,Consilvio2019,Birolini2023b}). Integer variables are nonetheless useful for modelling fundamentally discrete resource allocation decisions, as used among surveyed papers to pick the best demand-driven order of building a set of air taxi stations \cite{Sinha2023} or the best demand-driven denomination mix in ATMs \cite{Heide2020}, for instance.

More general problem classes than Linear Programming were also found among surveyed papers, including both \textit{Convex} \cite{Kumar2023,Qu2020,John2019,Sangwan2023,Shi2020,hassanzadeh2024conclude}, and \textit{Non-Convex Programming} \cite{Keskin2022,Goyal2016} techniques, which allow for more expressive, and possibly more accurate, problem models. For convex optimization, efficient solution algorithms based on, e.g., descent methods like gradient descent, exist \cite{kochenderfer2019algorithms}. For example, one paper solved a convex problem with gradient-based methods to make optimal B2B pricing decisions \cite{Qu2020}. As for general-case non-convex optimization, relying on approximations might be the only practical option. For example, one paper performed a grid search, evaluating a subset of all solution candidates in a discretized search space, to approximately solve a non-convex pricing and ordering problem \cite{Keskin2022}.  

\textit{Multi-Objective} and \textit{Stochastic Programming} are additional optimization extensions, introducing the complications of several objective functions and stochastic optimization models, respectively. Real-world problems often involve a number of trade-off factors in terms of solution quality, and several surveyed papers (e.g., \cite{Amiel2023,Kiaghadi2023,Bertsimas2021b,Jacquillat2022,oudani2023prescriptive}) thus use multi-objective programming. For instance, to allocate ventilators in the USA during COVID-19, while finding the best trade-off in terms of minimizing daily shortages and inter-state transfers of equipment, \cite{Bertsimas2021b}. Other papers used stochastic programming \cite{Shi2021,Birolini2023a,Galli2021,Williams2022}, e.g., in another COVID-19-themed paper, to perform nurse transshipment decisions based on stochastic workload forecasts \cite{Shi2021}. 

The closest contender to Linear Programming and extensions among surveyed papers were \textit{Heuristic Optimization} techniques, used in 11.5\% of papers. Most of these methods don't assume any particular underlying problem structure, unlike, e.g., Linear and Convex Programming methods, yet do so at the cost of generally less efficient solution algorithms with fewer analytical guarantees \cite{kochenderfer2019algorithms}. Well-known methods used among surveyed papers include \textit{Hill Climbing} \cite{Grzegorowski2022}, \textit{Evolutionary Algorithms} \cite{Ribeiro2023,Thammaboosadee2018,Yanta2021,oudani2023prescriptive}, \textit{Particle Swarm Optimization} \cite{Lee2022,Ravi2021,Suvarna2022}, and \textit{Simulated Annealing} \cite{Grzegorowski2022}. One prominent reason for preferring such methods is that the optimization problem in question is \textit{black-box}. That is, objective functions or constraints are not analytically available, and it is therefore impossible to run, e.g., the simplex method \cite{audet2017derivative}. Given how pervasive DM/ML is in DPSA (cf. Figure \ref{fig:method-bar}), a black-box optimization problem in which an ML model takes decision variables as input in the objective function is the typical use case for heuristic methods among DPSA applications, used to optimize delivery plans \cite{Grzegorowski2022} and biodiesel properties \cite{Suvarna2022}, among other objectives.

Other researchers opt for \textit{Problem-Specific Heuristic Methods} exploiting underlying problem structures in their solution strategy \cite{Huang2019,Kandula2021,Brandt2021,consilvio2024data}, which can be more efficient than relying on problem-agnostic methods. For instance, one paper used a polynomial-time heuristic to schedule E-commerce delivery routes while minimizing missed deliveries \cite{Kandula2021}.

\subsection{Probabilistic Modelling (PROB)}

One paper used a \textit{Hidden Markov Model} to forecast aircraft trajectories with respect to uncertain weather conditions \cite{Ayhan2018}. Other papers, used a \textit{Bayesian Model} \cite{karakaya2024sensor,Li2019}, forming predictions by Bayesian inference, e.g., for predicting unstable soil failure modes during building construction \cite{Li2019}.

Different \textit{Probability Distributions} \cite{Lee2022,Qu2020,Shi2020,Srinivas2018}, as shown in Figure \ref{fig:prediction-tree}, were also used to form predictions under uncertainty, a \textit{Triangular Distribution} to predict the quality of supplier raw materials for a solar cells \cite{Li2019}, and a logistic regression model with coefficients following a \textit{Multivariate Normal Distribution} to predict the demand curve for B2B goods \cite{Qu2020}, for instance.

Queueing Models were quite common in the set of surveyed papers \cite{Birolini2023a,Caro2019,Hauser2021,Notz2023,Srinivas2018,Tan2021,Jabr2023,Shi2021}. One paper used a queueing model to predict the distribution of incoming calls for an outpatient appointment system in a clinical setting \cite{Srinivas2018}, while another used an elaborate \textit{Stochastic Network Model}, involving several different stages and probability distributions to predict expected workloads at hospitals within a COVID-19 setting, from diagnosis to surgery and intensive care \cite{Shi2021}.

Other surveyed papers utilized \textit{Probabilistic Time Series Forecasting} for predictions, using a variety of different models \cite{Adulyasak2023,Mehrotra2020,Yu2021,Mandl2021,consilvio2024data}. Some papers for use a \textit{Decomposition Model}, aggregating predictions in terms of trends, cycles, seasonality and randomness \cite{Adulyasak2023,consilvio2024data}, to predict the demand over time for critical product groups, such as toilet paper, during the COVID-19 pandemic, where panic buying and shortages were a concern \cite{Adulyasak2023}. Another paper used \textit{Bayesian Structural Time Series} demand forecasting within the general retail setting of Walmart \cite{Mehrotra2020}.

Finally, a couple of surveyed papers utilized \textit{Problem-Specific Probabilistic} models that do not fit into any of the other categories mentioned \cite{Keskin2022,Consilvio2019}. One paper incorporated beta and geometric distributions into an exponential model to predict the changing demands for different kinds of groceries \cite{Keskin2022}, while the other, forming a hybrid approach with DM/ML, integrated outputs from a number of Support Vector Machines, corresponding to different railway circuits, in order to give an aggregate probability of defects in need of railway system maintenance \cite{Consilvio2019}.

\subsection{Domain Expertise (EXP)}\label{sc:app:methodtype:exp}

A couple of papers used EXP as a comprehensive approach to prediction (cf. Figure \ref{fig:prediction-tree}), with one of them using a \textit{Physics Model} of heat exchange to predict the lifetime of assets in the electric power grid \cite{Goyal2016}, and another using an \textit{Epidemiological Model} based on differential equations to forecast how the COVID-19 pandemic might develop on a state level in the USA \cite{Bertsimas2021b}.
\newline
\noindent For automated prescription, EXP is applied in a variety of ways (cf. Figure \ref{fig:prescription-tree}). Some surveyed papers used a \textit{Sorting Scheme}, which picks the top k items from an ordered list of results to obtain prescriptions \cite{Gubela2021,Yan2023,singh2023machine,zhu2024optimal}. For an advertisement campaign, one paper picked the top k customers most likely to engage by using the outputs of an ML model \cite{Gubela2021}, while another picked the top k ships most likely to have defects for inspection \cite{Yan2023}.

Imitating role models is a common social learning mechanism \cite{bandura1977social}. It is also a prescriptive approach followed in surveyed papers using \textit{Exemplar Matching} \cite{Brockett2019,Cho2015,Du2016}. These papers address employee retention \cite{Brockett2019} and academic guidance for individual students \cite{Du2016} or researchers \cite{Cho2015}, respectively. For prescriptions, they search for similar individuals to some input person of interest, where the former has achieved some desirable state (loyal employee, high academic achiever...), and then propose remedial actions based on the difference between the exemplar and query subject - a higher salary when the loyal exemplar gets paid more than the query subject at risk of attrition, for instance \cite{Brockett2019}.

Other papers provided prescriptions by applying a set of \textit{Work Scheduling Rules} based on how work is organized within the problem domain \cite{Oberdorf2021,Salah2022,Srinivas2018}, for, e.g., continuously scheduling patient appointments after having predicted the risk of no-shows \cite{Srinivas2018}.

Another prescriptive approach is based on executing a \textit{Hand-Made Decision Tree}, i.e., one manually encoded by consulting a domain expert and not an ML algorithm. This method was utilized for deciding whether a defective car part detected during production should be reworked or dropped, taking into consideration whether there is time for more reworks at present \cite{Vater2019,Vater2020}.

Surveyed papers doing uplift analysis for Advertising and Marketing (cf. Appendix \ref{sc:advertising-and-marketing}) typically used \textit{Qini Curve Analysis}, a problem-specific technique for determining which customers to target with a treatment after having predicted uplift potential for each \cite{Caigny2021,Devriendt2021,Sanisoglu2023}. In brief, customers are first sorted by their predicted uplift, and then the percentage of customers to target in this sorted order is plotted on the x-axis of a coordinate system with the cumulative uplift on the y-axis. Assuming an infinite campaign budget, the optimum percentage of customers to target in sorted order can then be found at the maximum point of the curve \cite{Caigny2021,Sanisoglu2023}. To explain this solution, recall that four types of customers are considered in uplift analysis, with campaign treatments for \textit{Persuadables} and \textit{Do-Not-Disturbs} respectively causing engagement and disengagement (cf. Table \ref{tab:uplift}). The cumulative uplift function of the Qini curve will not increase unless the campaign targets Persuadables, and will indeed \textit{decrease} if it targets Do-Not-Disturbs \cite{Devriendt2021}.

Finally, some papers utilized methods very closely tied to their particular problem domain, with one paper using \textit{Financial Analysis} formulas along with some thresholds for desired yields in order to provide bank loaning strategies \cite{kurniawan2023prescriptive}, and another calculating \textit{Proportional Anomaly Adjustment}s based on detected irregularities and material properties, in order to increase reliability of 3D printing in aerospace manufacturing \cite{reisch2023prescriptive}.

\subsection{Simulation (SIM)}\label{sc:app:methodtype:sim}

Providing prescriptions by evaluating different solution candidates in simulations, i.e., by using \textit{Simulation-Based Optimization} \cite{gosavi2015simulation}, is the strategy followed by these papers. One paper simulated different scenarios with respect to a stochastic model, in order to determine the minimum replenishment quantity needed for critical groceries to prevent retailer shortages during COVID-19 panic buying \cite{Adulyasak2023}. Using a similar \textit{what-if}-based approach, another paper trialled various inputs to a simulation model in order to identify which changes to existing ESG metrics would yield the highest overall ESG score increase for an enterprise \cite{sariyer2024predictive}. Yet another paper used Simulation-Based Optimization to maximize the joint probability of revisits and purchases in an online retail setting, by varying, e.g., categories and prices of highlighted products \cite{Jabr2023}. SIM has also been used to generate clinical schedules under uncertainty \cite{Srinivas2018,Tan2021}, and to evaluate possible search paths for shop floor workers to discover leaks in vacuum-packed part moulds, as used in, e.g., the automotive industry \cite{Stein2018}.

\section{Papers mapped to method combinations}\label{app:d}
How papers were associated with each method combination is summarized in Table \ref{tab:combinations}.

\begin{table*}[ht]
    \centering
    \footnotesize
    \resizebox{\linewidth}{!}{%
    \begin{tabular}{lll}
    \hline
    \textbf{Prediction}  & \textbf{Prescription} & \textbf{Papers} \\ \hline
    DM/ML                & OPT & \cite{Achenbach2018,Amiel2023,Sinha2023,Bahulkar2018,Birolini2023b,Borenstein2023,Brandt2021,Brandt2022,Ceselli2018,Ceselli2019,Chaudhry2023,Chen2022,Galli2021,Grzegorowski2022,Huang2019} \\
                         &     & \cite{Jacquillat2022,John2019,Kandula2021,Kiaghadi2023,Kumar2023,Lash2016,Pessach2020,Rakhmasari2018,Ravi2021,Ribeiro2023,Sangwan2023,Suvarna2022,Thammaboosadee2018} \\ 
                         &     & \cite{Tian2023a,Tian2023b,Heide2020,Bischhoffshausen2015,Williams2022,Yanta2021,oudani2023prescriptive,chen2024towards,hassanzadeh2024conclude} \\
    DM/ML                & DM/ML & \cite{Bertsimas2022,Bertsimas2021a,Cakir2023,Dash2022,Delen2021,Fang2019,Ferreira2022,Khan2021,Kuzyakov2020,Mandl2023,Meng2020,Mohan2023,Notz2022,Punia2020} \\ 
                         &     &   \cite{Ramannavar2018,Schwartz2017,Susnjak2023,Tham2023,Zhou2023,razgallah2024using,osakwe2024towards,ara2024collaborative,stratigakos2024interpretable,mohd2022prescriptive,han2024identifying} \\ 
    DM/ML                & EXP & \cite{Brockett2019,Cho2015,Caigny2021,Devriendt2021,Du2016,Gubela2021,Oberdorf2021,Salah2022,Sanisoglu2023,Vater2019,Vater2020,Yan2023,kurniawan2023prescriptive,singh2023machine,reisch2023prescriptive,zhu2024optimal} \\ 
    DM/ML + PROB         & OPT & \cite{Birolini2023a,Caro2019,Consilvio2019,Hauser2021,Qu2020,Shi2021,Shi2020,yang2024prescriptive,consilvio2024data} \\ 
    PROB                 & OPT & \cite{Ayhan2018,Keskin2022,Lee2022,Li2019,Mehrotra2020,Yu2021,karakaya2024sensor} \\ 
    DM/ML + PROB         & SIM & \cite{Adulyasak2023,Jabr2023} \\ 
    DM/ML + EXP          & OPT & \cite{Bertsimas2021b,Goyal2016} \\ 
    DM/ML + PROB         & DM/ML & \cite{Notz2023} \\ 
    DM/ML + PROB         & EXP + SIM & \cite{Srinivas2018} \\ 
    DM/ML                & OPT + SIM & \cite{Stein2018} \\ 
    DM/ML                & SIM & \cite{sariyer2024predictive} \\ 
    PROB                 & DM/ML & \cite{Mandl2021} \\ 
    PROB                 & SIM & \cite{Tan2021} \\ \hline
    \end{tabular}}
    \caption{Papers associated with each found method combination.}
    \label{tab:combinations}
\end{table*}




\bibliographystyle{elsarticle-num} 
\bibliography{library.bib}

\begin{thebibliography}{100}
\expandafter\ifx\csname url\endcsname\relax
  \def\url#1{\texttt{#1}}\fi
\expandafter\ifx\csname urlprefix\endcsname\relax\def\urlprefix{URL }\fi
\expandafter\ifx\csname href\endcsname\relax
  \def\href#1#2{#2} \def\path#1{#1}\fi

\bibitem{createthefuture}
Peter {D}rucker - the best way to predict the future is to create it, \url{https://www.brainyquote.com/quotes/peter_drucker_131600}, accessed: 2024-11-06 (2024).

\bibitem{knight1921risk}
F.~H. Knight, Risk, Uncertainty and Profit, Vol.~31, Houghton Mifflin, 1921.

\bibitem{holsapple2014unified}
C.~Holsapple, A.~Lee-Post, R.~Pakath, A unified foundation for business analytics, Decision Support Systems 64 (2014) 130--141.

\bibitem{davenport2017competing}
T.~Davenport, J.~Harris, Competing on Analytics: Updated, With a New Introduction: The New Science of Winning, Harvard Business Press, 2017.

\bibitem{lustig2010analytics}
I.~Lustig, B.~Dietrich, C.~Johnson, C.~Dziekan, The analytics journey, Analytics Magazine 3~(6) (2010) 11--13.

\bibitem{maoz2013should}
M.~Maoz, How it should deepen big data analysis to support customer-centricity, Gartner G00248980 (2013).

\bibitem{siksnys2018prescriptive}
L.~Siksnys, T.~B. Pedersen, Prescriptive analytics, in: Encyclopedia of Database Systems, Springer, 2018, pp. 2792--2793.

\bibitem{vsikvsnys2016solvedb}
L.~{\v{S}}ik{\v{s}}nys, T.~B. Pedersen, Solvedb: Integrating optimization problem solvers into sql databases, in: Proceedings of the 28th International Conference on Scientific and Statistical Database Management, 2016, pp. 1--12.

\bibitem{lepenioti2020prescriptive}
K.~Lepenioti, A.~Bousdekis, D.~Apostolou, G.~Mentzas, Prescriptive analytics: Literature review and research challenges, International Journal of Information Management 50 (2020) 57--70.

\bibitem{frazzetto2019prescriptive}
D.~Frazzetto, T.~D. Nielsen, T.~B. Pedersen, L.~{\v{S}}ik{\v{s}}nys, Prescriptive analytics: a survey of emerging trends and technologies, The VLDB Journal 28 (2019) 575--595.

\bibitem{sigmod2024callforpapers}
The 2025 {ACM SIGMOD/PODS} conference: Berlin, {Germany} - {Welcome}, \url{https://2025.sigmod.org/}, accessed: 2024-11-06 (2024).

\bibitem{davenport2007competing}
T.~Davenport, J.~Harris, Competing on Analytics: The New Science of Winning, Harvard Business Press, 2007.

\bibitem{stefani2018constituent}
K.~Stefani, P.~Zschech, Constituent elements for prescriptive analytics systems, in: 26th European Conference on Information Systems: Beyond Digitization - Facets of Socio-Technical Change, ECIS 2018, 2018, p.~39.

\bibitem{poornima2020survey}
S.~Poornima, M.~Pushpalatha, A survey on various applications of prescriptive analytics, International Journal of Intelligent Networks 1 (2020) 76--84.

\bibitem{raeesi2021prescriptive}
I.~Raeesi~Vanani, S.~Majidian, Prescriptive analytics in internet of things with concentration on deep learning, Introduction to Internet of Things in Management Science and Operations Research: Implemented Studies (2021) 31--54.

\bibitem{fox2022review}
H.~Fox, A.~C. Pillai, D.~Friedrich, M.~Collu, T.~Dawood, L.~Johanning, A review of predictive and prescriptive offshore wind farm operation and maintenance, Energies 15~(2) (2022) 504.

\bibitem{kubrak2022prescriptive}
K.~Kubrak, F.~Milani, A.~Nolte, M.~Dumas, Prescriptive process monitoring: Quo vadis?, PeerJ Computer Science 8 (2022) e1097.

\bibitem{soeffker2022stochastic}
N.~Soeffker, M.~W. Ulmer, D.~C. Mattfeld, Stochastic dynamic vehicle routing in the light of prescriptive analytics: A review, European Journal of Operational Research 298~(3) (2022) 801--820.

\bibitem{mishra2023prescriptive}
D.~B. Mishra, S.~Naqvi, A.~Gunasekaran, V.~Dutta, Prescriptive analytics applications in sustainable operations research: conceptual framework and future research challenges, Annals of Operations Research (2023) 1.

\bibitem{hall2024systematic}
S.~F. Hall, M.~Sage, C.~F. Scott, K.~Joseph, A systematic review of sophisticated predictive and prescriptive analytics in child welfare: Accuracy, equity, and bias, Child and Adolescent Social Work Journal 41~(6) (2024) 831--847.

\bibitem{mendoza2024prescriptive}
G.~E. Mendoza-Olgu{\'\i}n, M.~J. Somodevilla-Garc{\'\i}a, C.~P{\'e}rez-de Celis, Y.~Chavarri-Guerra, Prescriptive analytics-based methodologies for healthcare data: A systematic literature review, Computaci{\'o}n y Sistemas 28~(4) (2024).

\bibitem{niederhaus2024technical}
M.~Niederhaus, N.~Migenda, J.~Weller, W.~Schenck, M.~Kohlhase, Technical readiness of prescriptive analytics platforms: A survey, in: 2024 35th Conference of Open Innovations Association (FRUCT), IEEE, 2024, pp. 509--519.

\bibitem{wissuchek2024prescriptive}
C.~Wissuchek, P.~Zschech, Prescriptive analytics systems revised: a systematic literature review from an information systems perspective, Information Systems and e-Business Management (2024) 1--75.

\bibitem{wissuchek2023survey}
C.~Wissuchek, P.~Zschech, Survey and systematization of prescriptive analytics systems: Towards archetypes from a human-machine-collaboration perspective, in: 31st European Conference on Information Systems - Co-creating Sustainable Digital Futures, {ECIS} 2023, Kristiansan, Norway, June 11-16, 2023, 2023, p. 289.

\bibitem{lepenioti2019prescriptive}
K.~Lepenioti, A.~Bousdekis, D.~Apostolou, G.~Mentzas, Prescriptive analytics: a survey of approaches and methods, in: Business Information Systems Workshops: BIS 2018 International Workshops, Berlin, Germany, July 18--20, 2018, Revised Papers 21, Springer, 2019, pp. 449--460.

\bibitem{elmachtoub2022smart}
A.~N. Elmachtoub, P.~Grigas, {Smart ''Predict, then Optimize''}, Management Science 68~(1) (2022) 9--26.

\bibitem{soltanpoor2016prescriptive}
R.~Soltanpoor, T.~Sellis, Prescriptive analytics for big data, in: Databases Theory and Applications: 27th Australasian Database Conference, ADC 2016, Sydney, NSW, September 28-29, 2016, Proceedings 27, Springer, 2016, pp. 245--256.

\bibitem{hagerty2017planning}
J.~Hagerty, Planning guide for data and analytics, Gartner-Technical Professional Advice (2017) 1--27.

\bibitem{delen2013data}
D.~Delen, H.~Demirkan, Data, information and analytics as services, Decision Support Systems 55~(1) (2013) 359--363.

\bibitem{deshpande2019predictive}
P.~S. Deshpande, S.~C. Sharma, S.~K. Peddoju, P.~S. Deshpande, S.~C. Sharma, S.~K. Peddoju, Predictive and prescriptive analytics in big-bata era, Security and Data Storage Aspect in Cloud Computing (2019) 71--81.

\bibitem{oesterreich2020understanding}
T.~D. Oesterreich, C.~Fitte, A.~Behne, F.~Teuteberg, Understanding the role of predictive and prescriptive analytics in healthcare: A multi-stakeholder approach, in: Proceedings of the 28th European Conference on Information Systems, 2020, p. 102.

\bibitem{banihashem2022systematic}
S.~K. Banihashem, O.~Noroozi, S.~van Ginkel, L.~P. Macfadyen, H.~J. Biemans, A systematic review of the role of learning analytics in enhancing feedback practices in higher education, Educational Research Review (2022) 100489.

\bibitem{petropoulos2023operational}
F.~Petropoulos, G.~Laporte, E.~Aktas, S.~A. Alumur, C.~Archetti, H.~Ayhan, M.~Battarra, J.~A. Bennell, J.-M. Bourjolly, J.~E. Boylan, et~al., Operational research: Methods and applications, Journal of the Operational Research Society (2023) 1--195.

\bibitem{evans1998history}
J.~Evans, The History and Practice of Ancient Astronomy, Oxford University Press, 1998.

\bibitem{hillier2021introduction}
F.~S. Hillier, Introduction to Operations Research, 11th Edition, McGrawHill, 2021.

\bibitem{giesecke2018call}
K.~Giesecke, G.~Liberali, H.~Nazerzadeh, J.~G. Shanthikumar, C.~P. Teo, Call for papers — management science — special issue on data-driven prescriptive analytics, Management Science 64~(6) (2018) 2972--2972.

\bibitem{kleppmann2017designing}
M.~Kleppmann, Designing Data-intensive Applications: The Big Ideas Behind Reliable, Scalable, and Maintainable systems, " O'Reilly Media, Inc.", 2017.

\bibitem{zaharia2016apache}
M.~Zaharia, R.~S. Xin, P.~Wendell, T.~Das, M.~Armbrust, A.~Dave, X.~Meng, J.~Rosen, S.~Venkataraman, M.~J. Franklin, et~al., Apache spark: a unified engine for big data processing, Communications of the ACM 59~(11) (2016) 56--65.

\bibitem{chollet2021deep}
F.~Chollet, Deep Learning with Python, Simon and Schuster, 2021.

\bibitem{he2021automl}
X.~He, K.~Zhao, X.~Chu, Automl: A survey of the state-of-the-art, Knowledge-Based Systems 212 (2021) 106622.

\bibitem{page2021prisma}
M.~J. Page, J.~E. McKenzie, P.~M. Bossuyt, I.~Boutron, T.~C. Hoffmann, C.~D. Mulrow, L.~Shamseer, J.~M. Tetzlaff, E.~A. Akl, S.~E. Brennan, et~al., The prisma 2020 statement: an updated guideline for reporting systematic reviews, International Journal of Surgery 88 (2021) 105906.

\bibitem{dblp2024howto}
How to perform a search within the full-texts?, \url{https://dblp.org/faq/How+to+perform+a+search+within+the+full-texts.html}, accessed: 2024-11-06 (2024).

\bibitem{aub2024databases}
AUB, Databases, \url{https://kbdk-aub.primo.exlibrisgroup.com/discovery/dbsearch?vid=45KBDK_AUB:AUB\&lang=en}, accessed: 2024-11-06 (2024).

\bibitem{predatory2024thelist}
P.~Reports, Predatory reports - the list, \url{https://predatoryreports.org/the-list}, accessed: 2024-11-06 (2024).

\bibitem{bjork2018evolution}
B.-C. Björk, Evolution of the scholarly mega-journal, 2006--2017, PeerJ 6 (2018) e4357.

\bibitem{scimago2024ranking}
SCImago, {Scimago Journal \& Country Rank}, \url{https://www.scimagojr.com/}, accessed: 2024-11-06 (2024).

\bibitem{core2024ranking}
Core rankings portal, \url{https://www.core.edu.au/conference-portal}, accessed: 2024-11-06 (2024).

\bibitem{conference2024ranking}
Conference ranks, \url{http://www.conferenceranks.com/}, accessed: 2024-11-06 (2024).

\bibitem{siksnys2021solvedb+}
L.~Siksnys, T.~B. Pedersen, T.~D. Nielsen, D.~Frazzetto, {SolveDB+: SQL-Based Prescriptive Analytics}, in: Advances in Database Technology - 24th International Conference on Extending Database Technology, EDBT 2021, OpenProceedings.org, 2021, pp. 133--144.

\bibitem{osakwe2024towards}
I.~Osakwe, G.~Chen, Y.~Fan, M.~Rakovic, S.~Singh, L.~Lim, J.~Van Der~Graaf, J.~Moore, I.~Molenaar, M.~Bannert, et~al., Towards prescriptive analytics of self-regulated learning strategies: A reinforcement learning approach, British Journal of Educational Technology (2024).

\bibitem{ara2024collaborative}
S.~S. Ara, R.~Tanuja, S.~Manjula, Collaborative filtering based module recommendation to boost learners achievement, in: 2023 4th International Conference on Intelligent Technologies (CONIT), IEEE, 2024, pp. 1--6.

\bibitem{Susnjak2023}
T.~Susnjak, Beyond predictive learning analytics modelling and onto explainable artificial intelligence with prescriptive analytics and chatgpt, International Journal of Artificial Intelligence in Education (2023).

\bibitem{Yanta2021}
S.~Yanta, S.~Thammaboosadee, P.~Chanyagorn, R.~Chuckpaiwong, Probation status prediction and optimization for undergraduate engineering students, 2021 13th International Conference on Knowledge and Smart Technology (KST) (2021) 191--196.

\bibitem{Du2016}
F.~Du, C.~Plaisant, N.~Spring, B.~Shneiderman, Eventaction: Visual analytics for temporal event sequence recommendation, in: 2016 IEEE Conference on Visual Analytics Science and Technology (VAST), 2016, pp. 61--70.

\bibitem{Kiaghadi2023}
M.~Kiaghadi, P.~Hoseinpour, University admission process: a prescriptive analytics approach, Artificial Intelligence Review 56 (2023) 233--256.

\bibitem{Cho2015}
M.~Cho, S.-K. Song, J.~Weber, H.~Jung, M.~Lee, Prescriptive analytics for planning research-performance strategy, in: Computer Science and its Applications: Ubiquitous Information Technologies, 2015, pp. 1123--1129.

\bibitem{Jabr2023}
W.~Jabr, A.~Ghoshal, Y.~Cheng, P.~Pavlou, Maximizing online revisiting and purchasing: A clickstream-based approach to enhancing customer lifetime value, Journal of Management Information Systems 40 (2023).

\bibitem{Ferreira2022}
K.~J. Ferreira, S.~Parthasarathy, S.~Sekar, Learning to rank an assortment of products, Management Science 68 (2022).

\bibitem{Borenstein2023}
A.~Borenstein, A.~Mangal, G.~Perakis, S.~Poninghaus, D.~Singhvi, O.~S. Lami, J.~W. Lua, Ancillary services in targeted advertising: From prediction to prescription, Manufacturing \& Service Operations Management 25 (2023) 1258--1303.

\bibitem{Chaudhry2023}
A.~Chaudhry, C.~Heilman, P.~B. Seetharaman, Measuring the effects of customized targeted promotions on retailer profits: prescriptive analytics using basket-level econometric analysis, Journal of Marketing Analytics (2023).

\bibitem{Gubela2021}
R.~M. Gubela, S.~Lessmann, Uplift modeling with value-driven evaluation metrics, Decision Support Systems 150 (2021) 113648.

\bibitem{Caigny2021}
A.~D. Caigny, K.~Coussement, W.~Verbeke, K.~Idbenjra, M.~Phan, Uplift modeling and its implications for b2b customer churn prediction: A segmentation-based modeling approach, Industrial Marketing Management 99 (2021).

\bibitem{Devriendt2021}
F.~Devriendt, J.~Berrevoets, W.~Verbeke, Why you should stop predicting customer churn and start using uplift models, Information Sciences 548 (2021).

\bibitem{Sanisoglu2023}
M.~Sanisoglu, S.~Burnaz, T.~Kaya, A gateway toward truly responsive customers: using the uplift modeling to increase the performance of a b2b marketing campaign, Journal of Marketing Analytics (2023).

\bibitem{singh2023machine}
S.~S.~K. Singh, A.~K. Sinha, T.~N. Pandey, B.~M. Acharya, A machine learning approach to compare causal inference modelling strategies in the digital advertising industry, in: 2023 2nd International Conference on Ambient Intelligence in Health Care (ICAIHC), IEEE, 2023, pp. 1--7.

\bibitem{Fang2019}
X.~Fang, Y.~Gao, P.~J. Hu, A prescriptive analytics method for cost reduction in clinical decision making, MIS Quarterly (2019).

\bibitem{Meng2020}
F.~Meng, Y.~Sun, B.~H. Heng, M.~K.~S. Leow, Analysis via markov decision process to evaluate glycemic control strategies of a large retrospective cohort with type 2 diabetes: the ameliorate study, Acta Diabetologica 57 (2020).

\bibitem{Khan2021}
A.~Khan, A.~Ghose, H.~Dam, Decision support for knowledge intensive processes using rl based recommendations, in: Business Process Management Forum: BPM Forum 2021, Rome, Italy, September 06–10, 2021, Proceedings 19, 2021, pp. 246--262.

\bibitem{Bertsimas2022}
D.~Bertsimas, A.~R.~A. Borenstein, A.~Dauvin, A.~Orfanoudaki, Ensemble machine learning for personalized antihypertensive treatment, Naval Research Logistics 69 (2022).

\bibitem{Bertsimas2021a}
D.~Bertsimas, A.~Borenstein, L.~Mingardi, O.~Nohadani, A.~Orfanoudaki, B.~Stellato, H.~Wiberg, P.~Sarin, D.~J. Varelmann, V.~Estrada, C.~Macaya, I.~J. Gil, Personalized prescription of acei/arbs for hypertensive covid-19 patients, Health Care Management Science 24 (2021).

\bibitem{Zhou2023}
T.~Zhou, Y.~Wang, L.~L. Yan, Y.~Tan, Spoiled for choice? personalized recommendation for healthcare decisions: A multiarmed bandit approach, Information Systems Research (2023).

\bibitem{Galli2021}
L.~Galli, T.~Levato, F.~Schoen, L.~Tigli, Prescriptive analytics for inventory management in health care, Journal of the Operational Research Society 72 (2021) 2211--2224.

\bibitem{Williams2022}
E.~Williams, D.~Gartner, P.~Harper, Linking predictive and prescriptive analytics of elderly and frail patient hospital services, in: 2022 IEEE 10th International Conference on Healthcare Informatics (ICHI), 2022, p.~1.

\bibitem{Tan2021}
K.~W. Tan, B.~K. Goh, A.~Gunawan, Redesigning patient flow in paediatric eye clinic for pandemic using simulation, in: 2021 IEEE International Smart Cities Conference, ISC2 2021, 2021, pp. 1--7.

\bibitem{Shi2021}
P.~Shi, J.~E. Helm, C.~Chen, J.~Lim, R.~P. Parker, T.~Tinsley, J.~Cecil, Operations (management) warp speed: Rapid deployment of hospital-focused predictive/prescriptive analytics for the covid-19 pandemic, Production and Operations Management (2021).

\bibitem{Bertsimas2021b}
D.~Bertsimas, L.~Boussioux, R.~Cory-Wright, A.~Delarue, V.~Digalakis, A.~Jacquillat, D.~L. Kitane, G.~Lukin, M.~Li, L.~Mingardi, O.~Nohadani, A.~Orfanoudaki, T.~Papalexopoulos, I.~Paskov, J.~Pauphilet, O.~S. Lami, B.~Stellato, H.~T. Bouardi, K.~V. Carballo, H.~Wiberg, C.~Zeng, From predictions to prescriptions: A data-driven response to covid-19, Health Care Management Science 24 (2021).

\bibitem{Srinivas2018}
S.~Srinivas, A.~R. Ravindran, Optimizing outpatient appointment system using machine learning algorithms and scheduling rules: A prescriptive analytics framework, Expert Systems with Applications 102 (2018) 245--261.

\bibitem{Salah2022}
H.~Salah, S.~Srinivas, Predict, then schedule: Prescriptive analytics approach for machine learning-enabled sequential clinical scheduling, Computers \& Industrial Engineering 169 (2022) 108270.

\bibitem{Pessach2020}
D.~Pessach, G.~Singer, D.~Avrahami, H.~C. Ben-Gal, E.~Shmueli, I.~Ben-Gal, Employees recruitment: A prescriptive analytics approach via machine learning and mathematical programming, Decision Support Systems 134 (2020) 113290.

\bibitem{Ramannavar2018}
M.~Ramannavar, N.~S. Sidnal, {A Proposed Contextual Model for Big Data Analysis Using Advanced Analytics}, in: Big Data Analytics: Proceedings of CSI 2015, 2018, pp. 329--339.

\bibitem{zhu2024optimal}
Y.~Zhu, I.~O. Ryzhov, Optimal data-driven hiring with equity for underrepresented groups, Production and Operations Management (2024) 10591478231224942.

\bibitem{Notz2022}
P.~M. Notz, R.~Pibernik, Prescriptive analytics for flexible capacity management, Management Science 68 (2022) 1756--1775.

\bibitem{Notz2023}
P.~M. Notz, P.~K. Wolf, R.~Pibernik, Prescriptive analytics for a multi-shift staffing problem, European Journal of Operational Research 305 (2023) 887--901.

\bibitem{Bischhoffshausen2015}
J.~K.~V. Bischhoffshausen, M.~Paatsch, M.~Reuter, G.~Satzger, H.~Fromm, An information system for sales team assignments utilizing predictive and prescriptive analytics, in: 2015 IEEE 17th Conference on Business Informatics, Vol.~1, 2015, pp. 68--76.

\bibitem{Brockett2019}
N.~Brockett, C.~Clarke, M.~Berlingerio, S.~Dutta, A system for analysis and remediation of attrition, in: Proceedings - 2019 IEEE International Conference on Big Data, Big Data 2019, 2019, pp. 2016--2019.

\bibitem{John2019}
I.~John, R.~Karumanchi, S.~Bhatnagar, Predictive and prescriptive analytics for performance optimization: Framework and a case study on a large-scale enterprise system, in: 2019 18th IEEE International Conference On Machine Learning And Applications (ICMLA), 2019, pp. 876--881.

\bibitem{Ceselli2018}
A.~Ceselli, M.~Fiore, A.~Furno, M.~Premoli, S.~Secci, R.~Stanica, Prescriptive analytics for mec orchestration, in: 2018 IFIP Networking Conference (IFIP Networking) and Workshops, 2018, pp. 1--9.

\bibitem{Ceselli2019}
A.~Ceselli, M.~Fiore, M.~Premoli, S.~Secci, Optimized assignment patterns in mobile edge cloud networks, Computers \& Operations Research 106 (2019) 246--259.

\bibitem{Brandt2021}
T.~Brandt, S.~Wagner, D.~Neumann, Prescriptive analytics in public-sector decision-making: A framework and insights from charging infrastructure planning, European Journal of Operational Research 291 (2021) 379--393.

\bibitem{Sinha2023}
A.~A. Sinha, S.~Rajendran, Study on facility location of air taxi skyports using a prescriptive analytics approach, Transportation Research Interdisciplinary Perspectives 18 (2023).

\bibitem{Li2019}
X.~Li, L.~Zhang, T.~Xiao, S.~Zhang, C.~Chen, Learning failure modes of soil slopes using monitoring data, Probabilistic Engineering Mechanics 56 (2019) 50--57.

\bibitem{Goyal2016}
A.~Goyal, E.~Aprilia, G.~Janssen, Y.~Kim, T.~Kumar, R.~Mueller, D.~Phan, A.~Raman, J.~Schuddebeurs, J.~Xiong, et~al., Asset health management using predictive and prescriptive analytics for the electric power grid, IBM Journal of Research and Development 60 (2016) 1--4.

\bibitem{stratigakos2024interpretable}
A.~Stratigakos, S.~Pineda, J.~M. Morales, G.~Kariniotakis, Interpretable machine learning for dc optimal power flow with feasibility guarantees, IEEE Transactions on Power Systems (2024).

\bibitem{chen2024towards}
X.~Chen, Y.~Liu, L.~Wu, Towards improving unit commitment economics: An add-on tailor for renewable energy and reserve predictions, IEEE Transactions on Sustainable Energy (2024).

\bibitem{Achenbach2018}
A.~Achenbach, S.~Spinler, Prescriptive analytics in airline operations: Arrival time prediction and cost index optimization for short-haul flights, Operations Research Perspectives 5 (2018) 265--279.

\bibitem{Ayhan2018}
S.~Ayhan, P.~Costas, H.~Samet, Prescriptive analytics system for long-range aircraft conflict detection and resolution, in: Proceedings of the 26th ACM SIGSPATIAL international conference on advances in geographic information systems, 2018, pp. 239--248.

\bibitem{Jacquillat2022}
A.~Jacquillat, Predictive and prescriptive analytics toward passenger-centric ground delay programs, Transportation Science 56 (2022) 265--298.

\bibitem{Birolini2023a}
S.~Birolini, A.~Jacquillat, Day-ahead aircraft routing with data-driven primary delay predictions, European Journal of Operational Research (2023).

\bibitem{Birolini2023b}
S.~Birolini, A.~Jacquillat, P.~Schmedeman, N.~Ribeiro, Passenger-centric slot allocation at schedule-coordinated airports, Transportation Science 57 (2023).

\bibitem{Tian2023a}
X.~Tian, R.~Yan, S.~Wang, G.~Laporte, Prescriptive analytics for a maritime routing problem, Ocean and Coastal Management 242 (2023).

\bibitem{Yan2023}
R.~Yan, S.~Wang, L.~Zhen, An extended smart “predict, and optimize” (spo) framework based on similar sets for ship inspection planning, Transportation Research Part E: Logistics and Transportation Review 173 (2023).

\bibitem{yang2024prescriptive}
Y.~Yang, R.~Yan, S.~Wang, Prescriptive analytics models for vessel inspection planning in maritime transportation, Computers \& Industrial Engineering 190 (2024) 110012.

\bibitem{oudani2023prescriptive}
M.~Oudani, A.~Sebbar, K.~Zkik, A.~Belhadi, A prescriptive analytics approach for port logistics planning, in: 2023 9th International Conference on Control, Decision and Information Technologies (CoDIT), IEEE, 2023, pp. 77--81.

\bibitem{Tian2023b}
X.~Tian, R.~Yan, Y.~Liu, S.~Wang, A smart predict-then-optimize method for targeted and cost-effective maritime transportation, Transportation Research Part B: Methodological 172 (2023).

\bibitem{mohd2022prescriptive}
M.~E. Mohd~Adha, M.~Masdi, H.~Hilmi, Prescriptive analytics for dynamic risk-based naval vessel maintenance decision-making, in: International Conference on Renewable Energy and E-mobility, Springer, 2022, pp. 345--359.

\bibitem{Amiel2023}
E.~Amiel, M.~Anastasopoulos, G.~Chevaleyre, A.~Consilvio, On applying artificial intelligence techniques to maximise passengers comfort and infrastructure reliability in urban railway systems, in: 2023 8th International Conference on Models and Technologies for Intelligent Transportation Systems (MT-ITS), 2023, pp. 1--6.

\bibitem{Consilvio2019}
A.~Consilvio, P.~Sanetti, D.~Anguita, C.~Crovetto, C.~Dambra, L.~Oneto, F.~Papa, N.~Sacco, Prescriptive maintenance of railway infrastructure: From data analytics to decision support, in: 2019 6th International Conference on Models and Technologies for Intelligent Transportation Systems (MT-ITS), 2019, pp. 1--10.

\bibitem{consilvio2024data}
A.~Consilvio, G.~Vignola, P.~L{\'o}pez~Ar{\'e}valo, F.~Gallo, M.~Borinato, C.~Crovetto, A data-driven prioritisation framework to mitigate maintenance impact on passengers during metro line operation, European Transport Research Review 16~(1) (2024) 6.

\bibitem{Grzegorowski2022}
M.~Grzegorowski, A.~Janusz, S.~Łażewski, M.~Świechowski, M.~Jankowska, Prescriptive analytics for optimization of fmcg delivery plans, in: International Conference on Information Processing and Management of Uncertainty in Knowledge-Based Systems, 2022, pp. 44--53.

\bibitem{Kandula2021}
S.~Kandula, S.~Krishnamoorthy, D.~Roy, A prescriptive analytics framework for efficient e-commerce order delivery, Decision Support Systems 147 (2021) 113584.

\bibitem{Kuzyakov2020}
O.~N. Kuzyakov, M.~A. Andreeva, Applying case-based reasoning method for decision making in iiot system, in: 2020 International Multi-Conference on Industrial Engineering and Modern Technologies, FarEastCon 2020, 2020, pp. 1--5.

\bibitem{Oberdorf2021}
F.~Oberdorf, N.~Stein, C.~M. Flath, Analytics-enabled escalation management: System development and business value assessment, Computers in Industry 131 (2021).

\bibitem{Stein2018}
N.~Stein, J.~Meller, C.~M. Flath, Big data on the shop-floor: sensor-based decision-support for manual processes, Journal of Business Economics 88 (2018) 593--616.

\bibitem{Vater2019}
J.~Vater, P.~Schamberger, A.~Knoll, D.~Winkle, Fault classification and correction based on convolutional neural networks exemplified by laser welding of hairpin windings, in: 2019 9th International Electric Drives Production Conference (EDPC), 2019, pp. 1--8.

\bibitem{Vater2020}
J.~Vater, M.~Kirschning, A.~Knoll, Closing the loop: Real-time error detection and correction in automotive production using edge-/cloud-architecture and a cnn, in: 2020 International Conference on Omni-layer Intelligent Systems (COINS), 2020, pp. 1--7.

\bibitem{Tham2023}
C.~K. Tham, N.~Sharma, J.~Hu, Model-based and model-free prescriptive maintenance on edge computing nodes, in: IEEE Vehicular Technology Conference, Vol. 2023-June, 2023, pp. 1--6.

\bibitem{Mohan2023}
S.~P. Mohan, S.~J. Nirmala, A prescriptive analytics approach for tool wear monitoring using machine learning techniques, in: ICSCCC 2023 - 3rd International Conference on Secure Cyber Computing and Communications, 2023, pp. 228--233.

\bibitem{karakaya2024sensor}
{\c{S}}.~Karakaya, M.~Yildirim, N.~Gebraeel, T.~Xia, A sensor-driven operations and maintenance planning approach for large-scale leased manufacturing systems, International Journal of Production Research (2024) 1--18.

\bibitem{Lee2022}
C.~Y. Lee, Y.~L. Lin, S.~H. Lin, T.~Yang, Virtual material quality investigation system, IEEE Transactions on Engineering Management (2022).

\bibitem{Thammaboosadee2018}
S.~Thammaboosadee, P.~Wongpitak, An integration of requirement forecasting and customer segmentation models towards prescriptive analytics for electrical devices production, in: 2018 International Conference on Information Technology (InCIT), 2018, pp. 1--5.

\bibitem{Rakhmasari2018}
A.~A. Rakhmasari, D.~Anwar, An analysis and design of a virtual collaboration information system of the jamu supply chain network based on a fair adaptive contract, in: Emerald Reach Proceedings Series, Volume 1: Proceedings of MICoMS 2017, Vol.~1, 2018, pp. 539--545.

\bibitem{Yu2021}
Y.~Yu, T.~Wang, Y.~Shi, Analytics for multiperiod risk-averse newsvendor under nonstationary demands, Decision Sciences (2021).

\bibitem{Mandl2021}
C.~Mandl, Prescriptive analytics for commodity procurement applications, in: N.~Trautmann, M.~Gnägi (Eds.), Operations Research Proceedings 2021, Selected Papers of the International Conference of the Swiss, German and Austrian Operations Research Societies (SVOR/ASRO, GOR e.V., ÖGOR), University of Bern, Switzerland, August 31 - September 3, 2021, Springer, 2021, pp. 27--32.

\bibitem{Mandl2023}
C.~Mandl, S.~Minner, Data-driven optimization for commodity procurement under price uncertainty, Manufacturing \& Service Operations Management 25 (2023) 371--390.

\bibitem{Cakir2023}
F.~Cakir, B.~W. Thomas, W.~N. Street, Rollout-based routing strategies with embedded prediction: A fish trawling application, Computers \& Operations Research 150 (2023) 106055.

\bibitem{Kumar2023}
R.~Kumar, K.~S. Sangwan, C.~Herrmann, R.~Ghosh, Development of a cyber physical production system framework for smart tool health management, Journal of Intelligent Manufacturing (2023).

\bibitem{Sangwan2023}
K.~S. Sangwan, R.~Kumar, C.~Herrmann, D.~K. Sharma, R.~Patel, Development of a cyber physical production system framework for 3d printing analytics, Applied Soft Computing 146 (2023).

\bibitem{reisch2023prescriptive}
R.~T. Reisch, L.~Janisch, J.~Tresselt, T.~Kamps, A.~Knoll, Prescriptive analytics-a smart manufacturing system for first-time-right printing in wire arc additive manufacturing using a digital twin, Procedia CIRP 118 (2023) 759--764.

\bibitem{Ribeiro2023}
R.~Ribeiro, A.~L. Pilastri, C.~Moura, J.~Morgado, P.~Cortez, A data-driven intelligent decision support system that combines predictive and prescriptive analytics for the design of new textile fabrics, Neural Computing and Applications 35 (2023) 17375--17395.

\bibitem{Suvarna2022}
M.~Suvarna, M.~I. Jahirul, W.~H. Aaron-Yeap, C.~V. Augustine, A.~Umesh, M.~G. Rasul, M.~E. Günay, R.~Yildirim, J.~Janaun, Predicting biodiesel properties and its optimal fatty acid profile via explainable machine learning, Renewable Energy 189 (2022).

\bibitem{Dash2022}
B.~B. Dash, S.~Banerjee, T.~Samant, T.~Swain, M.~K. Rath, Large scale follower recommendation in instagram, in: 2022 3rd International Conference for Emerging Technology, INCET 2022, 2022, pp. 1--6.

\bibitem{razgallah2024using}
H.~Razgallah, M.~Vlachos, A.~Ajalloeian, N.~Liu, J.~Schneider, A.~Steinmann, Using neural and graph neural recommender systems to overcome choice overload: Evidence from a music education platform, ACM Transactions on Information Systems 42~(4) (2024) 1--26.

\bibitem{hassanzadeh2024conclude}
A.~Hassanzadeh, M.~Hosseini, J.~G. Turner, How to conclude a suspended sports league?, Manufacturing \& Service Operations Management (2024).

\bibitem{Shi2020}
Y.~Shi, T.~Wang, L.~C. Alwan, Analytics for cross-border e-commerce: Inventory risk management of an online fashion retailer, in: Decision Sciences, Vol.~51, 2020, pp. 1347--1376.

\bibitem{Heide2020}
L.~M. van~der Heide, L.~C. Coelho, I.~F.~A. Vis, R.~G. van Anholt, Replenishment and denomination mix of automated teller machines with dynamic forecast demands, Computers \& Operations Research 114 (2020) 104828.

\bibitem{Keskin2022}
N.~B. Keskin, Y.~Li, J.~S. Song, Data-driven dynamic pricing and ordering with perishable inventory in a changing environment, Management Science 68 (2022).

\bibitem{Punia2020}
S.~Punia, S.~P. Singh, J.~K. Madaan, From predictive to prescriptive analytics: A data-driven multi-item newsvendor model, Decision Support Systems 136 (2020) 113340.

\bibitem{Adulyasak2023}
Y.~Adulyasak, O.~Benomar, A.~Chaouachi, M.~C. Cohen, W.~K. am~nuai, Using ai to detect panic buying and improve products distribution amid pandemic, AI and Society (2023).

\bibitem{Caro2019}
F.~Caro, F.~Babio, F.~Peña, Coordination of inventory distribution and price markdowns for clearance sales at zara, in: S.~Gallino, A.~Moreno (Eds.), Operations in an Omnichannel World, Vol.~8 of Springer Series in Supply Chain Management, Springer International Publishing, 2019, Ch.~13, pp. 311--339.

\bibitem{Chen2022}
X.~Chen, Z.~Owen, C.~Pixton, D.~Simchi-Levi, A statistical learning approach to personalization in revenue management, Management Science 68 (2022) 1923--1937.

\bibitem{Qu2020}
H.~Qu, I.~O. Ryzhov, M.~C. Fu, E.~Bergerson, M.~Kurka, L.~Kopacek, Learning demand curves in b2b pricing: A new framework and case study, Production and Operations Management 29 (2020).

\bibitem{Lash2016}
M.~T. Lash, K.~Zhao, Early predictions of movie success: The who, what, and when of profitability, Journal of Management Information Systems 33 (2016) 874--903.

\bibitem{Huang2019}
T.~Huang, D.~Bergman, R.~Gopal, Predictive and prescriptive analytics for location selection of add-on retail products, Production and Operations Management 28 (2019) 1858--1877.

\bibitem{Mehrotra2020}
P.~Mehrotra, L.~Pang, K.~Gopalswamy, A.~Thangali, T.~Winters, K.~Gupte, D.~Kulkarni, S.~Potnuru, S.~Shastry, H.~Vuyyuri, Price investment using prescriptive analytics and optimization in retail, in: Proceedings of the 26th ACM SIGKDD International Conference on Knowledge Discovery \& Data Mining, 2020, pp. 3136--3144.

\bibitem{han2024identifying}
S.~Han, L.~Chen, Z.~Su, S.~Gupta, U.~Sivarajah, Identifying a good business location using prescriptive analytics: Restaurant location recommendation based on spatial data mining, Journal of Business Research 179 (2024) 114691.

\bibitem{Hauser2021}
M.~Hauser, C.~M. Flath, F.~Thiesse, Catch me if you scan: Data-driven prescriptive modeling for smart store environments, European Journal of Operational Research 294 (2021) 860--873.

\bibitem{Ravi2021}
V.~Ravi, V.~Madhav, Optimizing the reliability of a bank with logistic regression and particle swarm optimization, in: Data Management, Analytics and Innovation: Proceedings of ICDMAI 2021, Volume 1, Vol.~70, 2021, pp. 91--107.

\bibitem{kurniawan2023prescriptive}
A.~Kurniawan, N.~Iriawan, A.~Choiruddin, A prescriptive analytics framework for banks' loan strategy development: Indonesian banks' case study, in: 2023 IEEE International Conference on Computing (ICOCO), IEEE, 2023, pp. 345--350.

\bibitem{Brandt2022}
T.~Brandt, O.~Dlugosch, A.~Abdelwahed, P.~L. van~den Berg, D.~Neumann, Prescriptive analytics in urban policing operations, Manufacturing \& Service Operations Management 24 (2022) 2463--2480.

\bibitem{Bahulkar2018}
A.~Bahulkar, N.~O. Baycik, T.~Sharkey, Y.~Shen, B.~Szymanski, W.~Wallace, Integrative analytics for detecting and disrupting transnational interdependent criminal smuggling, money, and money-laundering networks, in: 2018 IEEE International Symposium on Technologies for Homeland Security (HST), 2018, pp. 1--6.

\bibitem{Delen2021}
D.~Delen, H.~M. Zolbanin, D.~Crosby, D.~Wright, To imprison or not to imprison: an analytics model for drug courts, Annals of Operations Research 303 (2021) 101--124.

\bibitem{sariyer2024predictive}
G.~Sariyer, S.~K. Mangla, S.~Chowdhury, M.~E. Sozen, Y.~Kazancoglu, Predictive and prescriptive analytics for esg performance evaluation: A case of fortune 500 companies, Journal of Business Research 181 (2024) 114742.

\bibitem{Schwartz2017}
I.~M. Schwartz, P.~York, E.~Nowakowski-Sims, A.~Ramos-Hernandez, Predictive and prescriptive analytics, machine learning and child welfare risk assessment: The broward county experience, Children and Youth Services Review 81 (2017) 309--320.

\bibitem{osterwalder2010business}
A.~Osterwalder, Y.~Pigneur, Business Model Generation: a Handbook for Visionaries, Game Changers, and Challengers, Vol.~1, John Wiley \& Sons, 2010.

\bibitem{han2022data}
J.~Han, J.~Pei, H.~Tong, Data Mining: Concepts and Techniques, Morgan Kaufmann, 2022.

\bibitem{kochenderfer2019algorithms}
M.~J. Kochenderfer, T.~A. Wheeler, Algorithms for Optimization, Mit Press, 2019.

\bibitem{genova2011linear}
K.~Genova, V.~Guliashki, {Linear Integer Programming Methods and Approaches - A Survey}, Journal of Cybernetics and Information Technologies 11~(1) (2011).

\bibitem{danilova2022recent}
M.~Danilova, P.~Dvurechensky, A.~Gasnikov, E.~Gorbunov, S.~Guminov, D.~Kamzolov, I.~Shibaev, Recent theoretical advances in non-convex optimization, in: High-Dimensional Optimization and Probability: With a View Towards Data Science, Springer, 2022, pp. 79--163.

\bibitem{denning1978operational}
P.~J. Denning, J.~P. Buzen, The operational analysis of queueing network models, ACM Computing Surveys (CSUR) 10~(3) (1978) 225--261.

\bibitem{hanley2019more}
J.~A. Hanley, {A more intuitive and modern way to compute a small-sample confidence interval for the mean of a Poisson distribution}, Statistics in Medicine 38~(26) (2019) 5113--5119.

\bibitem{theng2024feature}
D.~Theng, K.~K. Bhoyar, Feature selection techniques for machine learning: a survey of more than two decades of research, Knowledge and Information Systems 66~(3) (2024) 1575--1637.

\bibitem{gosavi2015simulation}
A.~Gosavi, Simulation-Based Optimization, Springer, 2015.

\bibitem{harrison2010introduction}
R.~L. Harrison, {Introduction to Monte Carlo Simulation}, in: AIP Conference Proceedings, Vol. 1204, American Institute of Physics, 2010, pp. 17--21.

\bibitem{lex2014upset}
A.~Lex, N.~Gehlenborg, H.~Strobelt, R.~Vuillemot, H.~Pfister, Upset: Visualization of intersecting sets, IEEE Transactions on Visualization and Computer Graphics 20~(12) (2014) 1983--1992.

\bibitem{hansotia1980stochastic}
B.~J. Hansotia, Stochastic linear programming with recourse: A tutorial, Decision Sciences 11~(1) (1980) 151--168.

\bibitem{lundberg2017unified}
S.~M. Lundberg, S.-I. Lee, A unified approach to interpreting model predictions, Advances in Neural Information Processing Systems 30 (2017).

\bibitem{mothilal2020explaining}
R.~K. Mothilal, A.~Sharma, C.~Tan, Explaining machine learning classifiers through diverse counterfactual explanations, in: Proceedings of the 2020 conference on Fairness, Accountability, and Transparency, 2020, pp. 607--617.

\bibitem{sutton2020introduction}
R.~S. Sutton, A.~G. Barto, Introduction to Reinforcement Learning, 2nd Edition, MIT press Cambridge, 2020.

\bibitem{united2008international}
U.~N.~D. of~Economic, S.~A.~S. Division, International Standard Industrial Classification of All Economic Activities (ISIC), Rev. 4, United Nations Publications, 2008.

\bibitem{bengio2021machine}
Y.~Bengio, A.~Lodi, A.~Prouvost, Machine learning for combinatorial optimization: a methodological tour d’horizon, European Journal of Operational Research 290~(2) (2021) 405--421.

\bibitem{basu2013five}
A.~Basu, Five pillars of prescriptive analytics success, Analytics Magazine 8 (2013) 12.

\bibitem{thein2014apache}
K.~M.~M. Thein, {Apache Kafka: Next Generation Distributed Messaging System}, International Journal of Scientific Engineering and Technology Research 3~(47) (2014) 9478--9483.

\bibitem{chebotko2015big}
A.~Chebotko, A.~Kashlev, S.~Lu, {A Big Data Modeling Methodology for Apache Cassandra}, in: 2015 IEEE International Congress on Big Data, IEEE, 2015, pp. 238--245.

\bibitem{Ralphs2016}
T.~K. Ralphs, Y.~Shinano, T.~Berthold, T.~Koch, Parallel solvers for mixed integer linear programing, Handbook of Parallel Constraint Reasoning 74 (2016).

\bibitem{kreuzberger2023machine}
D.~Kreuzberger, N.~K{\"u}hl, S.~Hirschl, Machine learning operations (mlops): Overview, definition, and architecture, IEEE access 11 (2023) 31866--31879.

\bibitem{sculley2015hidden}
D.~Sculley, G.~Holt, D.~Golovin, E.~Davydov, T.~Phillips, D.~Ebner, V.~Chaudhary, M.~Young, J.-F. Crespo, D.~Dennison, Hidden technical debt in machine learning systems, Advances in neural information processing systems 28 (2015).

\bibitem{qin2011newsvendor}
Y.~Qin, R.~Wang, A.~J. Vakharia, Y.~Chen, M.~M. Seref, The newsvendor problem: Review and directions for future research, European Journal of Operational Research 213~(2) (2011) 361--374.

\bibitem{conn2009introduction}
A.~R. Conn, K.~Scheinberg, L.~N. Vicente, Introduction to Derivative-Free Optimization, SIAM, 2009.

\bibitem{snoek2012practical}
J.~Snoek, H.~Larochelle, R.~P. Adams, Practical bayesian optimization of machine learning algorithms, Advances in Neural Information Processing Systems 25 (2012).

\bibitem{forrester2009recent}
A.~I. Forrester, A.~J. Keane, Recent advances in surrogate-based optimization, Progress in Aerospace Sciences 45~(1-3) (2009) 50--79.

\bibitem{wolpert1997no}
D.~H. Wolpert, W.~G. Macready, No free lunch theorems for optimization, IEEE Transactions on Evolutionary Computation 1~(1) (1997) 67--82.

\bibitem{audet2017derivative}
C.~Audet, W.~Hare, Derivative-Free and Blackbox Optimization, Springer, 2017.

\bibitem{bandura1977social}
A.~Bandura, R.~H. Walters, Social Learning Theory, Englewood Cliffs Prentice Hall, 1977.

\end{thebibliography}

\end{document}